\documentclass[aps,pre,showpacs,twocolumn,superscriptaddress,amssymb]{revtex4-1}
\usepackage{graphicx,float,tabularx}   
\usepackage{amsmath}
\usepackage{ipa}
\usepackage{amssymb}
\usepackage{makeidx}
\usepackage{subfig}

\def\simge{\mathrel{%
    \rlap{\raise 0.511ex \hbox{$>$}}{\lower 0.511ex \hbox{$\sim$}}}}
\def\simle{\mathrel{
    \rlap{\raise 0.511ex \hbox{$<$}}{\lower 0.511ex \hbox{$\sim$}}}}

\newcommand \be{\begin{eqnarray}}
\newcommand \ee{\end{eqnarray}}

\newcommand{\tr}{{\rm tr}}
\newcommand{\ta}{\tilde{\alpha}}

\def\XXint#1#2#3{{\setbox0=\hbox{$#1{#2#3}{\int}$}
\vcenter{\hbox{$#2#3$}}\kern-.5\wd0}}

\begin{document}

\title{Signal from noise retrieval from one and two-point Green's function - comparison  }

\author{Zbigniew Drogosz} \email{zbigniew.drogosz@uj.edu.pl} \affiliation{M.
Smoluchowski Institute of Physics,  Jagiellonian University,
S. \L{}ojasiewicza 11, 30-348  Cracow, Poland}

\author{Jerzy Jurkiewicz }
\email{jerzy.jurkiewicz@uj.edu.pl} \affiliation{M. Smoluchowski Institute
of Physics and Mark Kac Center for Complex Systems Research,
Jagiellonian University, S. \L{}ojasiewicza 11, 30-348  Cracow, Poland}

\author{Grzegorz \L{}ukaszewski} \email{lukaszewski@th.if.uj.edu.pl} \affiliation{M.
Smoluchowski Institute of Physics,  Jagiellonian University, S. \L{}ojasiewicza 11, 30-348  Cracow, Poland
}

\author{Maciej A. Nowak}
\email{maciej.a.nowak@uj.edu.pl} \affiliation{M. Smoluchowski Institute
of Physics and Mark Kac Center for Complex Systems Research,
Jagiellonian University, S. \L{}ojasiewicza 11, 30-348  Cracow, Poland}

\date{\today}

\begin{abstract}  
We compare two methods of eigen-inference from large sets of data, based on the analysis of one-point and two-point Green's functions, respectively. Our analysis points at the superiority of eigen-inference based on one-point Green's function.  First, the applied by us method based  on Pad\'{e} approximants  is orders of magnitude faster comparing to the eigen-inference based on fluctuations (two-point Green's functions). Second, we  have identified the source of potential instability of the two-point Green's function method, as arising from the spurious zero and negative modes of the estimator  for  a variance operator of  the certain multidimensional Gaussian distribution, inherent for the two-point Green's function eigen-inference method.  Third, we have presented the cases of eigen-inference based on negative spectral  moments, for strictly positive spectra.   Finally, we have compared the cases of eigen-inference of  real-valued and complex-valued correlated Wishart distributions, reinforcing our conclusions  on an advantage of the one-point Green's function method.  
\end{abstract}

\maketitle
\section{Introduction}
In 1928, J. Wishart~\cite{WISHART}, studying the statistics of population dynamics,  has proposed a multidimensional generalization of the $\chi^2$ distribution. The Wishart matrix was a sample covariance matrix, constituting in this way the first historical ensemble of Gaussian random matrix theory. 
Explicitly, Wishart sample covariance matrix reads  $S=\frac{1}{T} XX^{\dagger}$, where   $X_{ia}$ is valued either in real or complex numbers space,   $i=1, ...N$ spans the size of the sample and $a=1,...T$ counts the number of measurements.  Since that time, Wishart ensemble (sometimes called also Laguerre ensemble) found broad applications in several branches of physics, ranging from chaotic  scattering~\cite{CHAOTIC}, through conductance in mesoscopic physics~\cite{MESOSCOPIC}, quantum information theory~\cite{RANEY}  to description of universal chiral properties of Quantum Chromodynamics~\cite{CHIRAL}.
With the  advent of computer era, original ideas of Wishart  met new challenges. Nowadays, speed of data acquisition  and feasibility  of massive data storage  caused that the sampled Wishart covariance matrices became huge, with $N$ and $T$ ranging easily from $10^3$ up to $10^9$.  This has triggered the need for new methodologies  going far beyond the classical multivariate statistical analysis. Random matrix theory  emerged as one of the most promising methods, since  the large dimensionality of the samples turned out to be actually beneficial,  due to the fast convergence of spectral properties of covariance matrices to the limiting distributions. This was secured by various central limit theorems,  exact in the limit when the dimension of matrices is infinite. This approach was strengthened by the development of free random variable calculus~\cite{VOICULESCU},  establishing the backbone of non-commutative (matrix-valued) probability calculus. Consequently,  random matrix analysis of huge samples has entered several new domains. 
In economy and financial engineering,  twinned papers ~\cite{BOUCHAUD,STANLEY}  launched massive response  for examining the role  of spectral properties of covariance matrix  for portfolio optimization~\cite{LIVAN,BOUCHAUDLACES,OURFIN}.  In telecommunication, papers~\cite{FOSCHINI,TELATAR} opened new theoretical and practical possibilities for Multiple Output Multiple Input wireless systems\cite{DEBBAHBOOK,TULINOBOOK}. Recently,  in genomics,  random matrix analysis gave hope to understand the mutation of the HIV virus~\cite{MITPAPER}. 

The estimation of true covariance from the measured data is a subtle problem. In  classical  papers~\cite{MP,BAISIL}, the authors have analyzed spectral properties of  the trivial true covariance matrix  
$\Sigma={\bf 1}_N$ in the limit when $N$ and $T$ are large, but their ratio (rectangularity $r \equiv N/T$) is fixed.  
The spectral distribution in this case is the unimodal (known) function,  located on the finite interval
$[l_-, l_+]$, where $l_{\pm}=(1\pm \sqrt{r})^2$. Only in the limit when $r$ approaches $0$, one can recover a Dirac delta-like  peak located at 1. So  the finiteness  of the number $T$ of measured samples  always introduces the distortion of the original spectrum of the true covariance matrix.   The second problem of   spectral inference comes from the statistical nature of random matrix ensembles. In many practical realizations, we do not have at our disposal
the whole ensemble of  estimators of the covariance matrix, but we have only a single measurement (although represented  usually by a large matrix). An obvious example is a particular stock market, when one cannot perform the averages over different realizations.  Both the above problems are interlinked, and the separation of the true signal from a noisy, single object represents a formidable challenge, addressed in numerous works~\cite{NUMEROUS}. Generally, majority of the methods of inference are related to the  analysis of {\it mean} spectral functions $\rho(\lambda)$, where $\lambda$ are eigenvalues of the estimator $S$. Recently,  a method based on the analysis of spectral {\it fluctuations} has been proposed~\cite{RAO,MIT}, where the central role is played by the two-point spectral function $\rho(\lambda, \lambda^{'})$.   In this work, we critically examine both methods and compare their computational efficiency. In Section~II,  we define basic one-point and two-point objects generating spectral single and double moments, establishing also the notation used for the whole of the paper.  In Section~III, we explain the methods of eigen-inference, based on previously defined quantities. Additionally,   
we introduce inverse spectral moments, which have been discussed in the literature only marginally. Section~IV presents results of several numerical studies comparing an analytical  method to a statistical one. Section~V discusses the speed of the algorithms and their accuracy. Section~VI identifies and elucidates some potential pitfalls of the statistical method.  Short section~VII discusses the similarities and differences  of the applied algorithms as a function  of  the  parameter $\beta$:   $\beta=2$ in the case of the  complex Wishart ensemble and $\beta=1$ in the case of the real Wishart ensemble.   Finally, section~VIII represents conclusions and recommendations and discusses further possibilities for improvement of eigen-inference.  We also explicitly highlight several novel aspects of our work. The paper is concluded with the  three appendices, explaining some technicalities  helpful  when   following  our analysis.  
\section{One and two-point Green's functions  for the Wishart ensemble.}
We recall main definitions and formulae for the case of the complex Wishart ensemble.  We populate  matrix $X_{ia}$, where $i=1, \ldots, N$ and $a=1, \ldots, T$ with independent, complex centered, Gaussian-distributed numbers for each pair $(i, a)$.   Complex-valued one-point Green's function $G_S(z)$ for the Wishart ensemble  $S=\frac{1}{T} XX^{\dagger}$  is defined as 
\be
G_S(z) =\frac{1}{N} \left<\tr \frac{1}{z{\bf{1}}_N -S}\right>
\label{1point}
\ee
where $<\ldots>$ denotes average with the respect of the Gaussian joint probability distribution. 
Discontinuities of the Green's function yield, on the basis of the Sochocki-Plemelj formula, the spectral distribution
\be
\rho_S(\lambda)=- \frac{1}{\pi} \lim_{\epsilon \rightarrow 0} \Im G_S(z)|_{z=\lambda +i\epsilon} 
\ee
At the vicinity of the point $z=\infty$ Green's function  serves as a generating function for spectral moments of the Wishart ensemble (defined as  $\alpha^S_i=\int_L \rho_S(\lambda)\lambda ^i d\lambda$, where $L$ denotes the support of eigenvalues $\lambda$): 
\be
G_S(z)=\sum_{i=0}^{\infty} \frac{1}{z^{i+1}} \left<  \frac{1}{N} \tr  S^i \right> \equiv \sum_{i=0}^{\infty} \frac{\alpha^S_i}{z^{i+1}} 
\label{1pointmom}
\ee
We consider the case  $ N, T \rightarrow \infty$ and $r\equiv N/T <1$.  Green's function for the Wishart ensemble is given in this case as 
\be
G_S(z)=\frac{1}{2rz}[r+z-1-\sqrt{(z-s_-)(z-s_+)}]
\label{greenpasmar}
\ee
where $s_{\pm}=(1\pm \sqrt{r})^2$ denote the ends of the eigenvalue spectrum. 
Corresponding spectral function is given by the Marcenko-Pastur formula
\be
\rho(\lambda)=\frac{1}{2 \pi r \lambda} \sqrt{(\lambda-s_-)(s_+-\lambda)}
\label{marpas}
\ee

For completeness we mention that the case $r > 1$ has the same spectral distribution, modulo $T-N$ trivial zero modes. The case $r=1$ is also singular at $z=0$ as visible from~(\ref{greenpasmar}). Since now we consider only the cases $r<1$, when the eigenvalues  are strictly positive, so $S^{-1}$ does exist.  

Note that since the spectrum of  the considered by us case of the asymptotic Wishart ensemble is strictly positive, we can define also {\it inverse spectral moments}, which we call dual moments.  Simple algebraic manipulation shows that the generating function for such moments $G_{S^{-1}}(1/z)$   can  be rephrased as expansion around $z=0$ for the Green's function $G_S(z)$, i.e. 
\be
G_{S^{-1}}(\frac{1}{z})&=& \left< \frac{1}{N}  {\rm tr}  \frac{1}{\frac{1}{z}{\bf 1}_N-S^{-1}} \right>  \nonumber \\
&=& z\left< \frac{1}{N} {\rm tr} \,  {\bf 1}_N \left(1-\frac{z}{z{\bf 1}_N-S}\right) \right> \nonumber \\ &=& z(1-z G_S(z))
\label{1greeninv}
\ee

One can easily generalize the definition of one-point Green's function for the case of two-point Green's  function, defined as 
\be
G_S(z,w)=\frac{1}{N^2} \left< \tr \frac{1}{z{\bf 1}_N- S} \tr \frac{1}{w{\bf 1}_N-S} \right>_c
\label{2pointgreen}
\ee
where  the subscript $c$ denotes the connected part, defined as  $<AB>_c \equiv <AB>-<A><B>$ for any $A, B$.  In analogy to the case of one-point Green's function, a double expansion in $z$ and $w$ around infinity yields double spectral moments  $\alpha^S_{i,j}=<\frac{1}{N} \tr S^i \frac{1}{N} \tr S^j>_c$.  Continuing the analogy, the properly taken discontinuities of the two-point Green's functions give the two-point spectral density function, yielding  the probability of finding  a pair of eigenvalues $\lambda, \lambda^{'}$, where the separation between the eigenvalues is of order $O(N^0)$ (so-called "wide" correlator). Explicitly
\be
\rho_c(\lambda, \lambda^{'})=-\frac{1}{4 \pi^2}(G_{++}-G_{+-}+G_{--}-G_{-+})
\ee
where the shorthand notation reflects double use of Sochocki-Plemejl formula, e.g.
$G_{+-}=\lim_{\epsilon \rightarrow 0} \lim_{\epsilon^{'} \rightarrow 0} G(z=\lambda +i \epsilon, w=\lambda^{'}-i\epsilon^{'})$. 

This quantity should not be confused which so-called universal (microscopic) kernel, representing similar function when the spacing between the eigenvalues is of order $1/N$.  
Remarkably, double spectral moments can be expressed in terms of usual spectral moments, which allows to infer 
the information on the spectral density $\rho(\lambda)$ from two-point Green's functions as well. 
This is a consequence of so-called AJM universality~\cite{AMBJURMAK}. In particular, in the case of Wishart ensemble the exact relation between two-point and one-point Green's function reads~\cite{BAISIL,MINGO,JURLUKNOW}
\be
G_S(z,w)&=&\frac{1}{N^2}\partial_z \partial_w \ln \left[  \frac{G_S(w)-G_S(z)}{z-w} \right] 
\nonumber \\
&=&\frac{1}{N^2} \left[\frac{\partial_zG_S (z) \partial_wG_S(w)}{[G_S(z)-G_S(w)]^2} -\frac{1}{(z-w)^2}  \right]
\label{2greenpasmar}
\ee
where the  second line comes after explicit differentiation of the first  formula and the corresponding Green's functions and their derivatives origin from~({\ref{greenpasmar}). 
Similarly, one can define two-point Green's function for the inverse matrix $S$, generating {\it double dual  spectral moments}   $\alpha^{S^{-1}}_{i,j}=<\frac{1}{N} \tr S^{-i} \frac{1}{N} \tr S^{-j}>_c$. 
Algebraic manipulations analogous to those we used  in (\ref{1greeninv}) lead to relation
\be
G_{S^{-1}}(1/z,1/w)=z^2w^2 G_S(z,w)
\label{2greenpasmarinv}
\ee
Combining the above expression  with the AJM universality we arrive at the explicit formula generating  double dual  spectral moments in terms of single dual spectral moments. 
We summarize   this section with the definitions of  following four generating functions for single moments, dual moments, double moments and dual double moments, respectively:
\be
M_S(1/x)&=&  \sum_{i=1} \alpha_i^S x^i \nonumber \\
 M_S(1/x,1/y)&= &\sum_{i,j=1}\alpha^S_{i,j} x^i y^j \nonumber \\ 
M_{S^{-1}}(1/x)&=&\sum_{i=1} \alpha_i^{S^{-1}} x^i \nonumber \\
 M_{S^{-1}}(1/x,1/y)&=&\sum_{i,j=1}\alpha^{S^{-1}}_{i,j} x^i y^j 
 \ee
 where
 \be
 xG_S(x)-1&=&M_S(x) \nonumber \\
 xG_{S^{-1}}(x)- 1&=&M_{S^{-1}}(x) \nonumber \\
 xyG_S(x,y)&=& M_S(x,y) \nonumber \\
 xyG_{S^{-1}}(x,y)&=& M_{S^{-1}}(x,y)
 \ee
 The definition of   $M_S(x,y)$ is  identical to the one introduced in~\cite{RAO,MIT}, for the purpose of easier  comparison of the results. As far as we know,  the broad  analysis of dual moments, both single and double, have not yet been published~\cite{PHDLUKA, DROGOSZ}, except for a  brief analysis of inverse moments in~\cite{JUREKZDZICH}.

\section{Signal retrieval from one-point and two-point  Green's functions}
We define an empirical covariance matrix $S=\frac{1}{T}XX^{\dagger}$, where $X_{ij}$  are standartized measurements.  The main goal is the eigen-inference, i.e. the extraction from the measured matrix $S$ of the "true", but unknown  spectral information on covariance matrix $\Sigma$.  Since now we  concentrate on hermitian matrices,  and we will later make a comment on the application of our formalism for the real ones.  Since matrix $\Sigma$ is hermitian,  and therefore diagonalizable  by the unitary  transformation $\Sigma = U\Lambda U^{\dagger}$, we parametrize unknown matrix $\Lambda$ as 
block-diagonal $\Lambda={\rm diag} (\Lambda_1 {\bf 1}_{n_1},\Lambda_2 {\bf 1}_{n_2}, \ldots, \Lambda_{m_{max}} {\bf 1}_{m_{max}})$, where $\sum_{n=1}^{n_{max}} n_i =N$.  
We reserved capital Greek letters for denoting the  eigenvalues of the "true" covariance matrix, whereas lowercase Greek letters denote the eigenvalues of the empirical estimator $S$. 
This means that we seek $m_{max}$ eigenvalues, each one with the multiplicity $n_i$. It is convenient to define the vector of the above spectral parameters as
\be
\Theta=(\Lambda_1, \ldots, \Lambda_{m_{max}}, p_1, \ldots, p_{m_{max}-1})
\label{Theta}
\ee
where $p_i=n_i/N$ with the obvious constraint $\sum_i p_i=1$.
\subsection{ Analytical estimator for single moments}
The cornerstone of the analytic method is the conformal mapping~\cite{JUREKZDZICH} between the generating functions for the "true" moments of matrix $\Sigma$ and the measured moments of the "estimator" $S$,
\be
M_S(z)=M_{\Sigma}(Z)
\ee
where $Z$ is related  to $z$ by 
\be
Z=\frac{1}{1+rM_S(z)}
\label{conformal}
\ee
The origin of conformal mapping is briefly explained in Appendix~A.  
Explicitly, 
\be
\sum_{k=1}^{\infty} \frac{\alpha_k^S}{z^k}=\sum_{k=1}^{\infty} \frac{\alpha^{\Sigma}_k}{z^k}\left(1+r\sum_{l=1}^{\infty} \frac{\alpha_l^S}{z^l}\right)^k
\ee
 Iteration of the above formula allows one to write down an infinite tower of algebraic relations between moments $\alpha_i^S$ and moments $\alpha_j^{\Sigma}$
 \be
 \alpha_1^S &=& \alpha_1^{\Sigma} \nonumber \\
 \alpha_2^S&=&\alpha_2^{\Sigma} +r (\alpha_1^{\Sigma})^2 \nonumber \\
 \alpha_3^S&=&\alpha_3^{\Sigma} +3r \alpha_1^{\Sigma}\alpha_2^{\Sigma} +r^2 (\alpha_1^{\Sigma})^3 \nonumber \\
 \ldots
 \label{moments}
 \ee
One can rephrase as well moments of $\Sigma$  in terms of moments of $S$, using backward iteration
 \be
 \alpha_1^{\Sigma} &=& \alpha_1^S \nonumber \\
 \alpha_2^{\Sigma}&=&\alpha_2^S -r (\alpha_1^S)^2 \nonumber \\
 \alpha_3^{\Sigma}&=&\alpha_3^S -3r \alpha_1^S\alpha_2^S +2 r^2 (\alpha_1^S)^3 \nonumber \\
 \ldots
 \label{backmoments}
 \ee
The algorithm of eigen-inference is as follows. 
First, we truncate the infinite tower of relations at some $K_{max}$. We calculate $K_{max}$ empirical moments and, using the above formulae, we rephrase them  in terms of 
moments $\alpha_i^{\Sigma}$, with $i=1, \ldots, K_{max}$. By definition, the unknown generating function for $\Sigma$ can be expressed   in terms of vector $\Theta$, 
\be
zG_{\Sigma}(z)=z\sum_{i=1}^{m_{max}}\frac{p_i}{z-\Lambda_i}=\sum_{i=1}^{m_{max}}\frac{p_i}{1-x\Lambda_i}
\ee
where we have introduced $x=1/z$. 
Second, we note, that by construction the above estimator is  the ratio of two polynomials in $x$, numerator $A_{m_{max}-1}(x)$ of order $m_{max}-1$ and denominator
$B_{m_{max}}(x)$ of order $m_{max}$. 
As the next step, we make an assumption on the value of $m_{max}$, and we approximate $zG_{\Sigma}(z)$ with the help of Pad\'{e} approximant ({\it cf.} Appendix~C), ideally suited  for an approximation of the unknown functions being the ratios of polynomials of fixed and known order.  Note that $K_{max}=2m_{max}-1$. 
Third,  we read from the approximant  the parameters of the vector $\Theta$. Eigenvalues $\Lambda_i$ correspond  to the poles  of the Green's function, therefore  corresponding to the inverse of the  zeroes of the denominator $B_{m_{max}}(x)$. Multiplicities $p_i$ correspond to the residues of the Green's functions, so can be easily found from the relation
\be
p_i=-\frac{1}{x}\frac{A(x)}{B'(x)}|_{x=\frac{1}{\Lambda_i}}
\ee
Lastly, we repeat the above procedure for other guesses of  $m_{\max}$ in order to choose the best estimator $\Theta$. 
If our guess is too small, we usually obtain eigenvalue estimations that are
between the ``true'' eigenvalues, as a kind of an average.
If, on the other hand, the tested $m_{max}$ is greater than the number of
different eigenvalues of the ``true'' spectrum, than
there appear spurious eigenvalue estimations,
 which 
either have an incorrect real value and a very small probability,
or are created in pairs while a real eigenvalue splits
into two complex conjugate eigenvalues with complex conjugate probabilities.
In most cases,
the choice of the best $m_{max}$ should be clear.

The above procedure is very fast, as we demonstrate on several examples presented in the following section. 

We can perform similar eigen-inference from the dual moments. The corresponding algorithm is as follows:
First, we calculate matrix $S^{-1}$, then moments $\alpha_{-k}^S$, and finally moments $\alpha_{-k}^{\Sigma}$, combining ({\ref{conformal}) with (\ref{1greeninv}), i.e  
\be
 \alpha_{-1}^{\Sigma} &=& (1-r) \alpha_1^S \nonumber \\
 \alpha_{-2}^{\Sigma}&=&(1-r)^2\alpha_{-2}^{S} -r(1-r) (\alpha_{-1}^{S})^2 \nonumber \\
 \alpha_{-3}^{\Sigma}&=&(1-r)^3\alpha_{-3}^S -3r (1-r)^2\alpha_{-1}^S \alpha_{-2}^S +r^2(1-r) (\alpha_{-1}^S)^3 \nonumber \\
 \ldots
 \label{invmoments}
 \ee
 Few lowest backward relations read
\be
 \alpha_{-1}^S &=&  \frac{1}{1-r}\alpha_{-1}^{\Sigma} \nonumber \\
 \alpha_{-2}^S&=&\frac{1}{(1-r)^2} \alpha_{-2}^{\Sigma} +\frac{r}{(1-r)^3} (\alpha_{-1}^{\Sigma})^2 \nonumber \\
 \alpha_{-3}^S&=&\frac{1}{(1-r)^3}\alpha_{-3}^{\Sigma} +\frac{3r}{(1-r)^4} \alpha_{-1}^{\Sigma}\alpha_{-2}^{\Sigma} +\frac{2r^2}{(1-r)^5} (\alpha_{-1}^{\Sigma})^3 \nonumber \\
 \ldots
 \label{invbackmoments}
 \ee 

Second, we calculate $zG_{\Sigma^{-1}}(1/z)$, and, assuming $K_{max}$, we get the Pad\'{e} approximant.  Finally, we infer the eigenvalues as zeroes of the denominator of 
Pad\'{e} approximant, and multiplicities as the corresponding residua.


\subsection{Statistical  estimator for double moments}
Let us consider an infinite vector of fluctuations for the matrix ensemble $S$, whose components are  defined as follows 
\be
v_j= {\rm tr} S^j-<{\rm tr} S^j>={\rm tr} S^j-N\alpha_j^S
\ee
The cornerstone of the statistical estimator is represented by the  following theorem~\cite{BAISIL,RAO,MIT}: The statistical distribution of vector $v$ is represented by the multidimensional Gaussian ensemble, where the elements of the dispersion matrix $Q$ are given by the   corresponding  double moments $\alpha_{ij}^S$. For sample estimator $\Theta$, we can write therefore probability distribution function (hereafter pdf)  for vector $v$ as 
\be 
f(v_{\Theta}) \sim \frac{1}{{\rm det} Q_{\theta}}
\exp -v_{\Theta}^{\dagger} Q^{-1}_{\Theta}  v_{\Theta}
\label{multigauss}
\ee
The maximum likelihood principle tells us that the desired estimator $\Theta$ is the maximizer  of  the pdf. Technically, it is easier to maximize the logarithm of the above  expression, since the logarithm is a  monotonic function  and is well defined due to the positivity of the pdf. Therefore the optimal estimator $\Theta$ is the minimizer of the following function 
\be
g_{\Theta}=v_{\Theta}^{\dagger} Q^{-1}_{\Theta}  v_{\Theta} +\ln {\rm det} Q_{\Theta}
\label{gmini}
\ee
 Note that double moments are not measured in an explicit way. We have however AJM universality~(\ref{2greenpasmar}), which allows us to express them in terms of single moments, which are directly related to the measurement.  Resulting formulae are lengthy due to the tangled  relation~(\ref{2greenpasmar}), but  can be easily generated numerically, as collected in the Appendix~B. 
For the simplest case of 2 by 2 covariance matrix $Q$   relations are as follows
\be
\alpha_{11}&=&-\alpha_1^2+\alpha_2 \nonumber \\
\alpha_{12}&=&\alpha_{21}=2\alpha_1^3-4\alpha_1 \alpha_2 +2\alpha_3 \nonumber \\ 
\alpha_{22}&=&-6\alpha_1^4 +16\alpha_1^2\alpha_2 -6 \alpha_2^2 -8\alpha_1 \alpha_3 + \alpha_4
\label{doublesingle}
\ee
where, for clarity, we have suppressed the index $S$.  Appendix~B lists higher double moments, up to 
$\alpha_{55}$. 

Finally, we note that similar construction can be performed for the dual double  moments.
For the simplest case of the 2 by 2 dispersion matrix, relation~(\ref{2greenpasmarinv}) yields
\be
\ta_{11} \ta_2^2&=&-\ta_3^2+\ta_2 \ta_4 \nonumber \\
\ta_{12} \ta_2^3&=&\ta_{21} \ta_2^3=2\ta_3^3-4\ta_2 \ta_3 \ta_4 +2\ta_2^2 \ta_5 \nonumber \\
\ta_{22} \ta_2^4&=&4\ta_2^3\ta_6-6\ta_3^4 +16\ta_2 \ta_3^2 \ta_4 -8 \ta_2^2 \ta_3 \ta_5 -6\ta_2^2 \ta_4^2 \nonumber \\
\label{doublesingleinv}
\ee
where, for clarity, we have suppressed the   index $S^{-1}$ and denoted  dual  moments by tilde, to avoid confusion with the relation (\ref{doublesingle}).  Higher double dual moments (up to $\ta_{55}$) are listed in the Appendix~B. 
 
\section{Data inference - analytical versus statistical method}

To make sure that the methods were implemented correctly, they were
tested on several ensembles of matrices that had  already been studied by
Rao, Mingo, Speicher and Edelman (Table 7 of \cite{MIT}).

\begin{table*}[ht]
\footnotesize
\caption{Comparison with the results in \cite{MIT}} \label{tab:aaa}
\begin{tabular}{|l|l|l|l|l|l|l|l|l|l|l|}
\hline
\multicolumn{11}{|l|}{Explanation of the symbols:}\\
\multicolumn{11}{|l|}{A/S - the analytical/ statistical method}\\
\multicolumn{11}{|l|}{AD - the analytical dual method}\\ 
\multicolumn{11}{|l|}{3x3 - the size of the matrix Q used }\\
\multicolumn{11}{|l|}{RMSE - results from the article \cite{MIT}}\\
\multicolumn{11}{|l|}{q - subjective assessment of the quality of the estimation (1 - 4 stars)}\\
\multicolumn{11}{|l|}{n - number of matrices with all parameters estimated as positive real}\\
\multicolumn{11}{|l|}{$\langle \dots \rangle$ - arithmetical mean of the estimations}\\
\multicolumn{11}{|l|}{$\sigma( \dots )$ - standard deviation of the estimations}\\
\multicolumn{11}{|l|}{$\eta$ - a parameter measuring the quality of the estimation (less is better)}\\
\multicolumn{11}{|l|}{n.d. - no data}\\
\hline
method & q & n & $\langle \lambda_1 \rangle $ & $ \sigma (\lambda _1)$ & $\langle \lambda_2 \rangle $ & $ \sigma (\lambda _2)$ & $\langle p_1^S \rangle $ & $ \sigma (p_1^S)$ & $\eta$ & time [s]\\
\hline
\multicolumn{11}{|c|}{100 matrices $80 \times 40$, $\Lambda_1=2$, $\Lambda_2=1$, $p_1=1/2$} \\
\hline
A   & ** & 96 & 2.1036 & 0.5307 & 0.6607 & 0.6757 & 0.5572 & 0.3208 & 0.8599 & 0.14\\
\hline
AD  & * & 0 & - & - & - & - & - & - & - & 0.14 \\
\hline
S 3x3  & ** & 100 & 2.0377 & 0.4955 & 0.6969 & 0.5212 & 0.5860 & 0.3163 & 0.7487 & 1076.9\\
\hline
S 3x3 RMSE & ** & n.d. & 2.0692 & 0.4968 & 0.7604 & 0.4751 & 0.5624 & 0.2965 & n.d. & n.d \\
(1000 matrices) &&&&&&&&&& \\
\hline
\multicolumn{11}{|c|}{100 matrices $ 320 \times 160$, $\Lambda_1=2$, $\Lambda_2=1$, $p_1=1/2$} \\
\hline
A & *** & 100 & 2.0179 & 0.1513 & 0.9698 & 0.1484 & 0.4940 & 0.1307 & 0.2426 & 0.13\\
\hline
AD & * & 0 & - & - & - & - & - & - & - & 0.14 \\
\hline
S 3x3 & *** & 100 & 2.0117 & 0.1499 & 0.9654 & 0.1496 & 0.5105 & 0.1307 & 0.2425 & 387.7\\
\hline
S 3x3 RMSE&  *** & n.d. & 2.0089 & 0.1398 & 0.9763 & 0.1341 & 0.5076 & 0.1239 & n.d. & n.d. \\
(1000 matrices) &&&&&&&&&& \\
\hline
\multicolumn{11}{|c|}{100 matrices $80 \times 82$, $\Lambda_1=2$, $\Lambda_2=1$, $p_1=1/2$} \\
\hline
A   & *** & 100 & 1.9698 & 0.2324 & 0.8819 & 0.2519 & 0.5671 & 0.1901 & 0.3759 & 0.16\\
\hline
AD  & * & 0 & - & - & - & - & - & - & - & 0.16 \\
\hline
S 3x3 & *** & 100 & 1.9461 & 0.2233 & 0.8630 & 0.2576 & 0.5834 & 0.1886 & 0.3738& 584.9\\
\hline
S 3x3 RMSE & *** & n.d. & 2.0021 & 0.2273 & 0.9287 & 0.2323 & 0.5310 & 0.1856 & n.d. & n.d. \\
(1000 matrices &&&&&&&&&& \\
$80 \times 80$) &&&&&&&&&&\\
\hline
\multicolumn{11}{|c|}{100 matrices $320 \times 322$, $\Lambda_1=2$, $\Lambda_2=1$, $p_1=1/2$} \\
\hline
A & *** & 100 &2.0119 & 0.0541 & 1.0065 & 0.0452 & 0.4916 & 0.0473 & 0.0826 & 0.13\\
\hline
AD & * & 0 & - & - & - & - & - & - & - & 0.16\\
\hline
S 3x3 & *** & 100 & 2.0101 & 0.0540 & 1.0055 & 0.0453 & 0.4929 & 0.0473 & 0.0826 & 779.4\\
\hline
S 3x3 RMSE & *** & n.d. & 2.0001 & 0.0548 & 0.9960 & 0.0469 & 0.5024 & 0.0492 & n.d. & n.d.\\
(1000 matrices &&&&&&&&&& \\
$320 \times 320$) &&&&&&&&&&\\
\hline
\multicolumn{11}{|c|}{100 matrices $80 \times 160$, $\Lambda_1=2$, $\Lambda_2=1$, $p_1=1/2$} \\
\hline
A & *** & 100 & 2.0110 & 0.0937 & 0.9977 & 0.0777 & 0.4977 & 0.0798 & 0.1407 & 0.14    \\
\hline
AD & * & 89 & - & - & - & - & - & - & - & 0.13\\
\hline
S 3x3 & *** & 100 & 2.0023 & 0.0924 & 0.9936 & 0.0783 & 0.5030 & 0.7990 & 0.1403 & 1188.3 \\
\hline
S 3x3 RMSE & *** & n.d. & 1.9925 & 0.0975 & 0.9847 & 0.0781 & 0.5116 & 0.0807 & n.d. & n.d.\\
\hline
\multicolumn{11}{|c|}{100 matrices $320 \times 640$, $\Lambda_1=2$, $\Lambda_2=1$, $p_1=1/2$} \\
\hline
A & *** & 100 & 2.0017 & 0.0220 & 1.0027 & 0.0180 & 0.4974 & 0.0192 & 0.0331 & 0.16\\
\hline
AD & * & 100 & 2.5733 & 1.8594 & 0.3746 & 6.1721 & 0.4918 & 0.2205 & 6.1437 & 0.17\\
\hline
S 3x3 & *** & 100 & 2.0011 & 0.0220 & 1.0024 & 0.0180 & 0.4978 & 0.0192 & 0.0331 & 1385.6 \\
\hline
S 3x3 RMSE & ***& n.d. & 1.9994 & 0.0232 & 0.9993 & 0.0178 & 0.5008 & 0.0193 & n.d. & n.d.\\
(1000 matrices) &&&&&&&&&& \\
\hline

\end{tabular}
\end{table*}

\begin{table*}[ht]
\footnotesize
\caption{Comparison of the methods of eigen-inference.}
\begin{tabular}{|l|l|l|l|l|l|l|l|l|l|l|}
\hline
\multicolumn{11}{|l|}{Explanation of the symbols:}\\
\multicolumn{11}{|l|}{A/S - the analytical/ statistical method}\\
\multicolumn{11}{|l|}{AD/SD - the analytical dual/ statistical dual method}\\ 
\multicolumn{11}{|l|}{3x3/4x4 - the size of the matrix Q used}\\
\multicolumn{11}{|l|}{w - with a starting point from the analytical method}\\
\multicolumn{11}{|l|}{q - subjective assessment of the quality of the estimation (1 - 4 stars)}\\
\multicolumn{11}{|l|}{n - number of matrices with all parameters estimated as positive real}\\
\multicolumn{11}{|l|}{$\langle \dots \rangle$ - arithmetical mean of the estimations}\\
\multicolumn{11}{|l|}{$\sigma( \dots )$ - standard deviation of the estimations}\\
\multicolumn{11}{|l|}{$\eta$ - a parameter measuring the quality of the estimation (less is better)}\\
\hline
method & q & n & $\langle \lambda_1 \rangle $ & $ \sigma (\lambda _1)$ & $\langle \lambda_2 \rangle $ & $ \sigma (\lambda _2)$ & $\langle p_1^S \rangle $ & $ \sigma (p_1^S)$ & $\eta$ & time [s]\\
\hline
\multicolumn{11}{|c|}{100 matrices $126 \times 180$, $\Lambda_1=0.5$, $\Lambda_2=1$, $p_1=1/3$} \\
\hline
A          &*** & 100 & 0.4910 & 0.0696 & 1.0026 & 0.0437 & 0.3341 & 0.0976 & 0.1253 & 0.13\\
\hline
AD          &* &60 & - & - & - & - & - & - & - & 0.14 \\
\hline
S 3x3       &*** & 100 & 0.4879 & 0.0700 & 0.9998 & 0.0431 & 0.3292 & 0.0968 & 0.1247 & 2336.1\\
\hline
S 3x3 w &*** & 100 & 0.4876 & 0.0700 & 0.9996 & 0.0431 & 0.3288 & 0.0968 & 0.1246 & 243.6\\
\hline
S 4x4      &**** & 100 & 0.5310 & 0.0521 & 1.0059 & 0.0354 & 0.3452 & 0.0762 & 0.0967 & 6603.5\\
\hline
S 4x4 w&**** & 100 & 0.5038 & 0.0521 & 1.0057 & 0.0353 & 0.3449 & 0.0761 & 0.0967 & 419.0\\
\hline
SD 3x3     &* & 100 & 0.1701 & 0.1999 & 2.3348 & 1.4128 & 0.0935 & 0.2406 & 1.4107 & 2048.7\\
\hline
SD 3x3 w& **& 100 & 0.5447 & 0.2018 & 64.052 & 80.604 & 0.6052 & 0.3644 & 80.200 & 214.5\\
\hline
\multicolumn{11}{|c|}{100 matrices $90 \times 9000$, $\Lambda_1=0.5$, $\Lambda_2=1$, $p_1=1/3$} \\
\hline
A         & ****& 100 & 0.5001 & 0.0010 & 1.0000 & 0.0015 & 0.3333 & 0.0012 & 0.0016 & 0.13 \\
\hline
AD        & ****& 100 & 0.5001 & 0.0009 & 1.0000 & 0.0015 & 0.3334 & 0.0007 & 0.0015 & 0.14 \\
\hline
S 3x3   & *& 100 & 0.4474 & 0.2007 & 0.9590 & 0.2102 & 0.3610 & 0.3317 & 0.3485 & 895.4\\
\hline
S 3x3 w & ***& 100 & 0.5003 & 0.0010 & 0.9999 & 0.0015 & 0.3332 & 0.0012 & 0.0016 &396.3\\
\hline
S 4x4    & *&100 & 0.2568 & 0.1379 & 0.9453 & 0.1109 & 0.2433 & 0.1611 & 0.1638 & 5173.8\\
\hline
S 4x4 w   &*** & 100 & 0.5036 & 0.0040 & 1.0009 & 0.0016 & 0.3381 & 0.0045 & 0.0059 & 533.1\\
\hline
SD 3x3  & *& 100 & 2.7383 & 0.3670 & 4.1544 & 0.7060 & 0.1558 & 0.2881 & 0.7523 & 2758.4\\
\hline
SD 3x3 w &*** & 100 & 0.5002 & 0.0010 & 0.9997 & 0.0020 & 0.3334 & 0.0013 & 0.0022 & 449.7\\
\hline
\end{tabular}\newline
\end{table*}

\begin{figure}[t]

\centering

\subfloat[Analytical method]{\includegraphics[width=8cm]{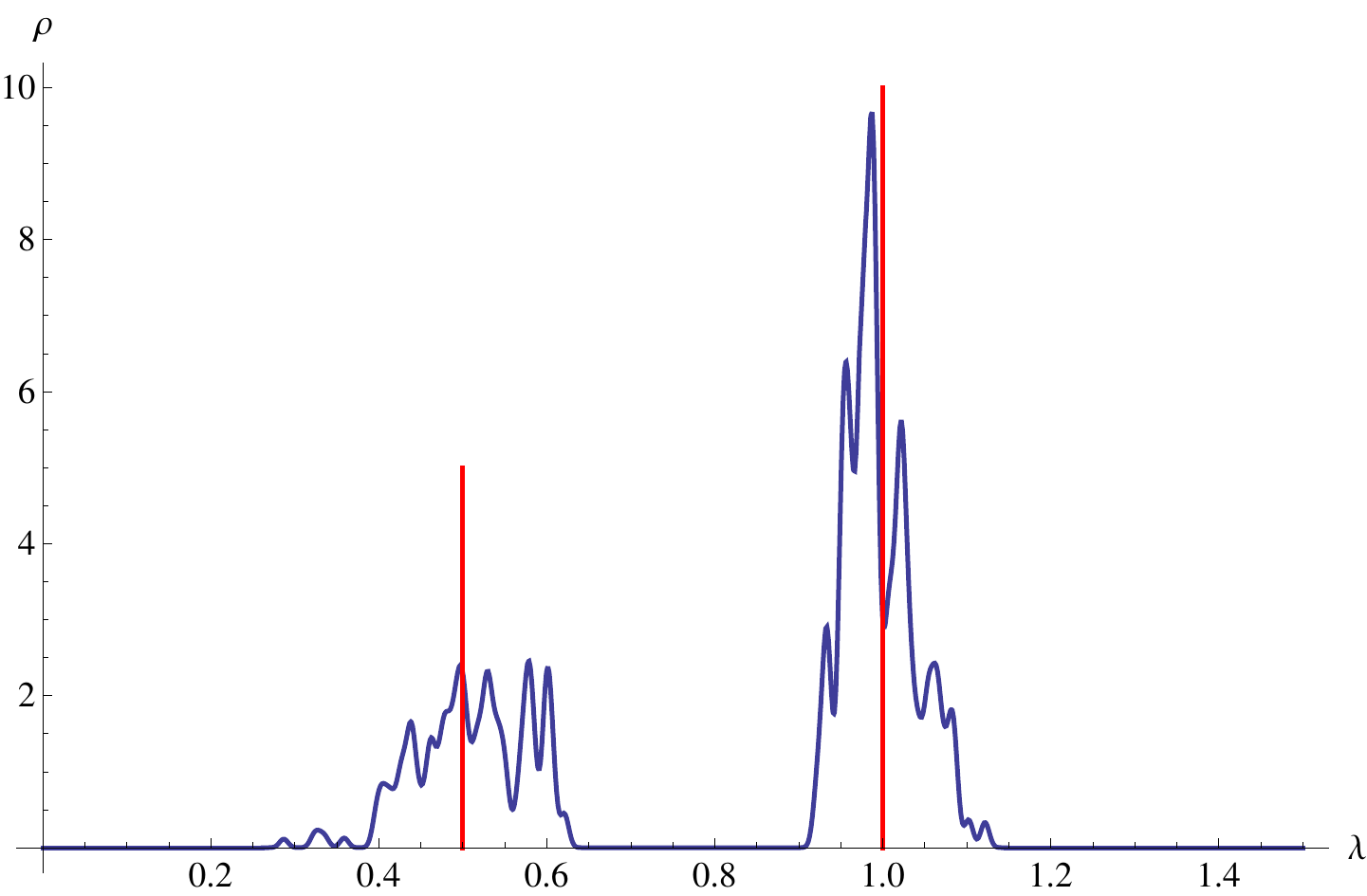}}

\begin{tabular}{cc}

\subfloat[Statistical $3 \times 3$ method]{\includegraphics[width=4cm]{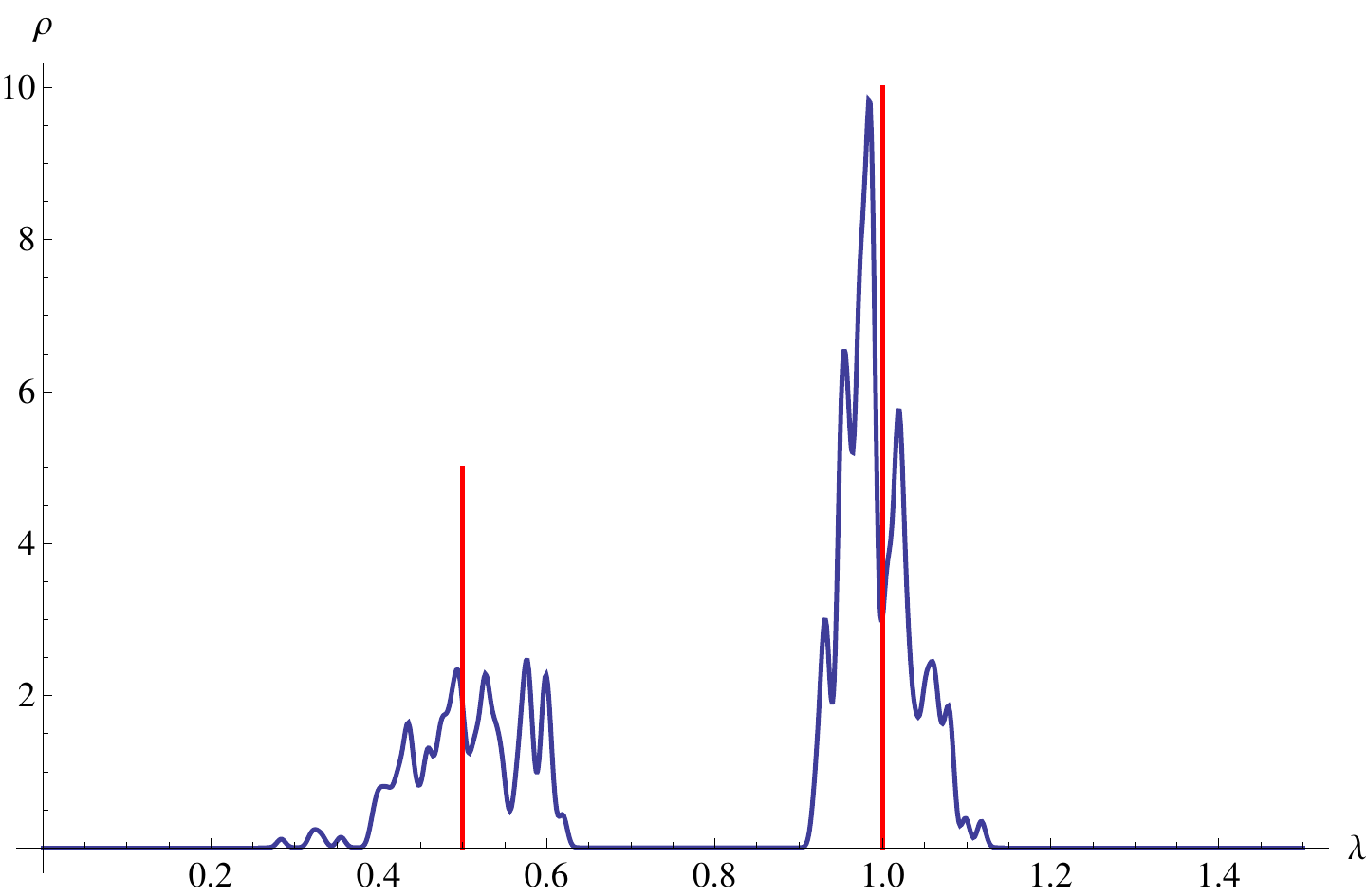}} 
   & \subfloat[Statistical $4 \times 4$ method]{\includegraphics[width=4cm]{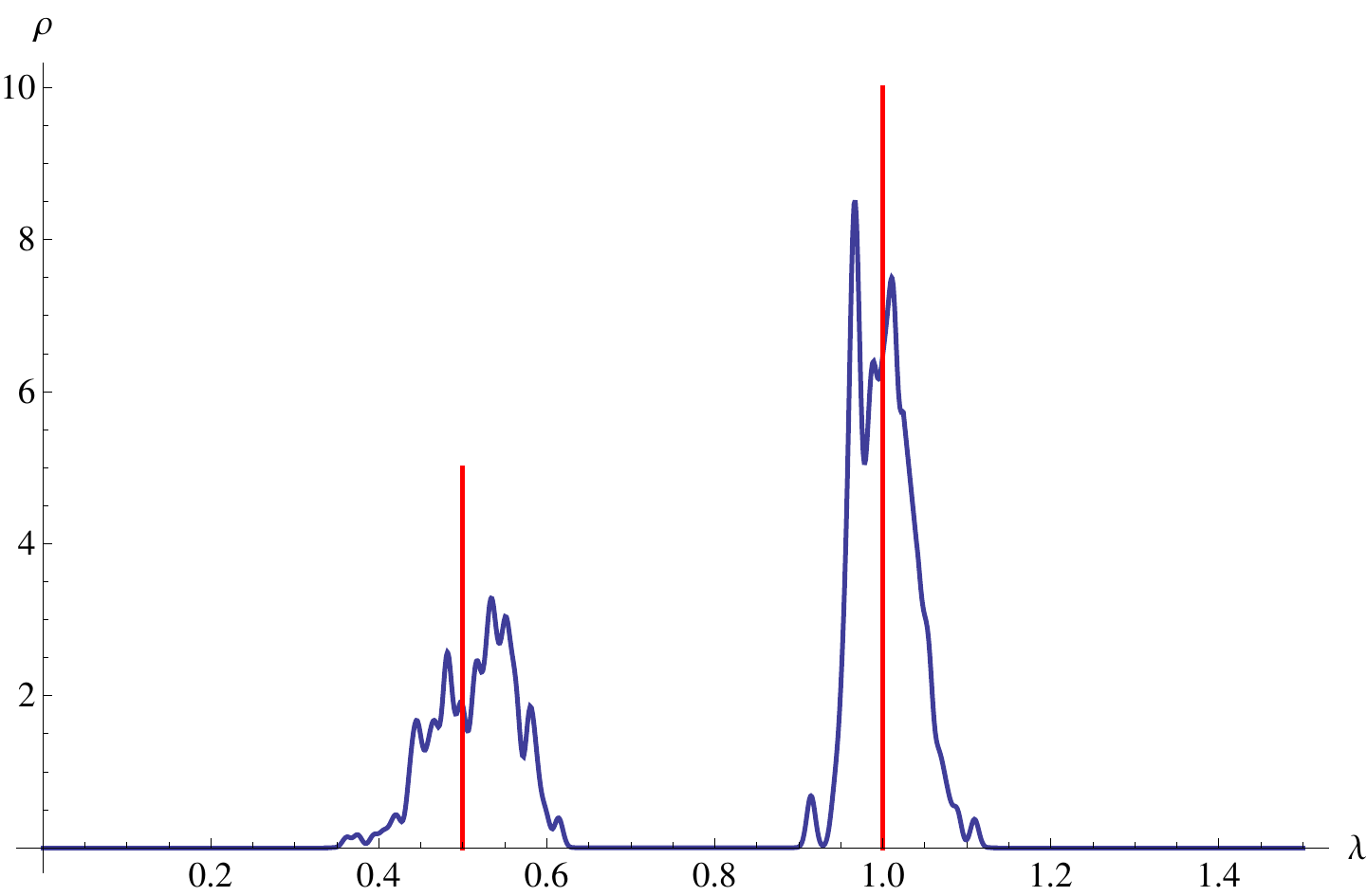}}\\

\subfloat[Analytical dual method]{\includegraphics[width=4cm]{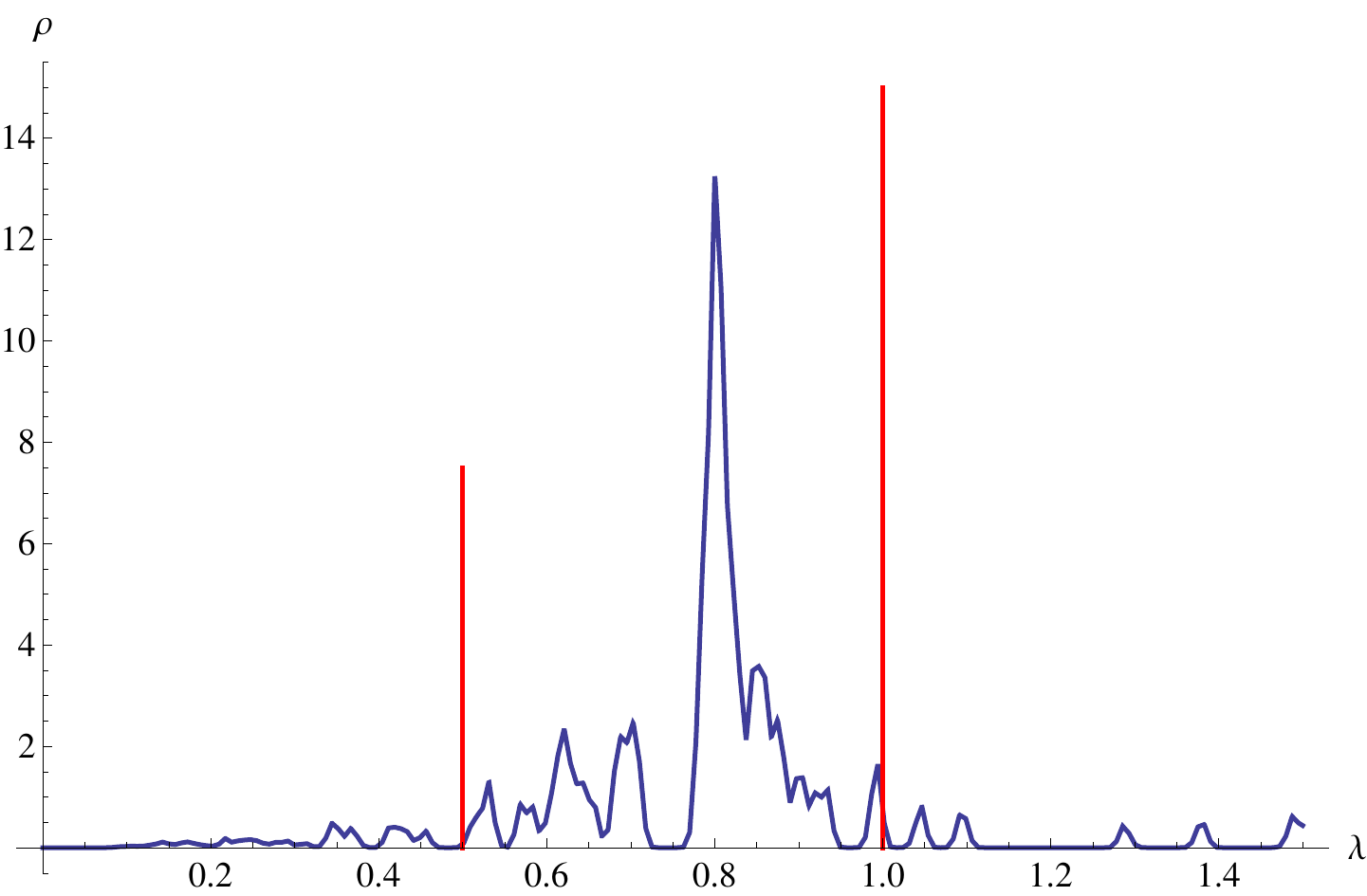}} 
   & \subfloat[Statistical dual $3 \times 3$ method]{\includegraphics[width=4cm]{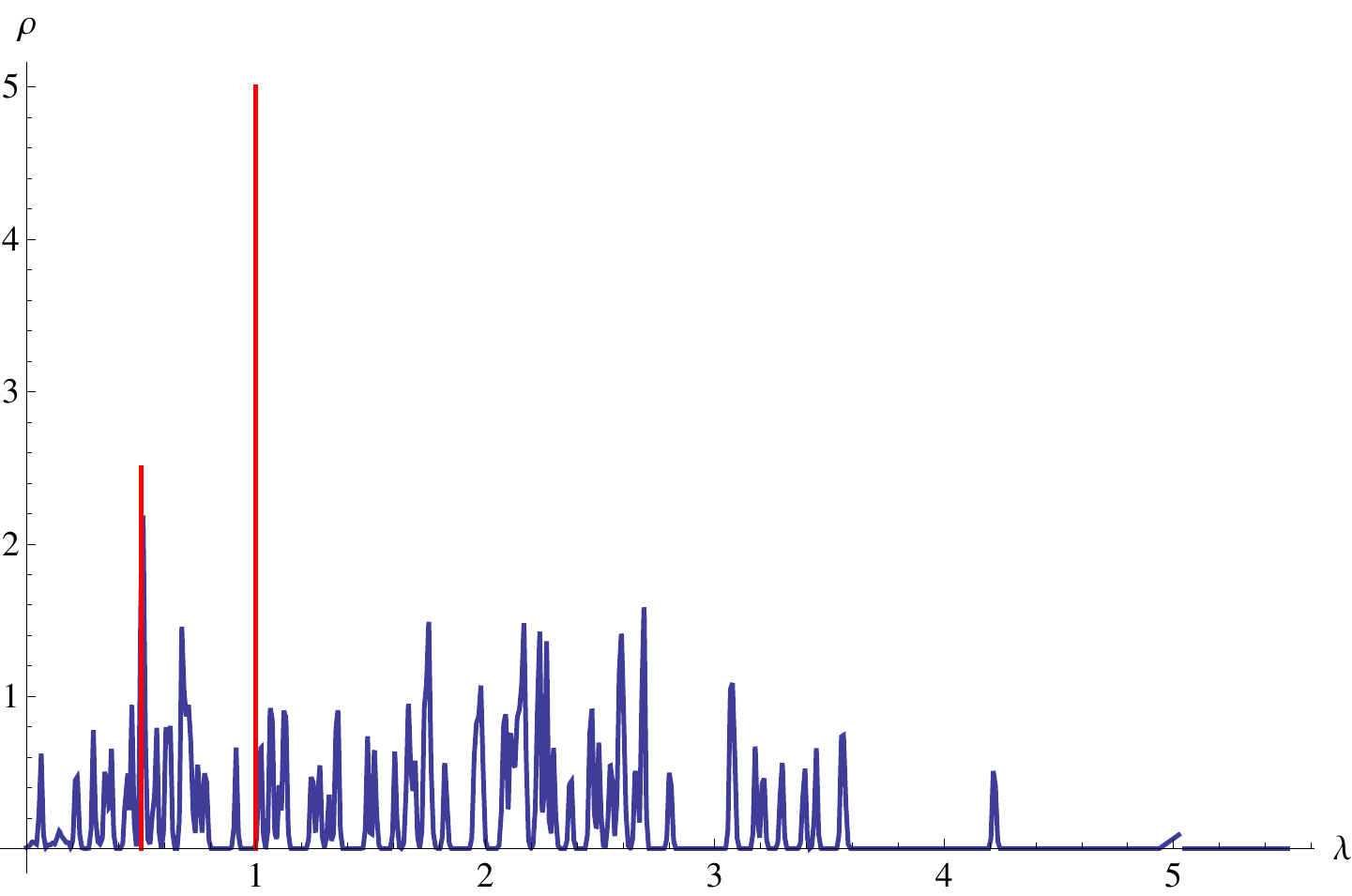}}\\
\end{tabular}
\caption{Estimated spectrum of the covariance matrix.
The underlying exact covariance matrix has eigenvalues $\mu_1 = 1$, $\mu_2 = 1/2$ (shown in red) with the degeneracies $p_1=2/3$, $p_2 = 1/3$.
100 empirical matrices $126 \times 180$ ($r=0.7$).}\label{foo}
\end{figure}

\begin{figure}[t]

\centering

\subfloat[Analytical method]{\includegraphics[width=8cm]{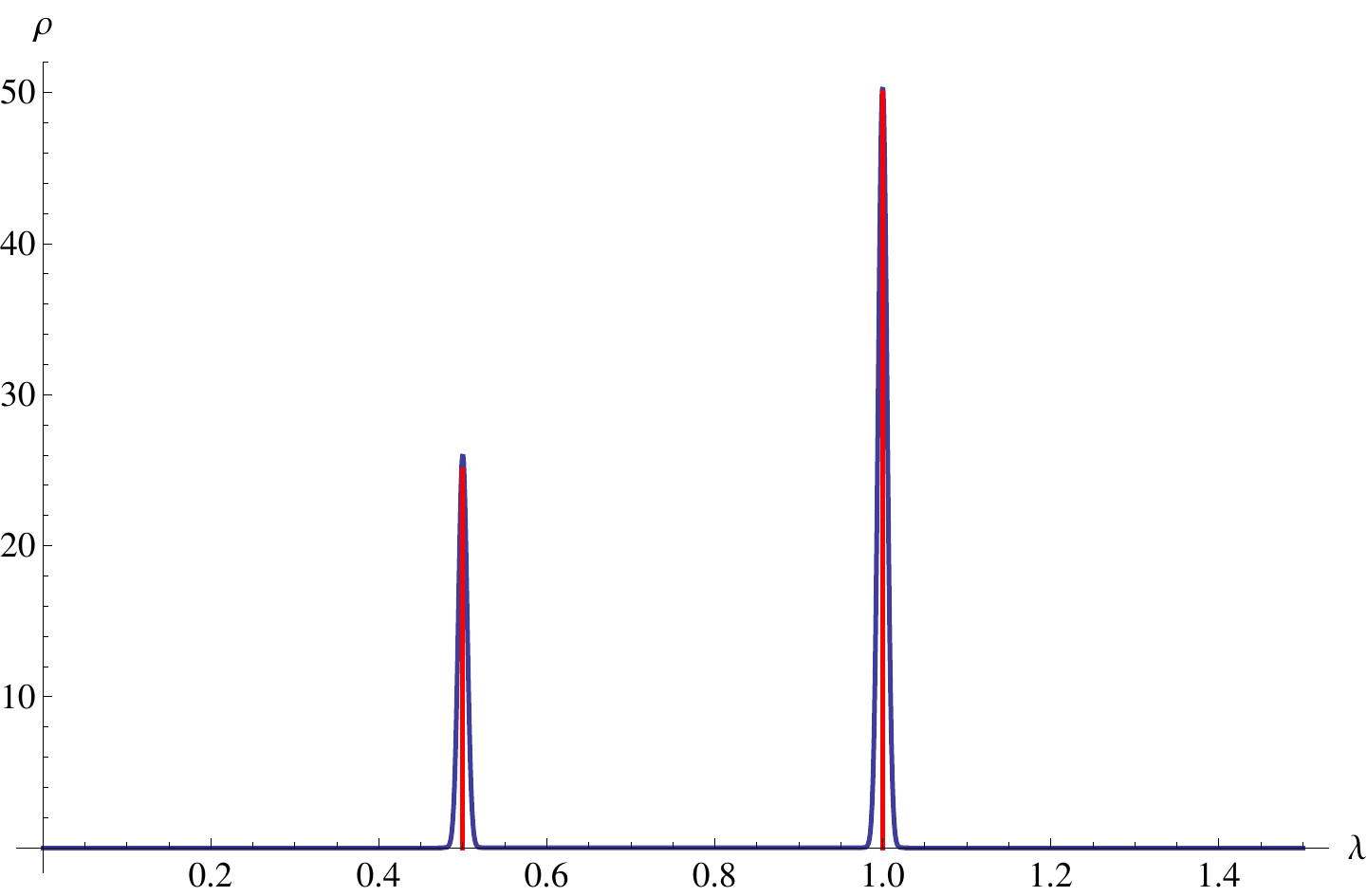}}

\begin{tabular}{cc}

\subfloat[Statistical $3 \times 3$ method]{\includegraphics[width=4cm]{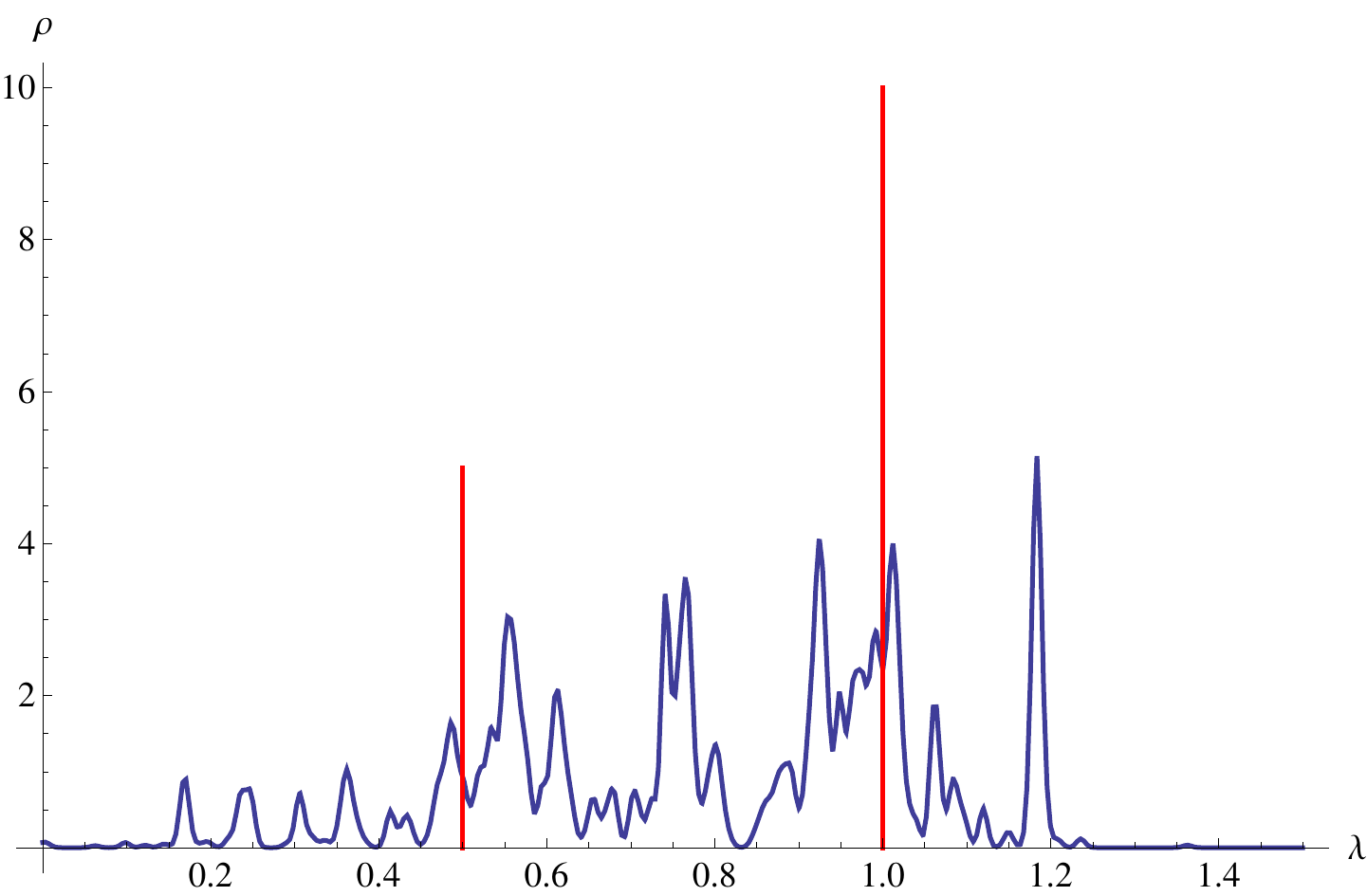}} 
   & \subfloat[Statistical $4 \times 4$ method]{\includegraphics[width=4cm]{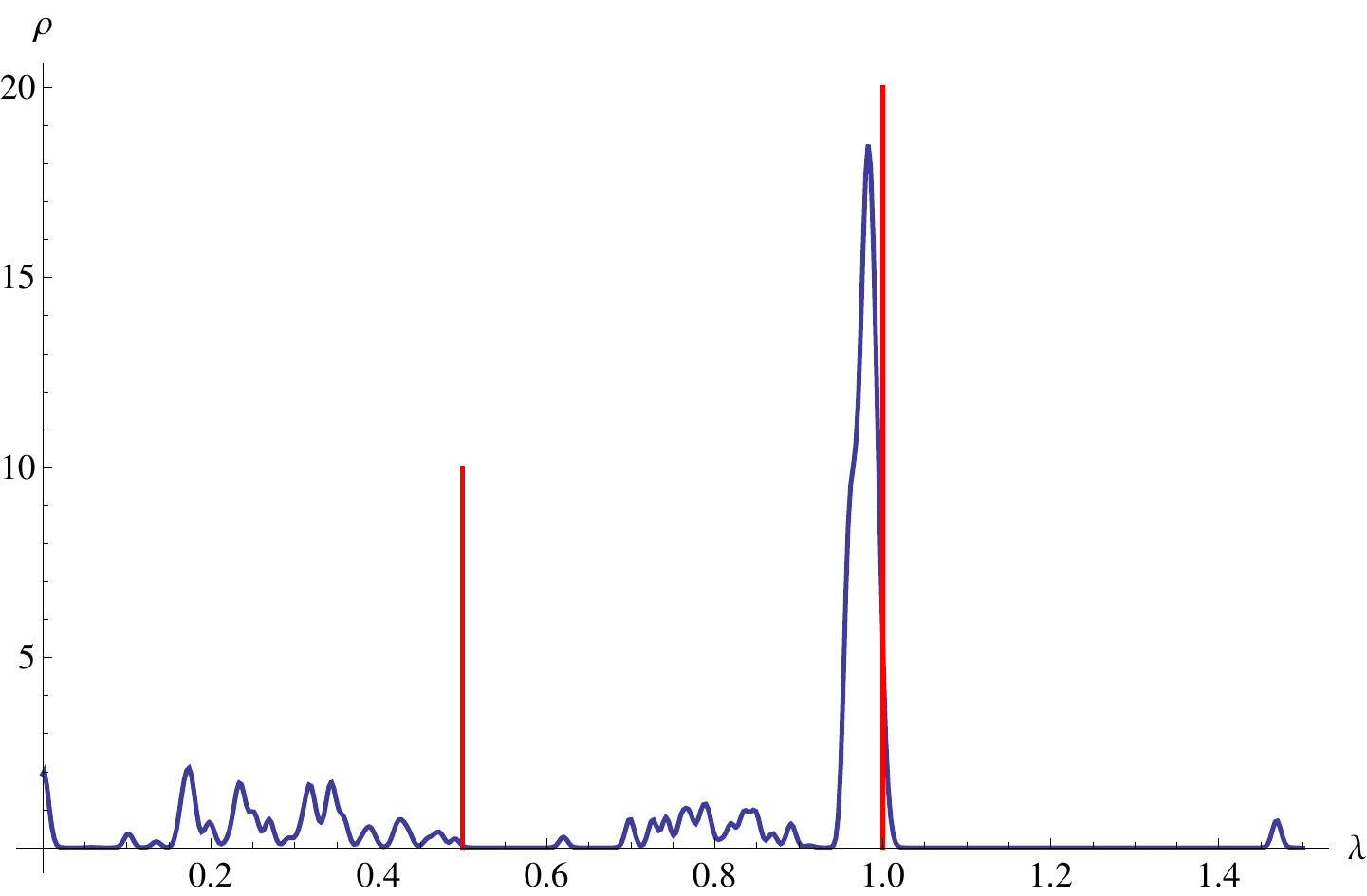}}\\

\subfloat[Analytical dual method]{\includegraphics[width=4cm]{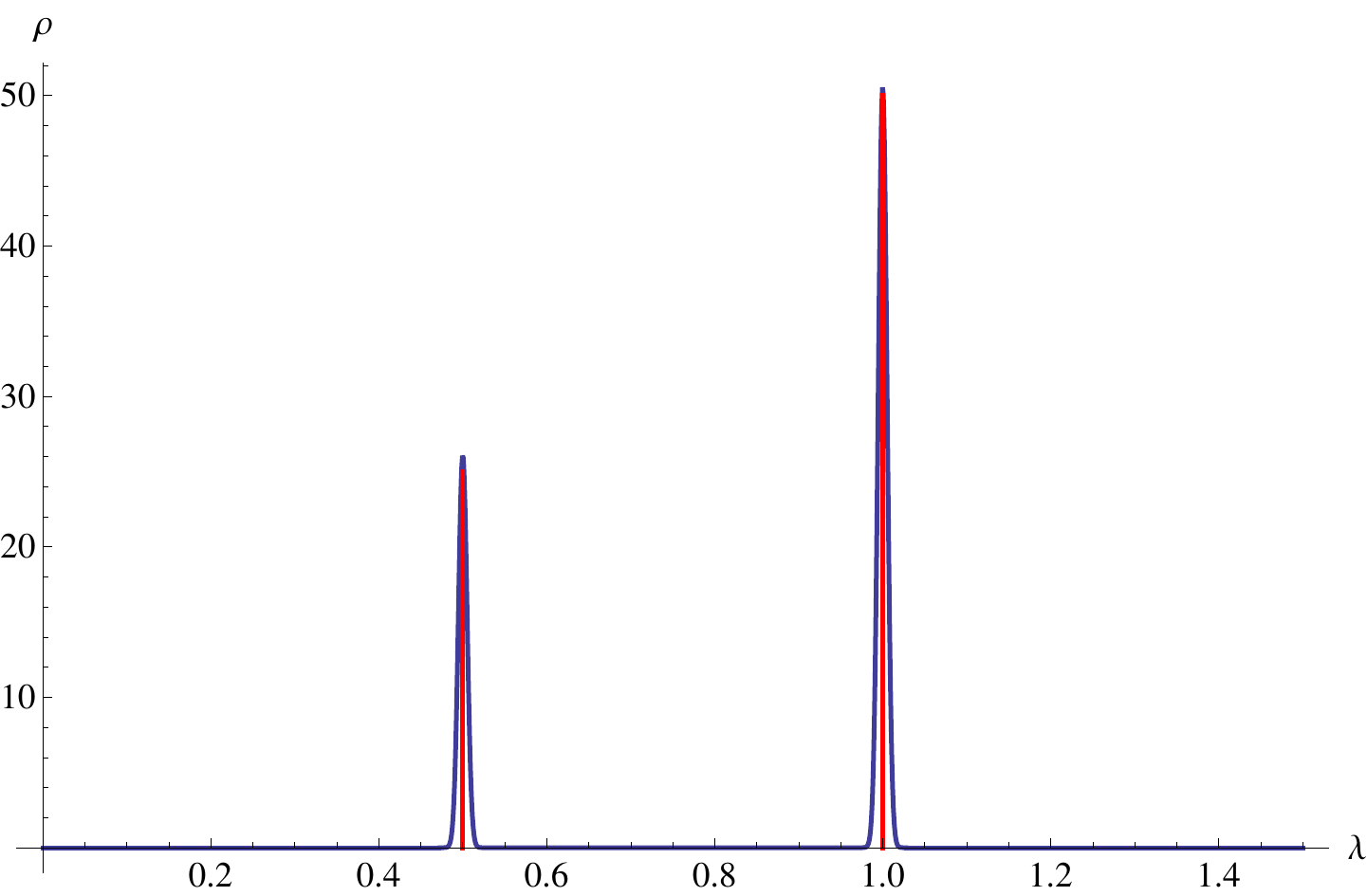}} 
   & \subfloat[Statistical dual $3 \times 3$ method]{\includegraphics[width=4cm]{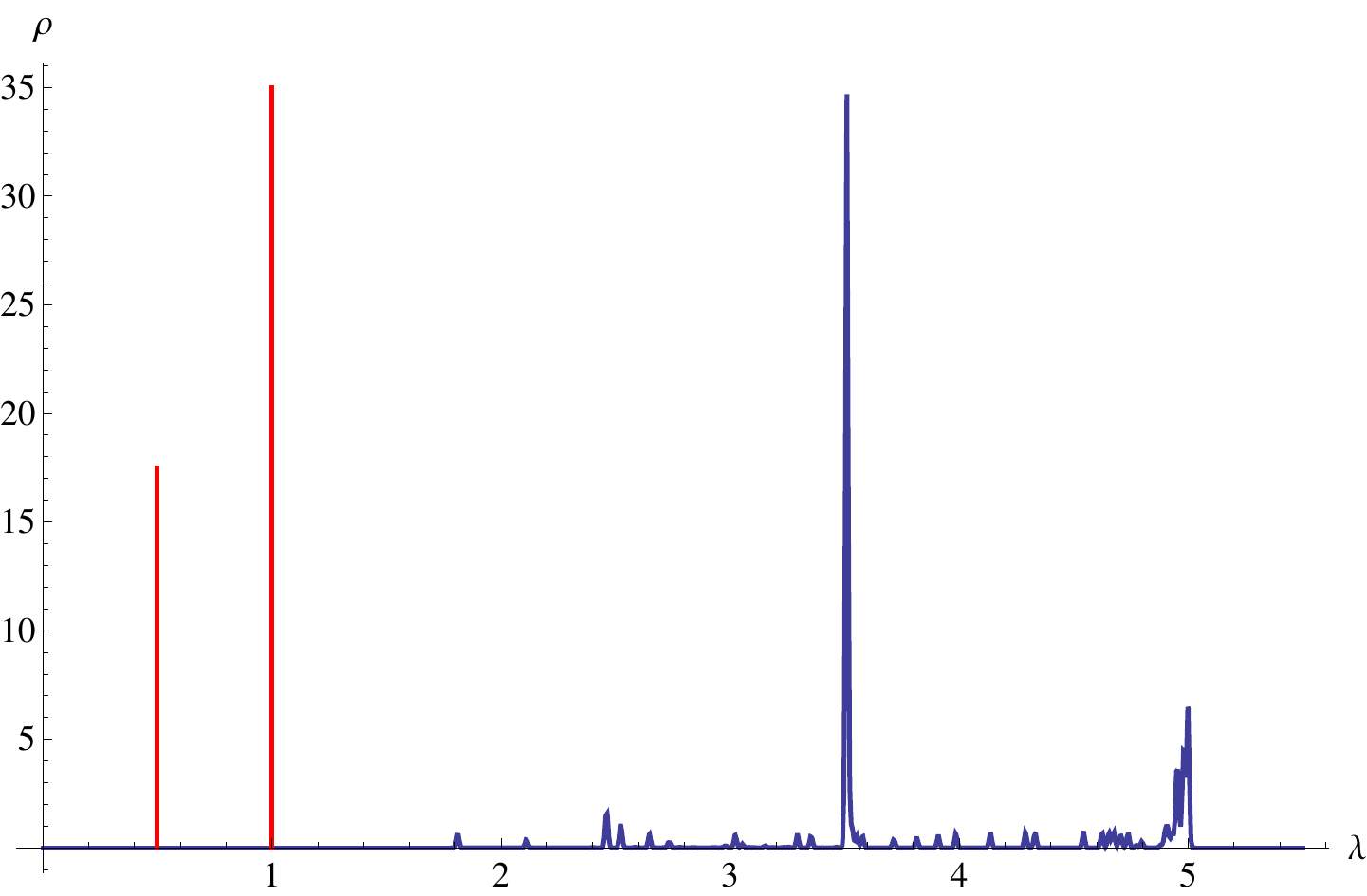}}\\
\end{tabular}
\caption{Estimated spectrum of the covariance matrix.
The underlying exact covariance matrix has eigenvalues $\mu_1 = 1$, $\mu_2 = 1/2$ (shown in red) with the degeneracies $p_1=2/3$, $p_2 = 1/3$.
100 empirical matrices $90 \times 9000$ ($r=0.01$).}\label{foo1}
\end{figure}

We started for the case of the exact covariance matrices $\Sigma$ with  two different eigenvalues,
$\Lambda_1=2$ and $\Lambda_2=1$ with the degeneracies $p_1=p_2=1/2$. 
In every test $L=100$  complex data
matrices $X$ were generated from a suitable ensemble,
and for each $X$ 
an experimental covariance matrix was calculated as $S=\frac{1}{T}XX^\dagger$. The size
$N \times T$ and the rectangularity $r=N/T$ of matrices $X$ 
varied from test to test. 

For each thus obtained $S$, eigen-inference was performed using 
the analytical, the analytical dual, and the statistical (for $3 \times 3$ matrix $Q$) method
to get the estimations $\lambda_1$, $\lambda_2$ and $p_1^S$ of the parameters
$\Lambda_1$, $\Lambda_2$ and $p_1$ (the calculation time was noted).
All obviously incorrect (non-real or real negative) estimations
of eigenvalues and degeneracies were rejected,
leaving $n$ sets of estimations. The arithmetical means and standard
deviations of real positive estimations of every parameter 
were calculated. Then $\eta$ (defined in the next section) was  calculated for every method. Lastly, a subjective
assessment $q$ of the quality of estimation, ranging from one to four stars, was
performed by the authors. Four stars might have been given
 for a method that would give
significantly better results than the other methods.

 The comparison is presented in Table \ref{tab:aaa}.

In the first place, it should be noted that in all of the cases the parameter $r$  was
large (equal to $2$, $1$ or $0.5$ - this means short samples of data). There were no cases with 
a smaller value of $r$ analyzed in \cite{MIT}.

For such large $r$, the analytical dual method
failed to give even remotely reasonable results (in most cases producing
negative or non-real eigenvalues or probabilities).

All the other methods behaved very similarly to each other.
The accuracy uniformly increased with increasing size of matrices and decreasing $r$. Our results agree with the results~\cite{MIT}. 

It can therefore be assumed 
that the analytical and statistical method were implemented correctly.
Most remarkably, the analytical method succeeded in producing almost
the same results as the statistical method, but several thousands times
faster. Its speed may be a key advantage in practical applications.

The analytical dual method is unsuitable
for large $r$. This section will show, however, that it works well when $r$ is small.

All the methods were tested on several large sets of matrices
with complex  entries. The real entries were tested as well, and the discussion of the comparison between real and complex Wishart ensembles is included in section~7.
The testing process 
included for the first time the statistical dual method and
the statistical method $4 \times 4$.
It was also tested whether
using the estimation from the analytical method as a starting point
for any version of the statistical method improves the precision and
decreases the computation time.

Figures~1  and 2 show in blue the
spectrum of the covariance matrix estimated by collecting
the results of eigen-inference from all the tested matrices
(smoothed so as to make the graph continuous instead of discrete).
The eigenvalues of the underlying exact covariance matrix are 
shown in red (scaled for the better presentation). 

Table II is organized similarly to Table I, but includes
results from a larger number of methods. 
The results for one set of matrices with large $r$ ($r=0.7$) and
another with small $r$ ($r=0.01$) are presented. The differences are
 manifest. In the first case the statistical method
$4 \times 4$ produced the most precise estimation (although it used
a lot of computation time), while the
analytical dual method sometimes failed
even to give real and positive estimations of parameters.
In the second case, however, the analytical dual
method offered the most accurate estimation in a short amount of time.  
The simple analytical method was the most robust, performing well in
all cases, and using always almost the same, little amount of time.

Since the analytical method is so fast, one might think it would be clever to 
use its result as a starting point for the 
minimization procedure of the statistical method. However, it was shown
that the possible gain of accuracy hardly recompenses the 
invested computation time. The results are almost the same as the results
of the analytical method alone.
In fact, if $r$ is small, the statistical method
 may \emph{reduce}
the accuracy of the estimation instead of improving it. 

The statistical dual method performed badly in all tests. Perhaps it is
because of the structure of the relations for double moments $\tilde{\alpha}_{i,j}^S$ (powers of $\alpha_{-2}^S$ in the denominator).

What seems especially puzzling is the fact that the statistical method
performed \emph{worse} when $r$ was small than when it was large.
 It is
unintuitive - small $r$ means that the experimental covariance matrix
is built from larger number of data, and hence the estimation should
be \emph{more} precise. 
For large $r$, the results of the analytical method and the statistical
$3 \times 3$ method are so similar that Figures 1a and 1b are almost
indistinguishable. The statistical $4 \times 4$ method gives estimations 
even better centered on the exact eigenvalues (Fig.\@ 1c).
However, for small $r$, whilst the analytical method reproduces the
spectrum of the exact covariance matrix almost perfectly (Figs.\@ 2a, 2d), neither
the statistical $3 \times 3$ method (Fig.\@ 2b) nor the statistical $4 \times 4$ method (Fig.\@ 2c) gives a correct estimation of the spectrum.


\section{Speed of the algorithms, complexity and quality measures}

The procedures written in Mathematica 9
for the present work succeeded in 
reducing the time needed 
to generate all the formulae for  
$\alpha_{j}^S$ and $\alpha_{j}^\Sigma$, $-1 \leq j \leq -10$,
from over 10000 seconds (as in the work \cite{PHDLUKA}, where
a computer with the processor
Inter Core \textsuperscript{TM} 2 Duo 6400 (2x2.0 GHz) and
4 GB RAM was used)
to 24 seconds
(on a comparable computer: Intel Core \textsuperscript{TM} i3-3227U (2x1.9 GHz) and
4 GB RAM).

Furthermore, the time for generating the formulae for $\alpha_{i,j}^S$, $i,j \leq 5$, was reduced from 1000 to 26 seconds.

With such fast algorithms, calculation  of the higher degree relations
becomes feasible.

\begin{figure}[b]

\begin{tabular}{cc}

\subfloat[$\alpha^S_i$ \lbrack and $\alpha^S_{-i}$\rbrack \ in terms of $\alpha^\Sigma_k$]{\includegraphics[width=4cm]{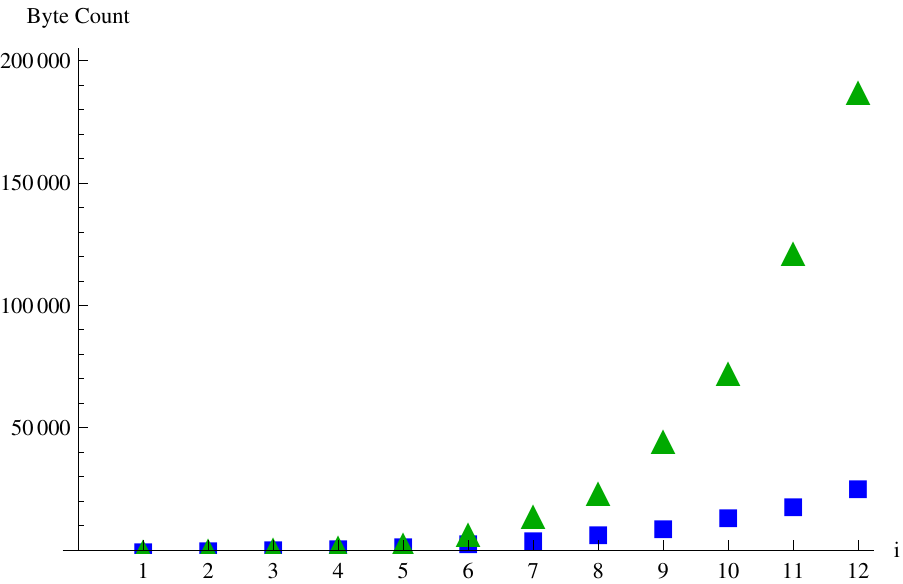}} 
   & \subfloat[$\alpha^S_{i,i}$ \lbrack and $\tilde{\alpha}^S_{i,i}$\rbrack \ in terms of $\alpha^S_k$]{\includegraphics[width=4cm]{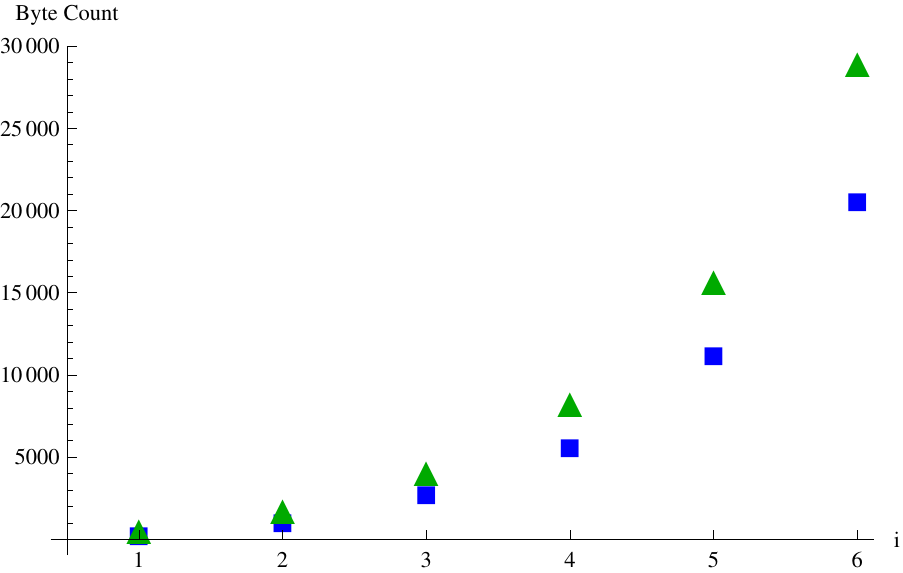}}\\
\end{tabular}

\centering

\subfloat[$Q_\Theta^{-1}$ in terms of $\Lambda_1$, $\Lambda_2$ and $p_1$ (two-eigenvalue case).]{\includegraphics[width=8cm]{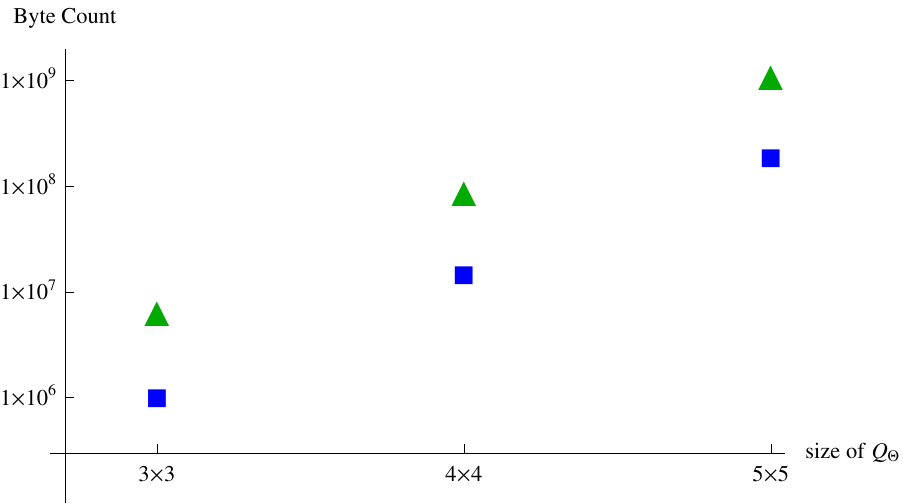}} 

\caption{Number of bytes used to store the formulae
appearing in the statistical (blue squares) and statistical dual method (green triangles). Note the logarithmic scale of the graph (c).}\label{byte}
\end{figure}

Figure \ref{byte} presents the complexity of the formulae
used in the statistical method. 
The size of the expressions for single moments of the empirical covariance
matrix in terms of single moments of the exact covariance matrix,
and for double moments in terms of single moments grow polynomially
with the value of the index $i$.
They are so complicated that it is better not
to perform symbolically all the algebraical calculations 
leading to the form (\ref{gmini}) of the function $g_\Theta$. 
This final form, which involves the inverse of the matrix $Q_\Theta$,
is unwieldy
and may use, depending on the size of the matrix $Q_\Theta$, megabytes or
even gigabytes of computer memory (Fig.\@3c). Byte count
apparently grows exponentially with the size of $Q_\Theta$.

Therefore, it is more appropriate
to make step-by-step numerical calculations:
when the minimization algorithm needs the value 
$g_\Theta$ for certain values of the parameters $\{ \Lambda_i, p_i \}$,
let it first calculate the values of
all the needed single moments $\alpha_{k}^\Sigma$,
then the single moments $\alpha_{k}^S$, then the double moments
$\alpha_{k,l}^S$, then construct the matrix $Q_\Theta$ and calculate its 
inverse, and only then calculate $g_\Theta$. 
This method of calculation uses much less memory.
Nevertheless, the minimization problem 
remains complex.

While the analytical method can be used
for any number of different eigenvalues of the exact covariance matrix,
the statistical method for four eigenvalues needs the matrix $Q_\Theta$ to
be of dimension at least $7 \times 7$, which fact, considering the
monstrous complexity of the expressions, makes the calculations for more than
three eigenvalues, as for now, well-nigh infeasible.

In each test each of the eigen-inference methods, numerous
data matrices were
generated from an ensemble described by a certain exact covariance 
matrix. For each data matrix, 
the experimental covariance matrix was calculated,
and then the parameters $\Lambda_i$, $p_i$ were estimated using the
eigen-inference methods. 
Among the simplest measures of the quality of estimation are 
the arithmetical means
(denoted $\langle \lambda _i \rangle $, $ \langle p_i \rangle $) 
and standard deviations (denoted $\sigma(\lambda_i)$, $\sigma(p_i)$)
of all the parameter estimations. The mean estimations 
$\langle \lambda _i \rangle $, $ \langle p_i \rangle $
should be close to the exact values of the parameters $\Lambda_i$, $p_i$.
The standard deviations $\sigma(\lambda_i)$, $\sigma(p_i)$ should be small.
The disadvantage of this approach to estimation quality assessment is the difficulty of comparing two 
methods 
if some of these numbers are better for the first of them while the others
for the second.

A useful idea is to introduce a 
measure of the quality of estimation that during a test
produces a single number for each eigen-inference method.
The parameter $\eta$ used in \cite{PHDLUKA} and in the present
work, although defined rather arbitrarily, serves this purpose. 
The estimations taken from all the data matrices may themselves be written
as a matrix $E$ of dimension $(2K-1) \times L$, where $K$ is, as in previous sections, 
the number of different eigenvalues of the exact covariance matrix, and
$L$ is the number of generated data matrices. The parameter $\eta$
is defined as the square root of the largest eigenvalue
of the covariance matrix built from $E$:
\begin{equation}
\eta=[\mathrm{Max} (\mathrm{Eig} (\mathrm{Cov} (E)))]^{\frac{1}{2}}
\end{equation}

It, one might say,
measures the 
width of the cloud of estimations in a $2K-1$-dimensional space.
The smaller it is, the better the estimation.

\section{Lack of positivity condition in the statistical method}
The process of minimization of function (\ref{gmini}) depends crucially on the "entropic" term $\ln {\rm det} Q_{\Theta}$.  In the limit, when the dimension of vector $\Theta$ tends to infinity, the limiting spectral distribution of $Q$ tends to the $\Lambda$, therefore is positive defined. However, this might not be true in the case when we approximate
the exact result by a truncated, finite dimensional vector  $\Theta$. In this case, as a result of truncation,  ${\rm det}{Q}$ can reach zero and can become negative for some range of the parameters. This pathology can be    demonstrated  even in the case of a very simple spectrum,  consisting of two distinct eigenvalues $ \Lambda_1$ and $\Lambda_2$  occurring with
the probabilities  $p_1$ and $p_2$ respectively.  In order to present simple, two-dimensional plots, we rescale  the eigenvalues,  so $\Lambda_s=\Lambda_1/\Lambda_2$ and the second eigenvalue is always fixed to 1. Then, we plot the sign of ${\rm det} Q$ as a function of
 $\Lambda_s$ and $p=p_1$ (note that $p_2=1-p_1$, so is not an independent variable). 

\begin{figure}[ht]

\centering

\begin{tabular}{ccc}
\subfloat[r=0.99]{\includegraphics[width=2.5cm]{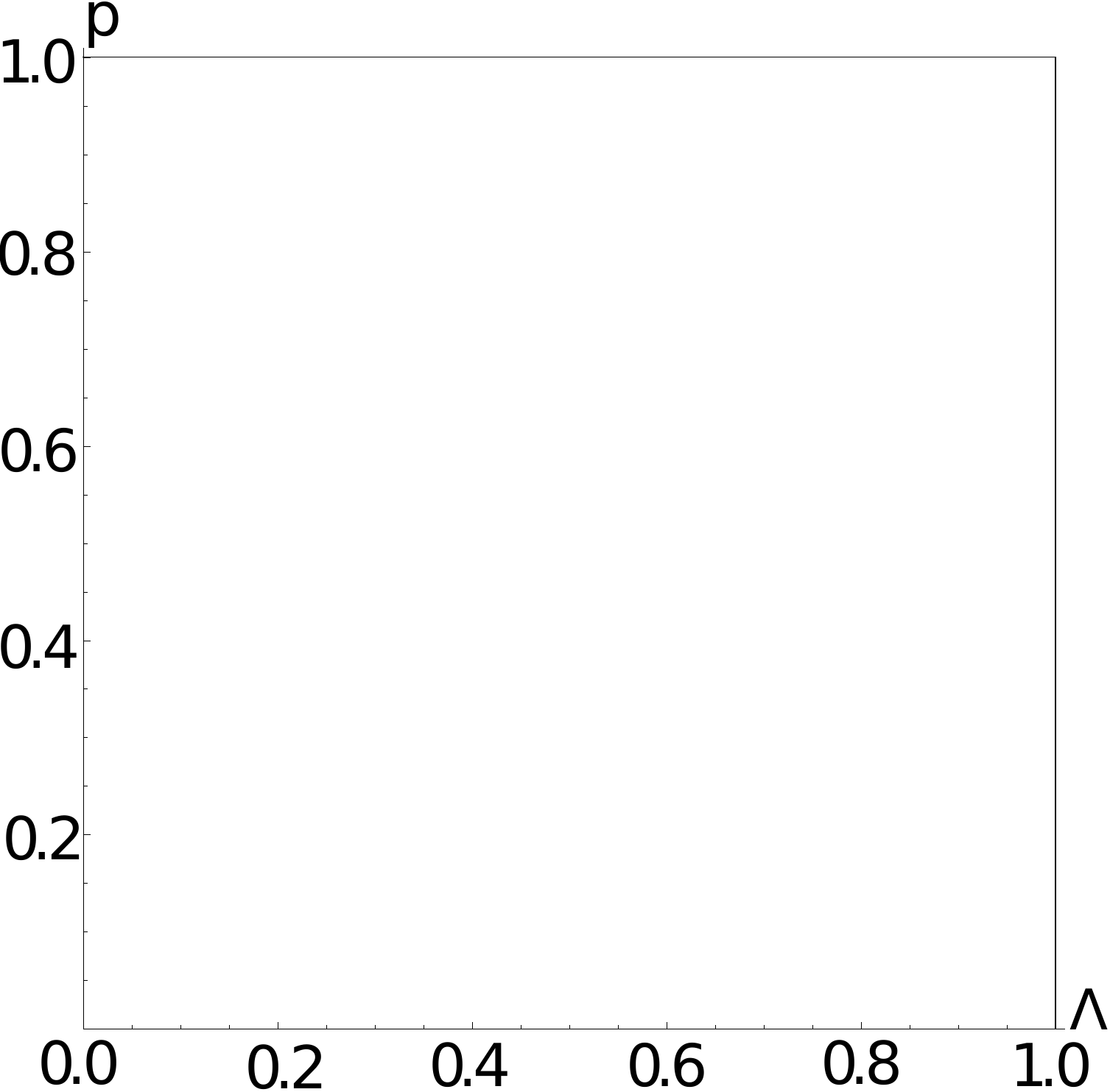}} 
   & \subfloat[r=0.9]{\includegraphics[width=2.5cm]{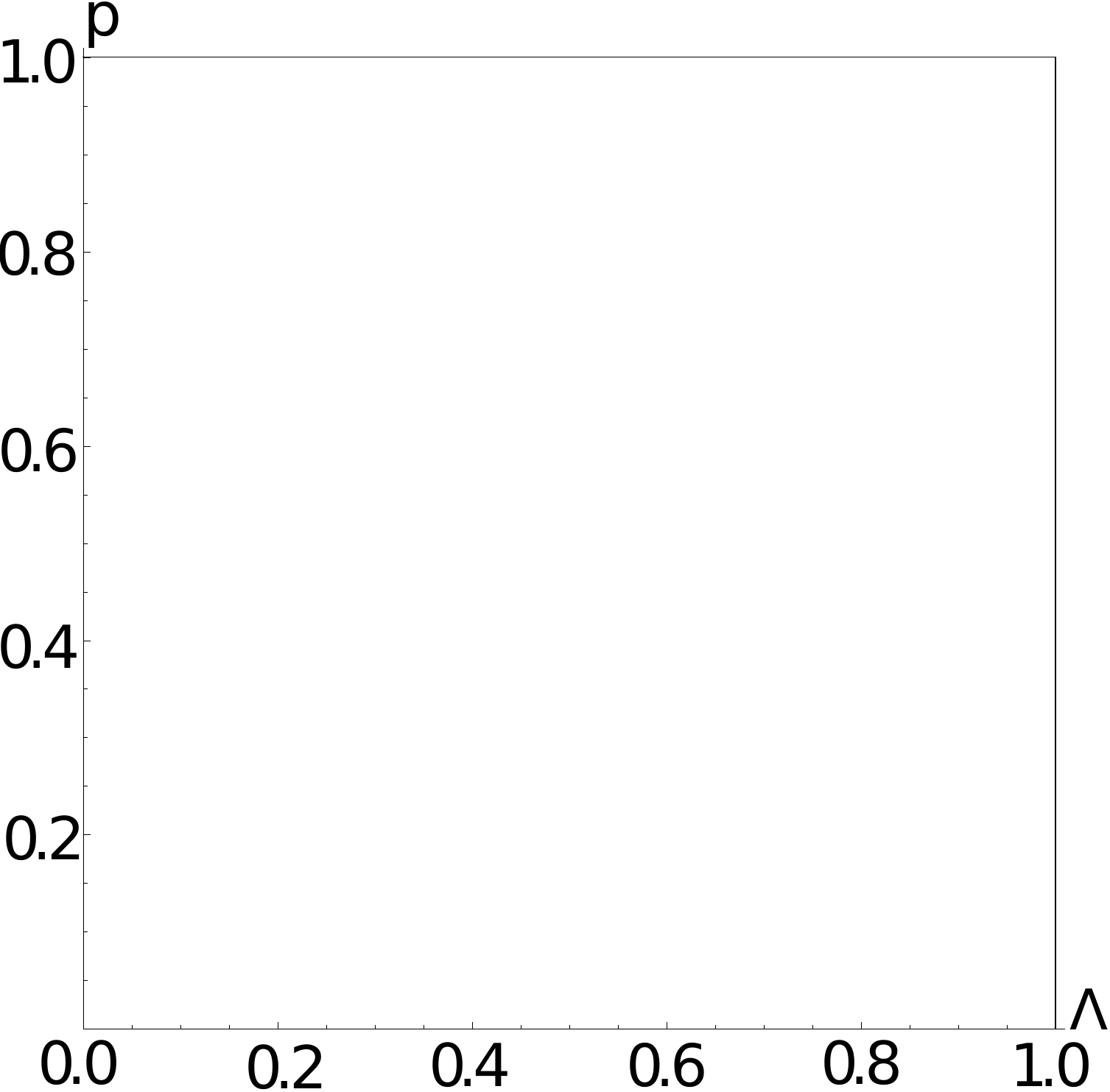}}
& \subfloat[r=0.7]{\includegraphics[width=2.5cm]{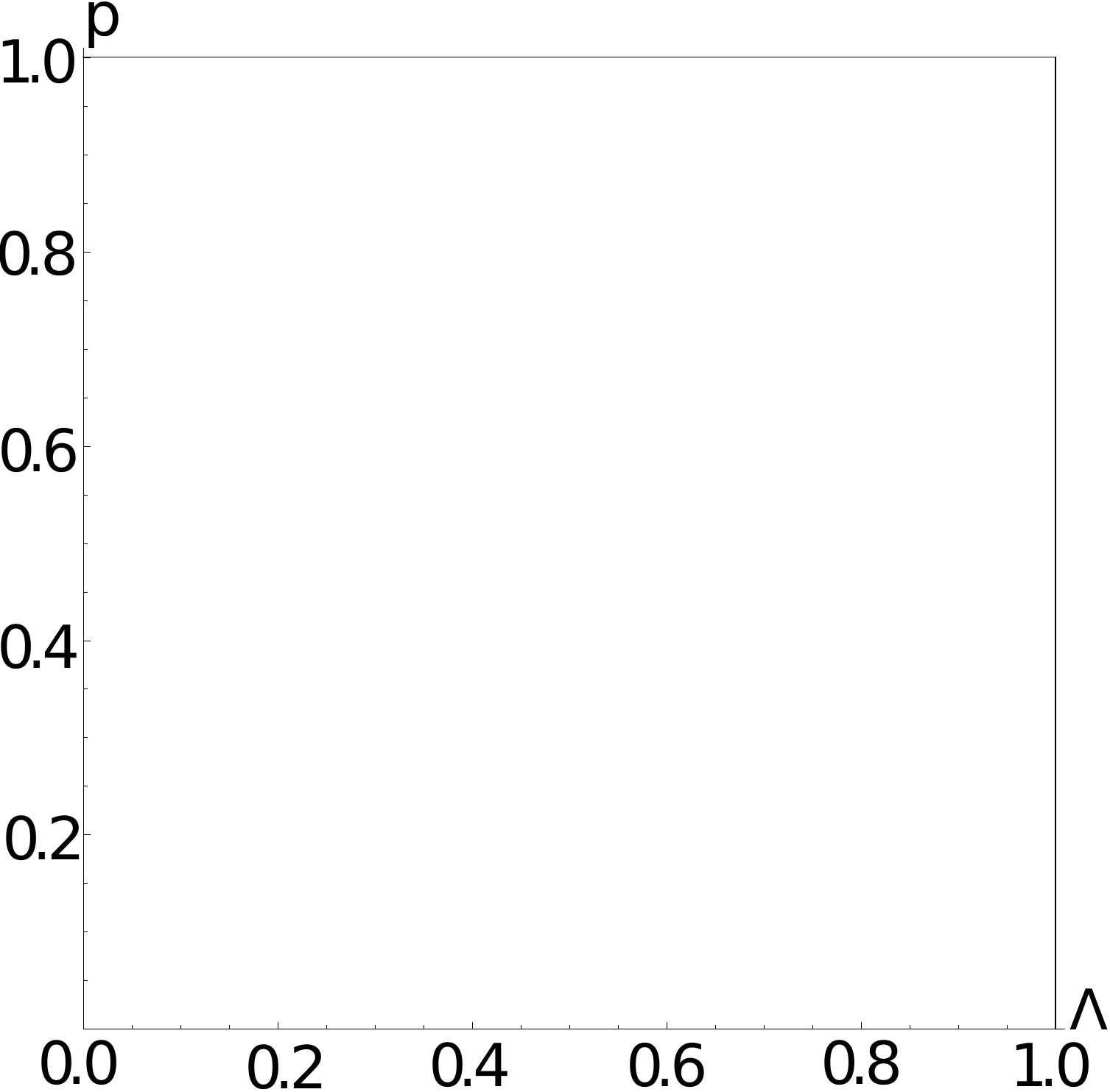}}\\

\subfloat[r=0.5]{\includegraphics[width=2.5cm]{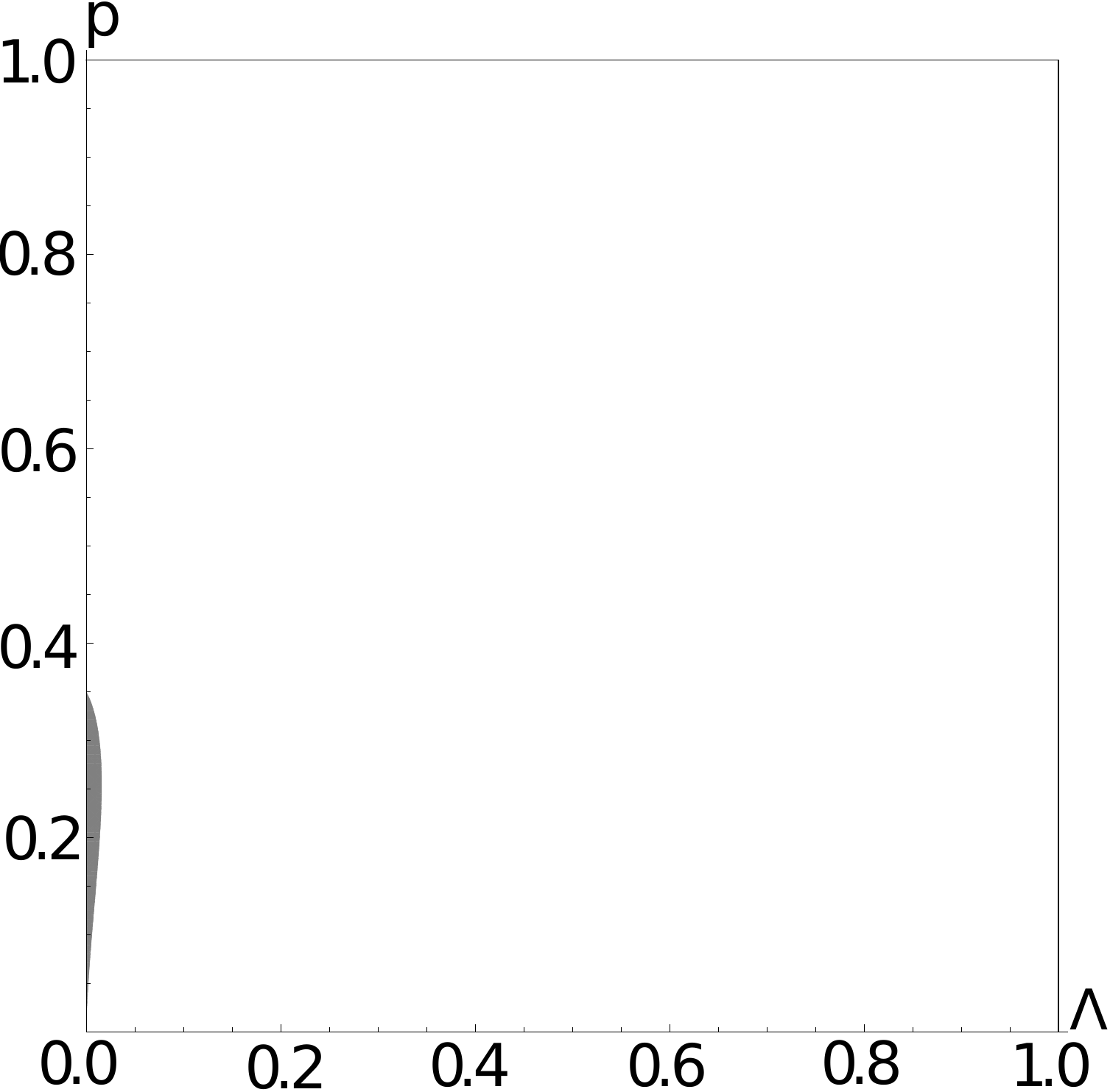}} 
   & \subfloat[r=0.3]{\includegraphics[width=2.5cm]{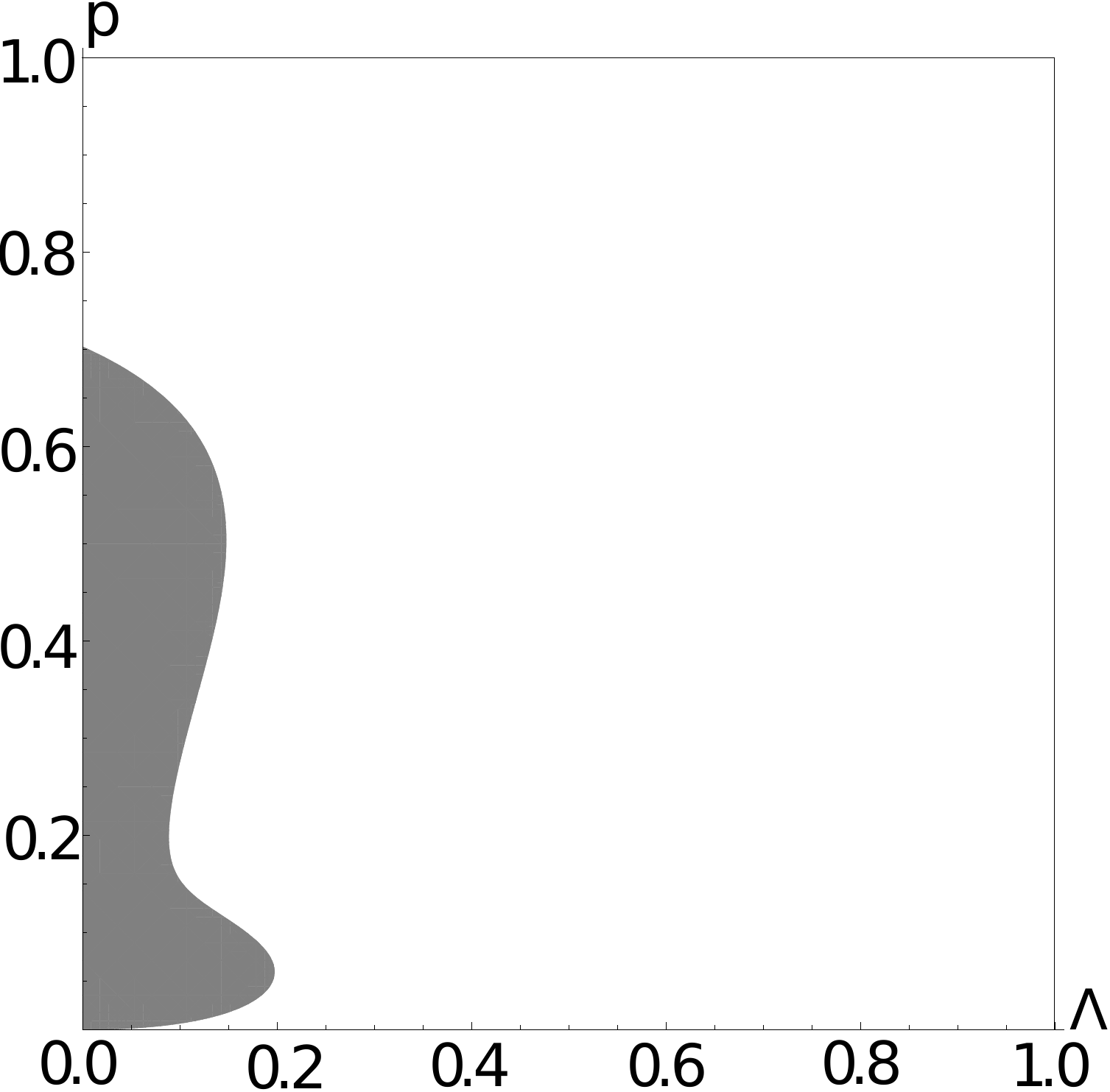}}
& \subfloat[r=0.1]{\includegraphics[width=2.5cm]{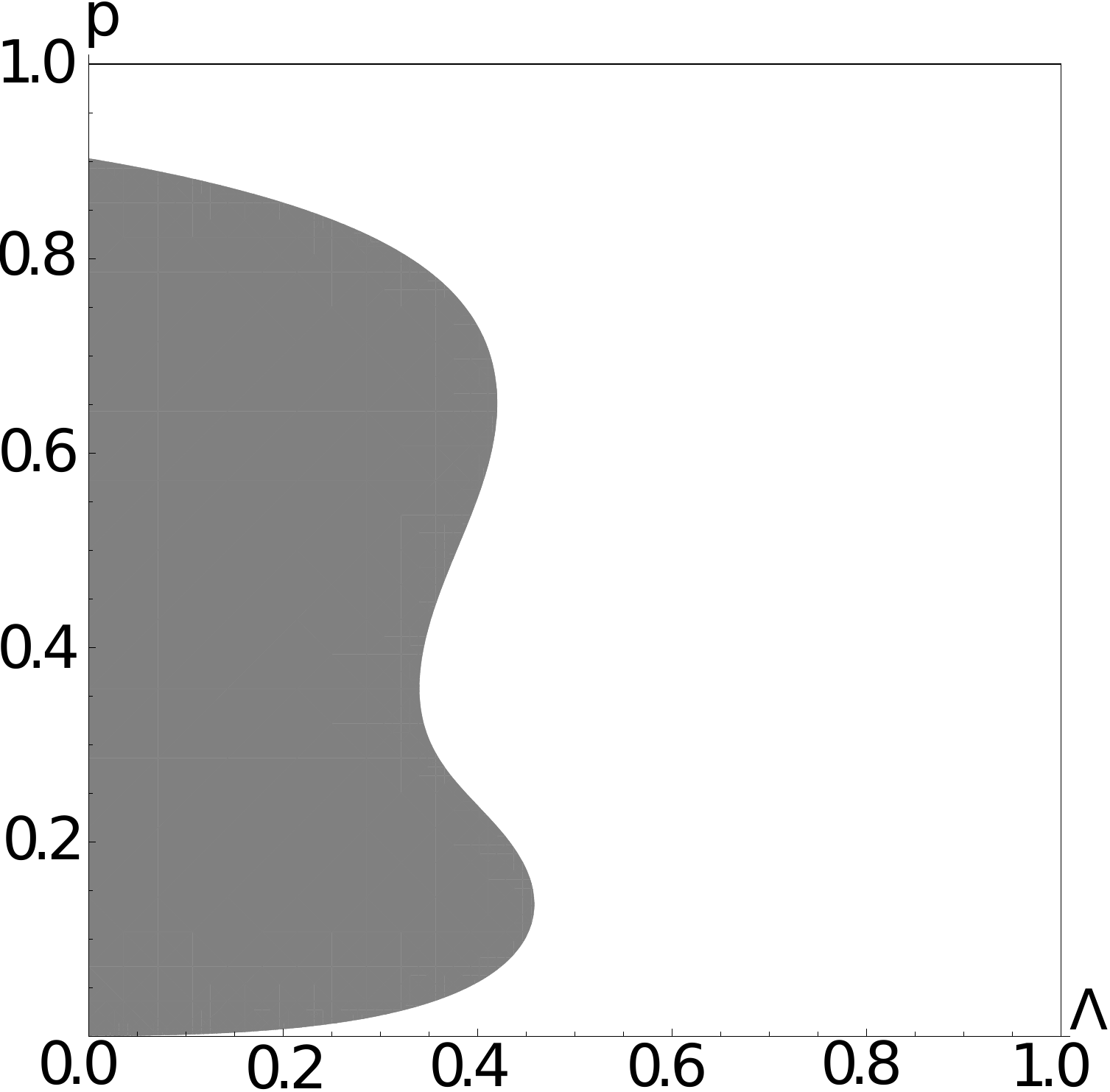}}\\

\subfloat[r=0.05]{\includegraphics[width=2.5cm]{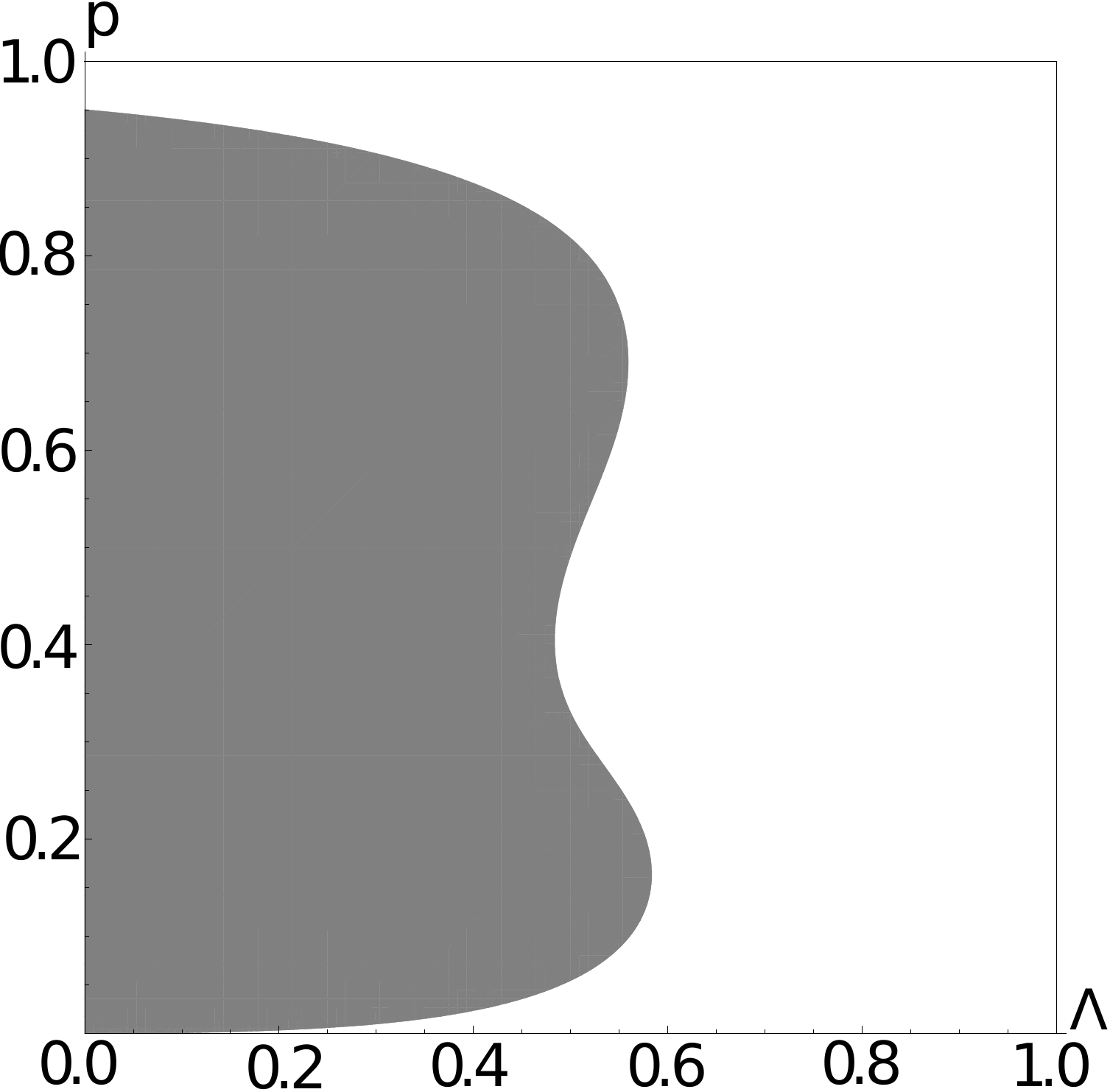}} 
   & \subfloat[r=0.01]{\includegraphics[width=2.5cm]{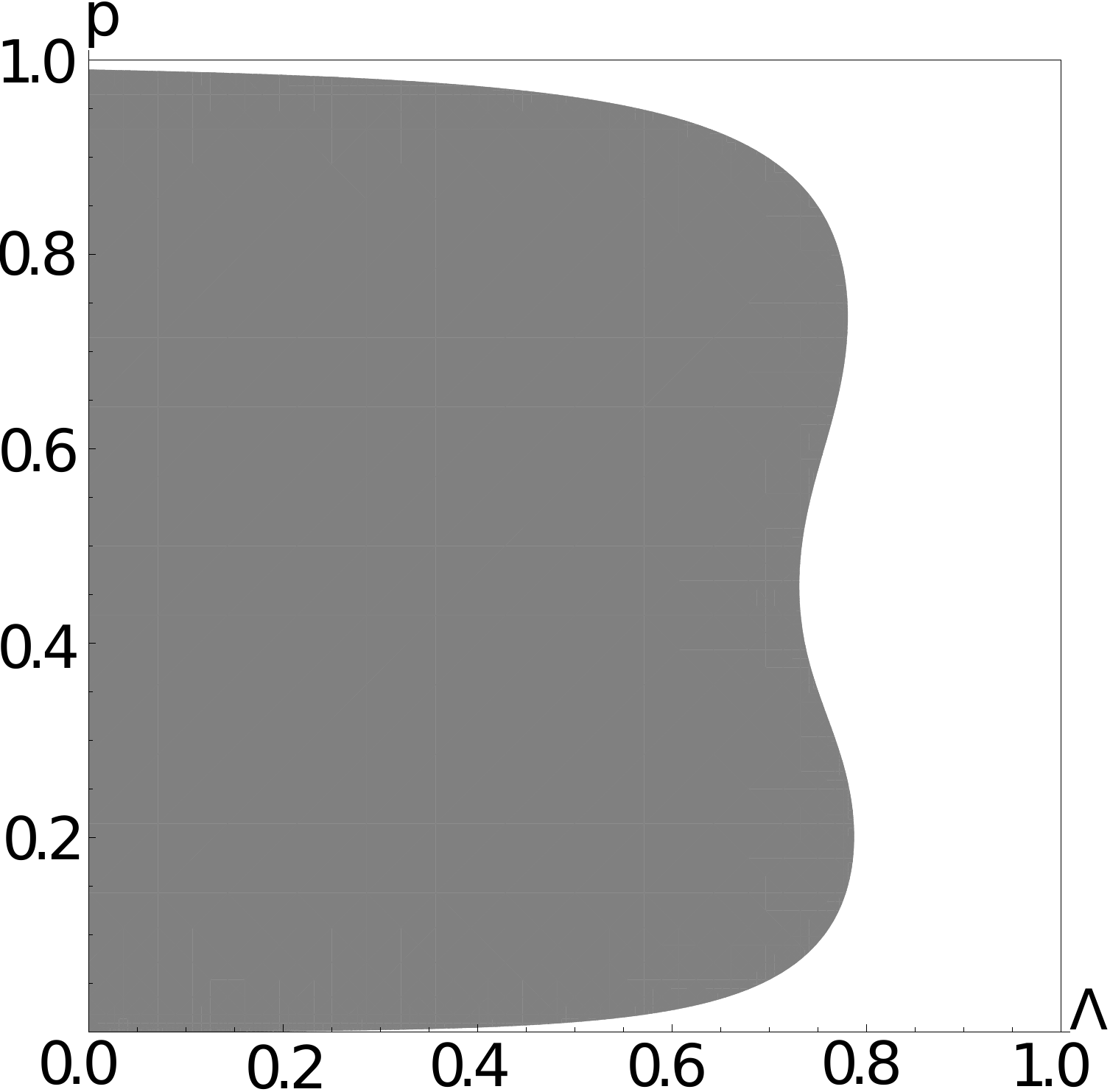}}
& \subfloat[r=0.001]{\includegraphics[width=2.5cm]{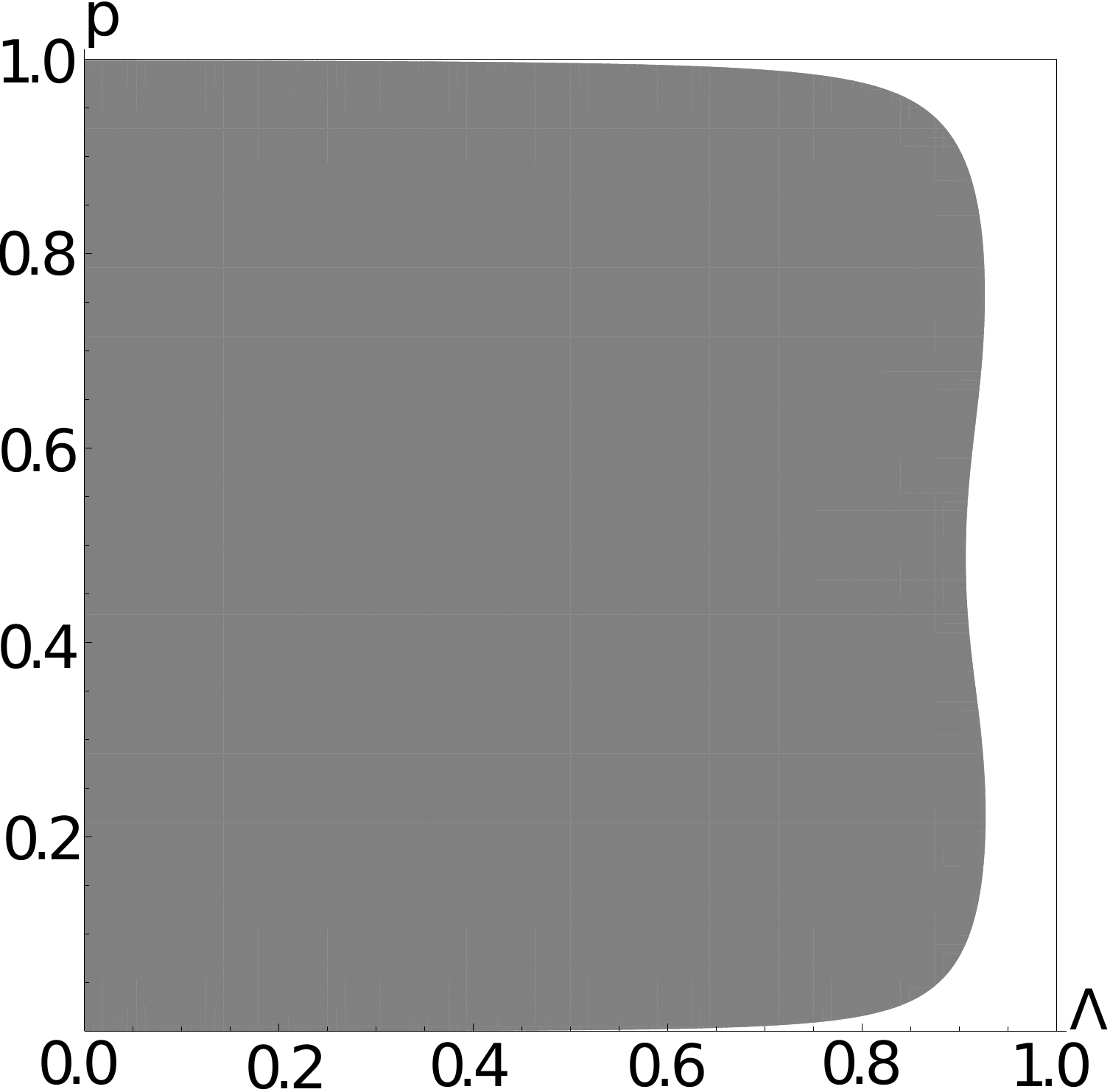}}\\
\end{tabular}

\caption{
Dark regions correspond to non-positivity of $\det Q$
 for the case when $Q$ is approximated by the $3\times 3$ matrix built of double moments.
}
\label{fig33}
\end{figure}

\begin{figure}[]

\centering

\begin{tabular}{ccc}
\subfloat[r=0.99]{\includegraphics[width=2.5cm]{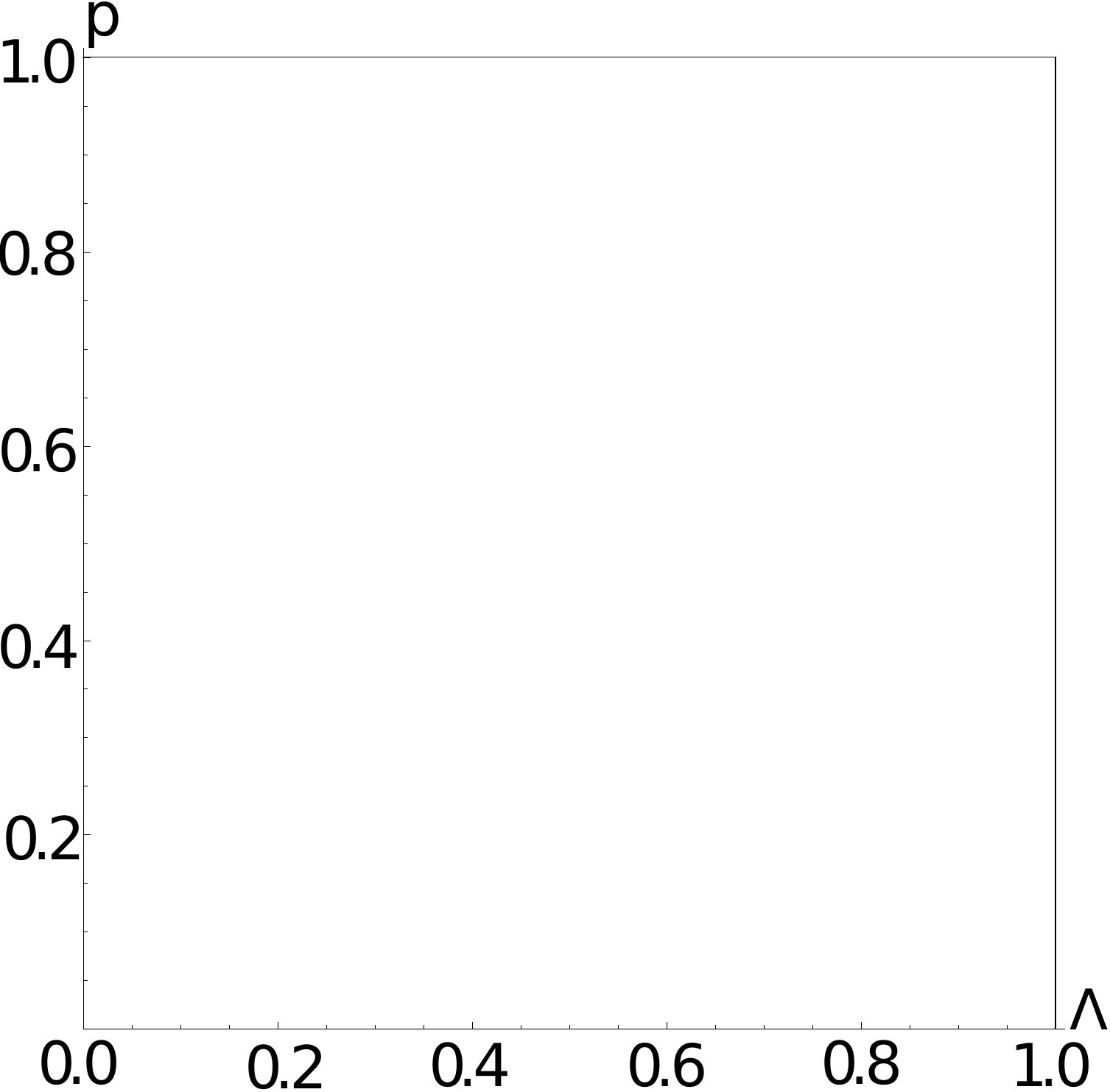}} 
   & \subfloat[r=0.9]{\includegraphics[width=2.5cm]{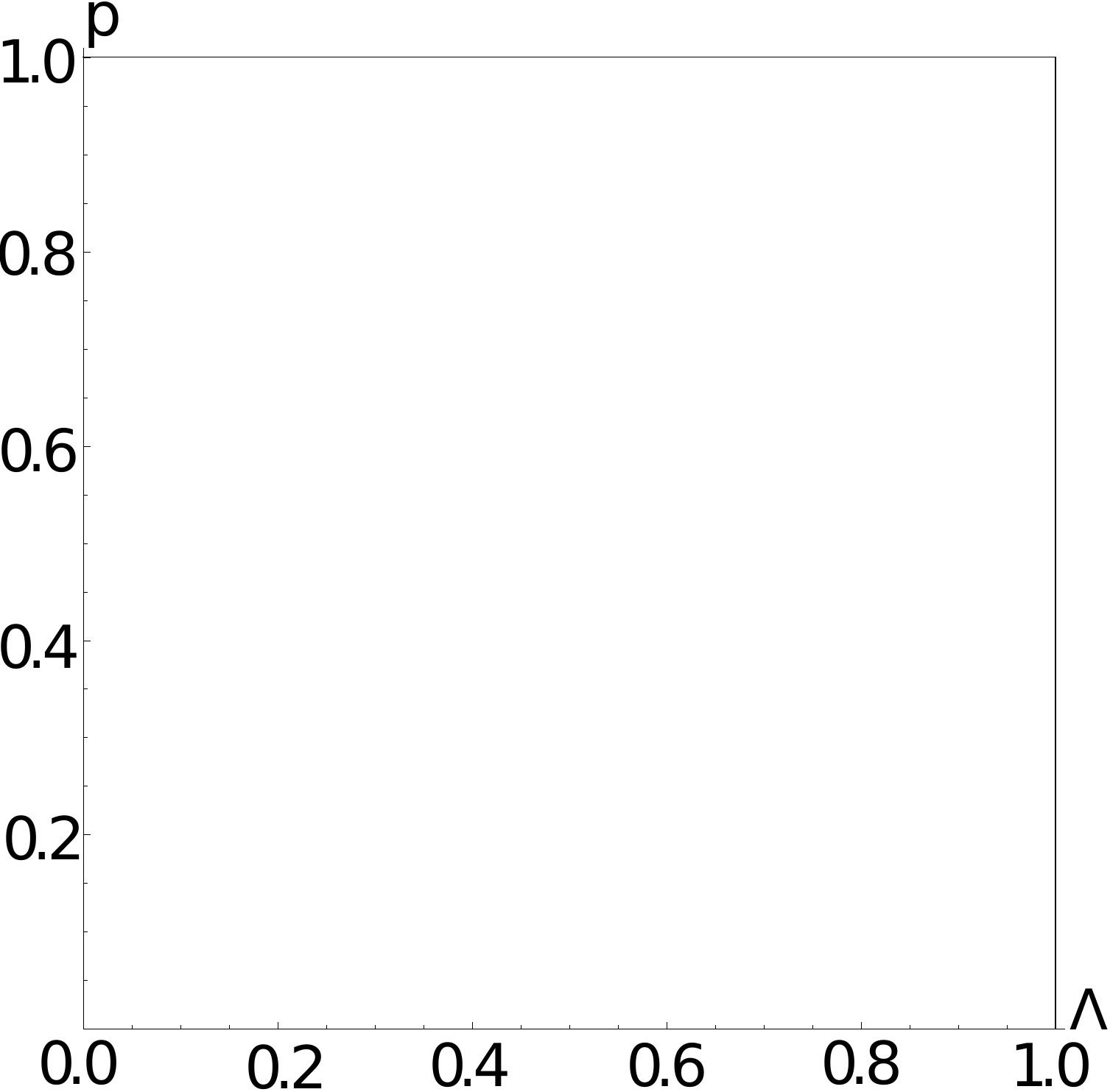}}
& \subfloat[r=0.7]{\includegraphics[width=2.5cm]{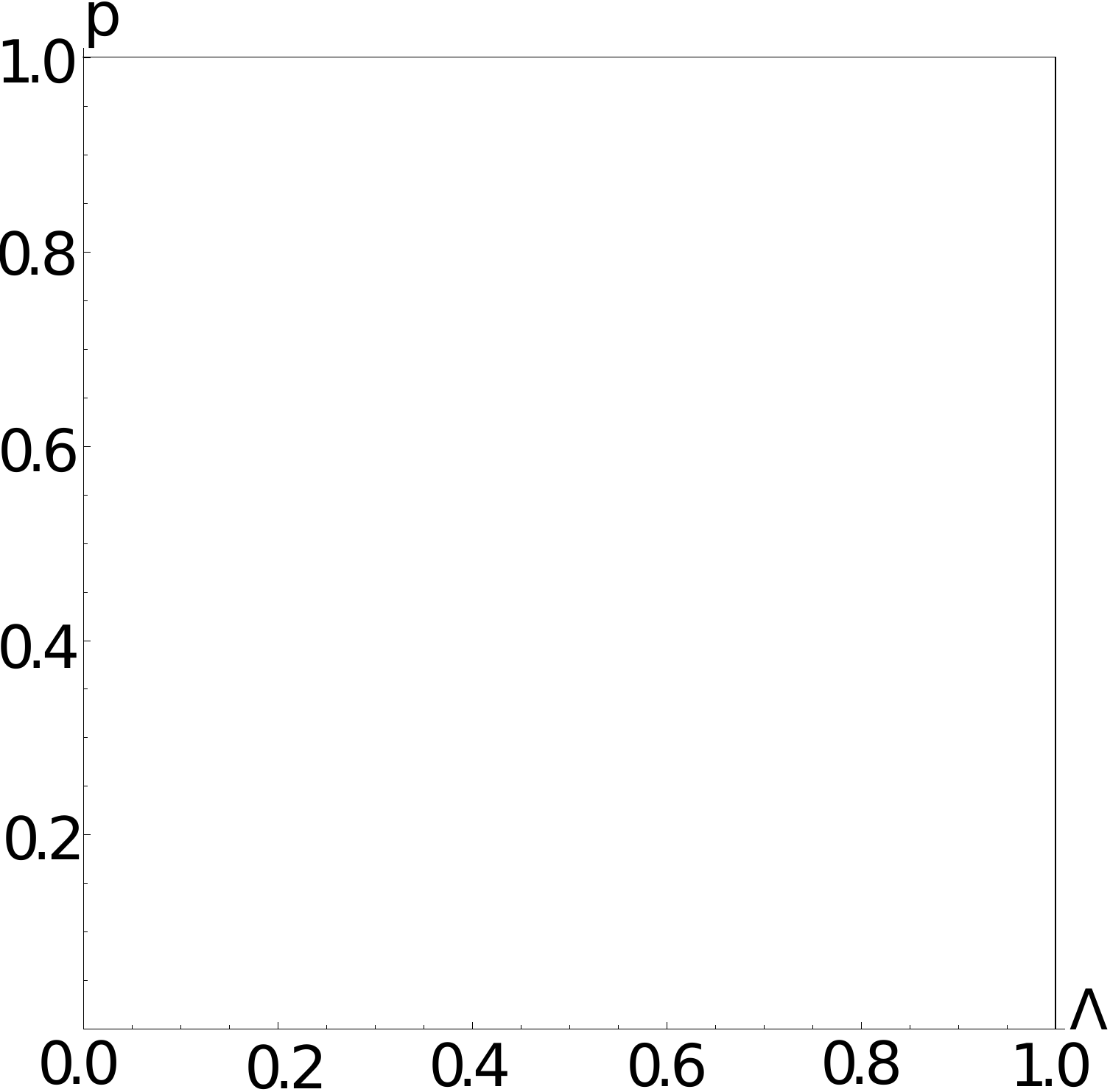}}\\

\subfloat[r=0.5]{\includegraphics[width=2.5cm]{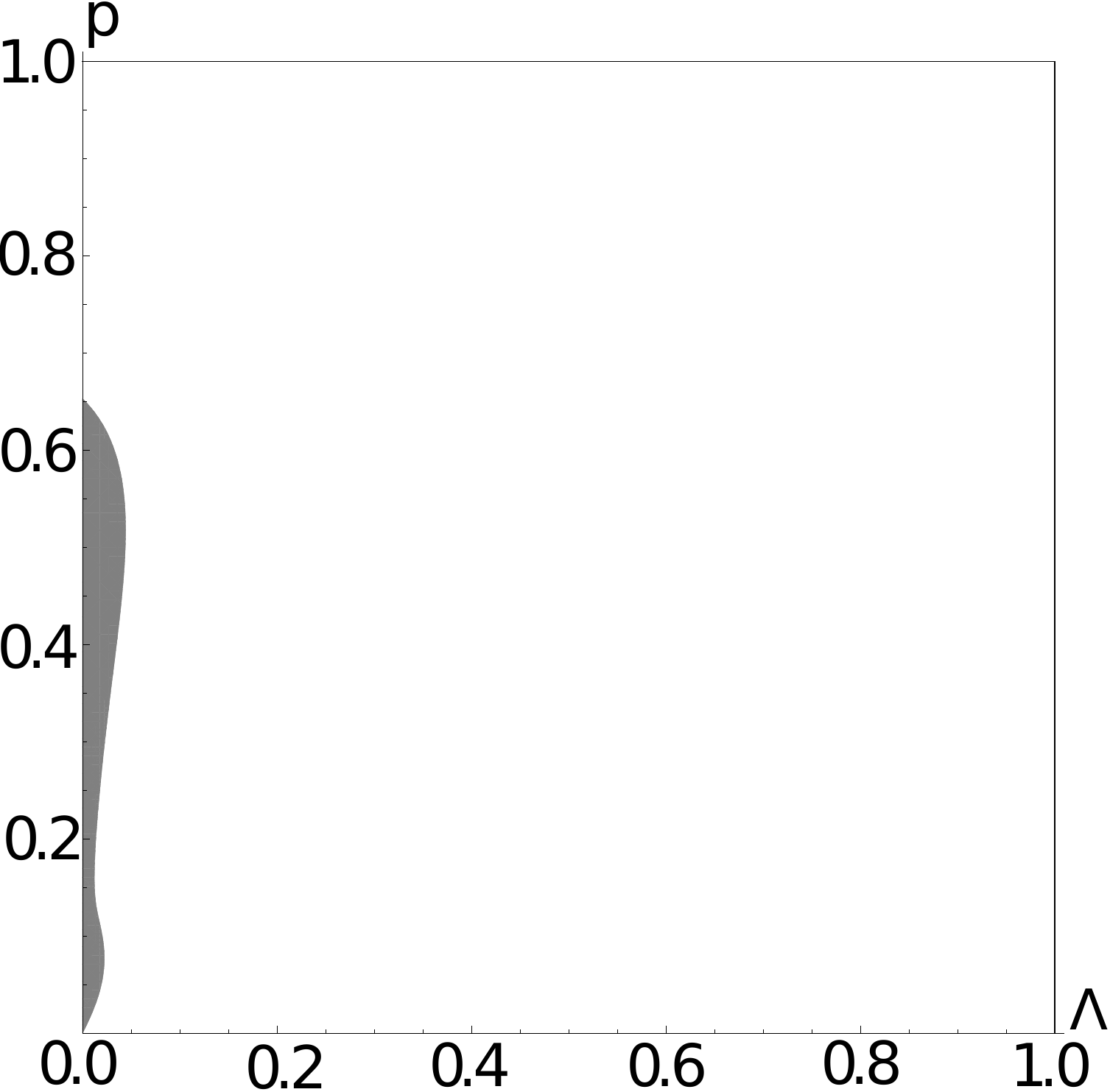}} 
   & \subfloat[r=0.3]{\includegraphics[width=2.5cm]{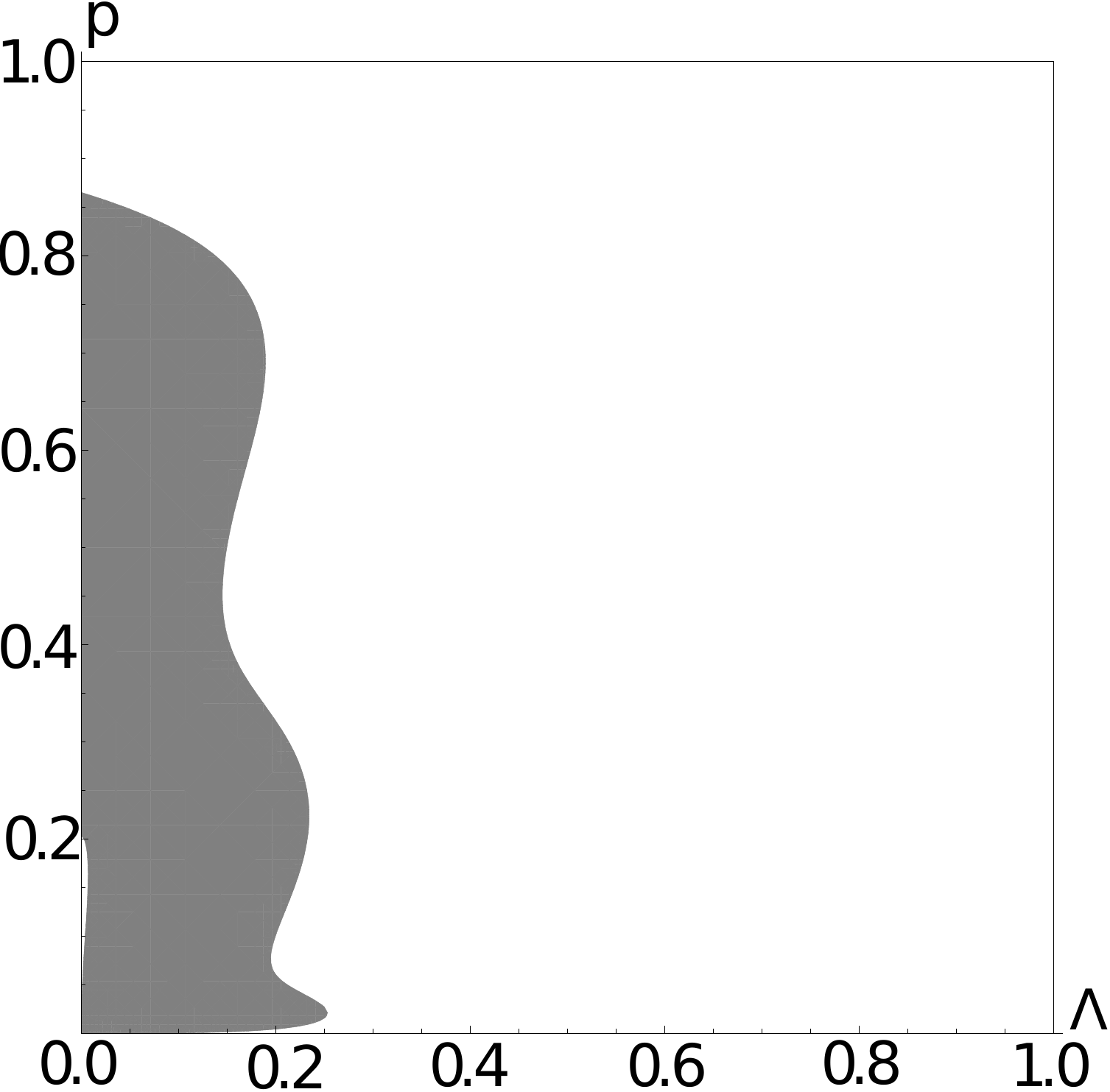}}
& \subfloat[r=0.1]{\includegraphics[width=2.5cm]{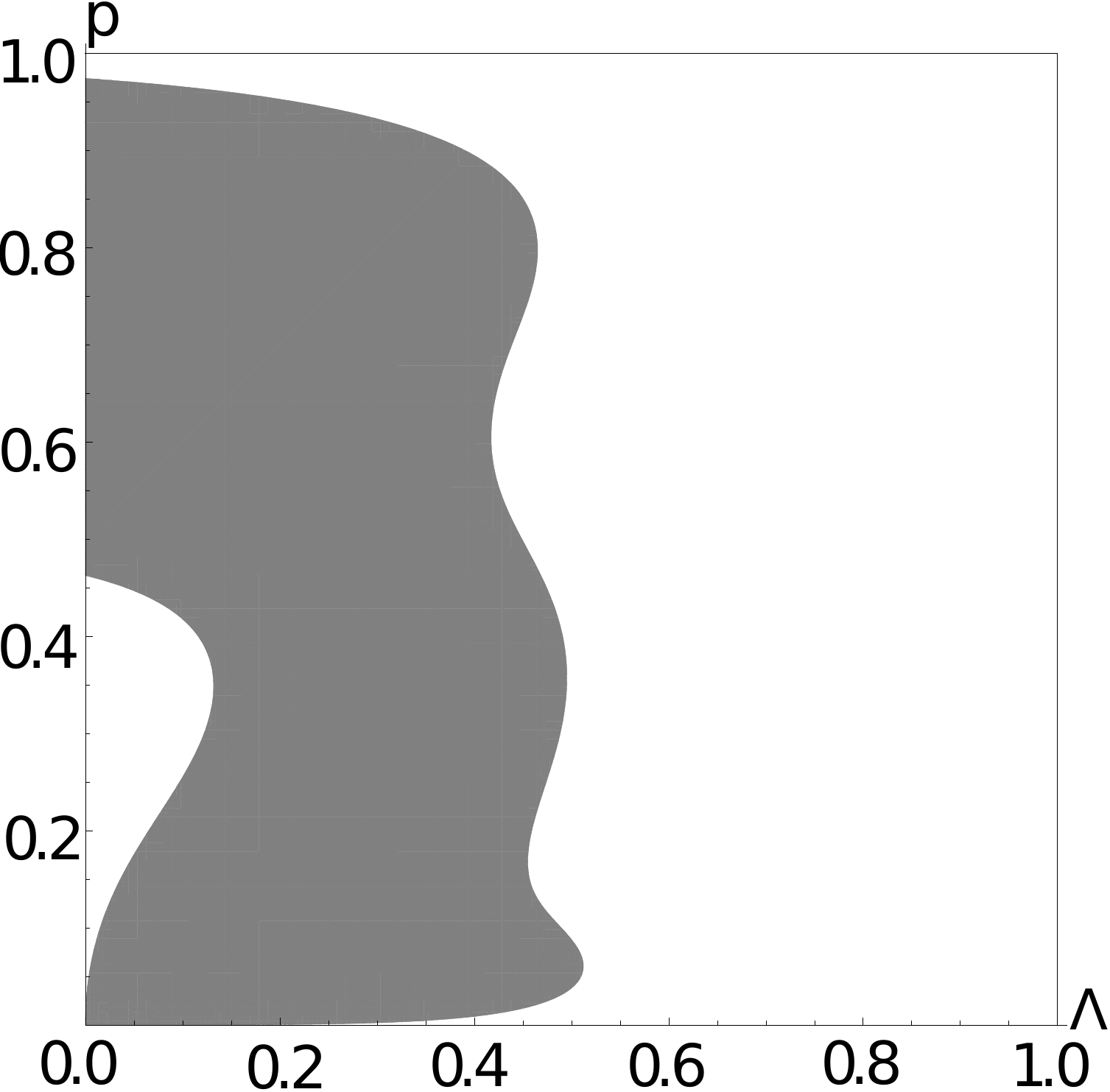}}\\

\subfloat[r=0.05]{\includegraphics[width=2.5cm]{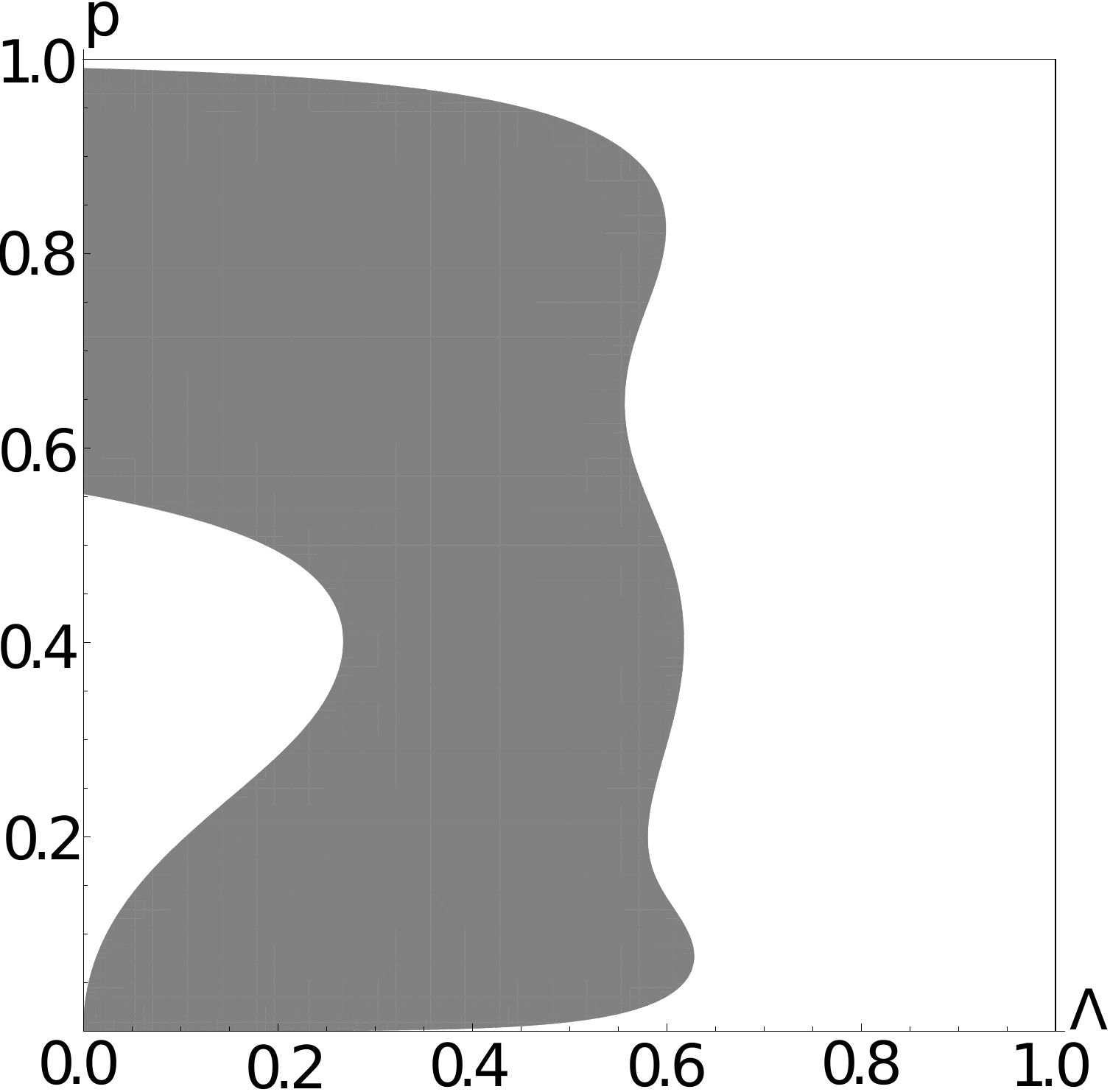}} 
   & \subfloat[r=0.01]{\includegraphics[width=2.5cm]{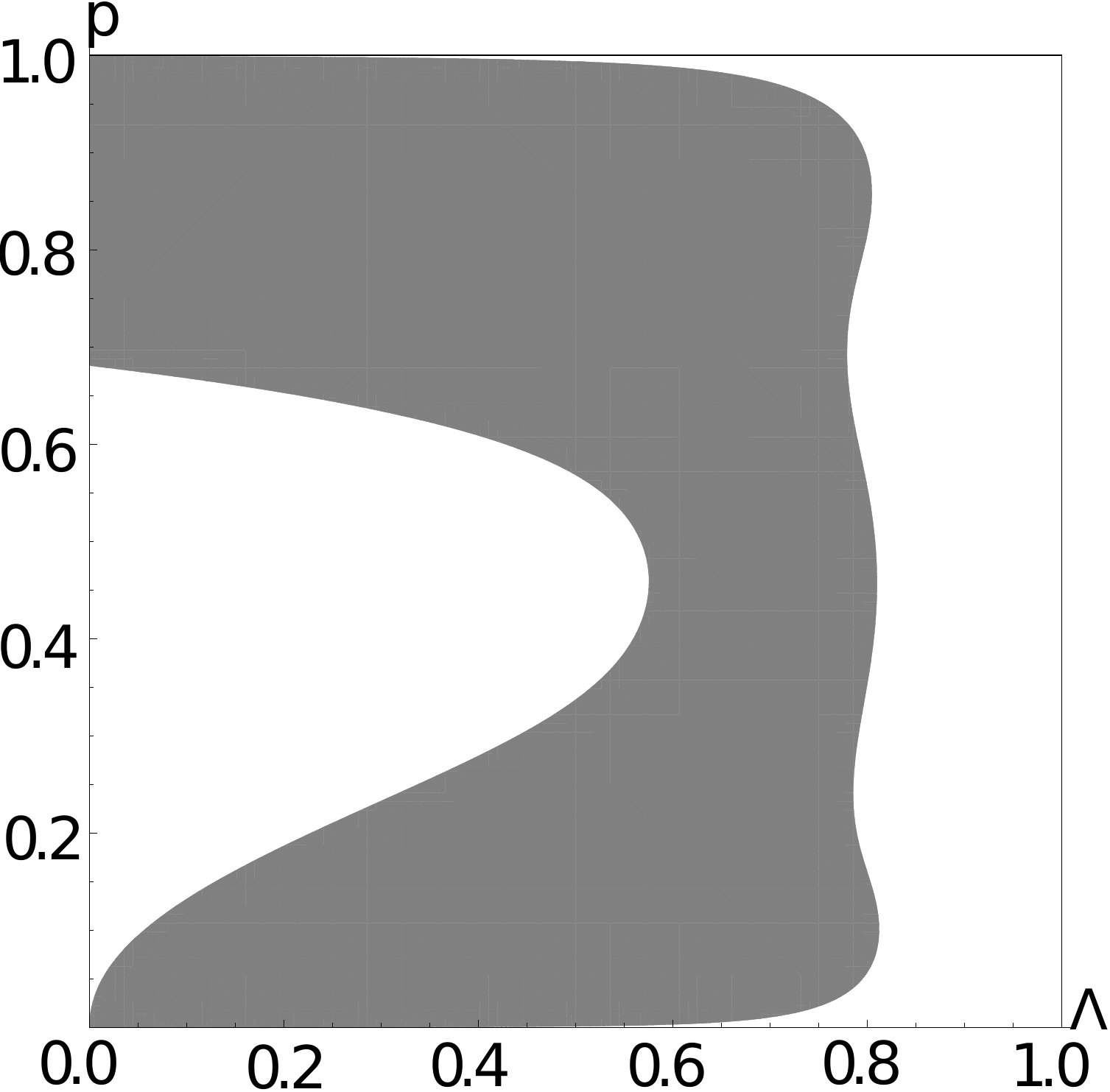}}
& \subfloat[r=0.001]{\includegraphics[width=2.5cm]{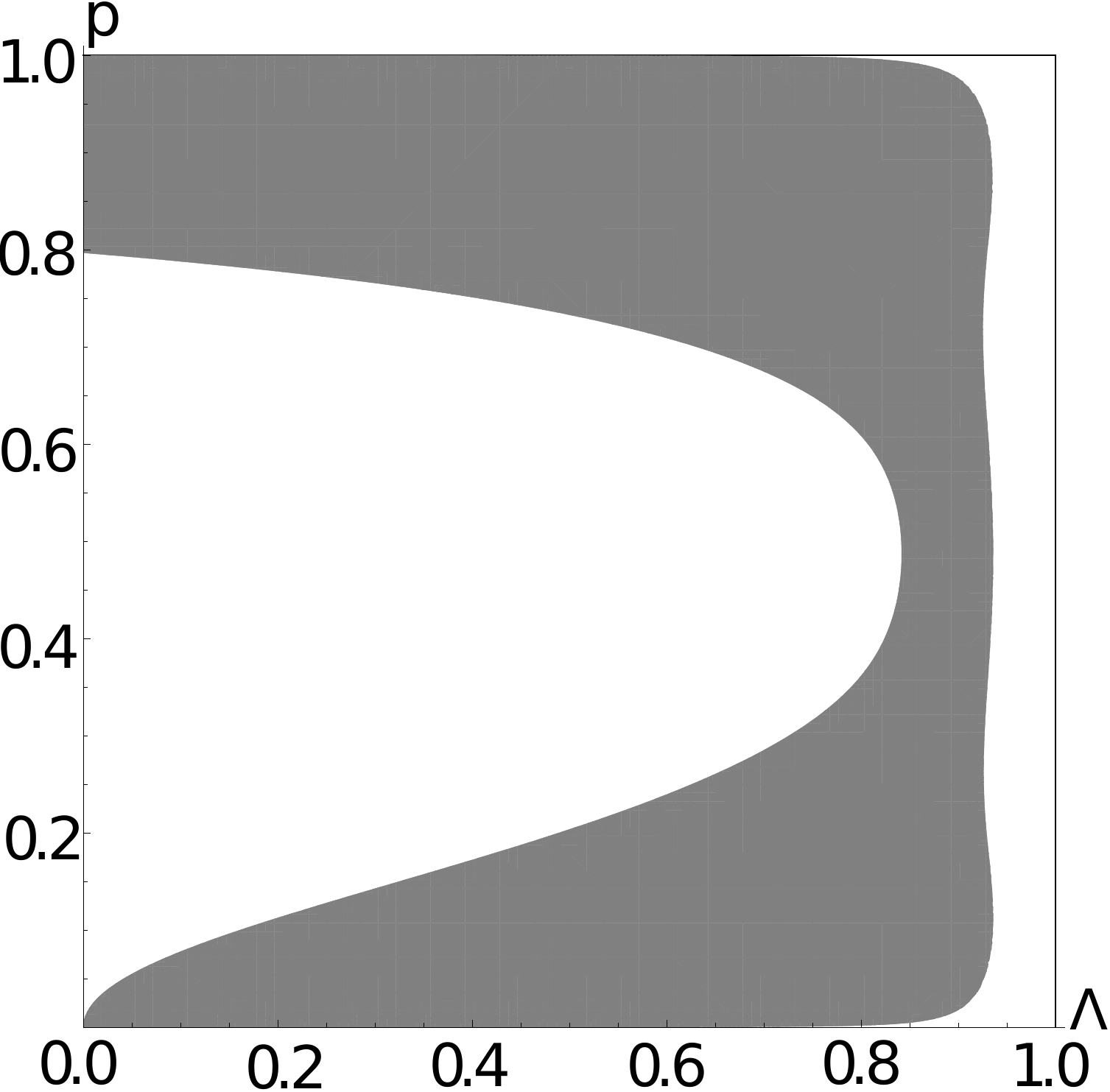}}\\
\end{tabular}

\caption{
Dark regions correspond to non-positivity of $\det Q$
 for the case when $Q$ is approximated by the $4\times 4$ matrix built of double moments.
}
\label{fig44}
\end{figure}

\begin{figure}[]

\centering

\begin{tabular}{ccc}
\subfloat[r=0.99]{\includegraphics[width=2.5cm]{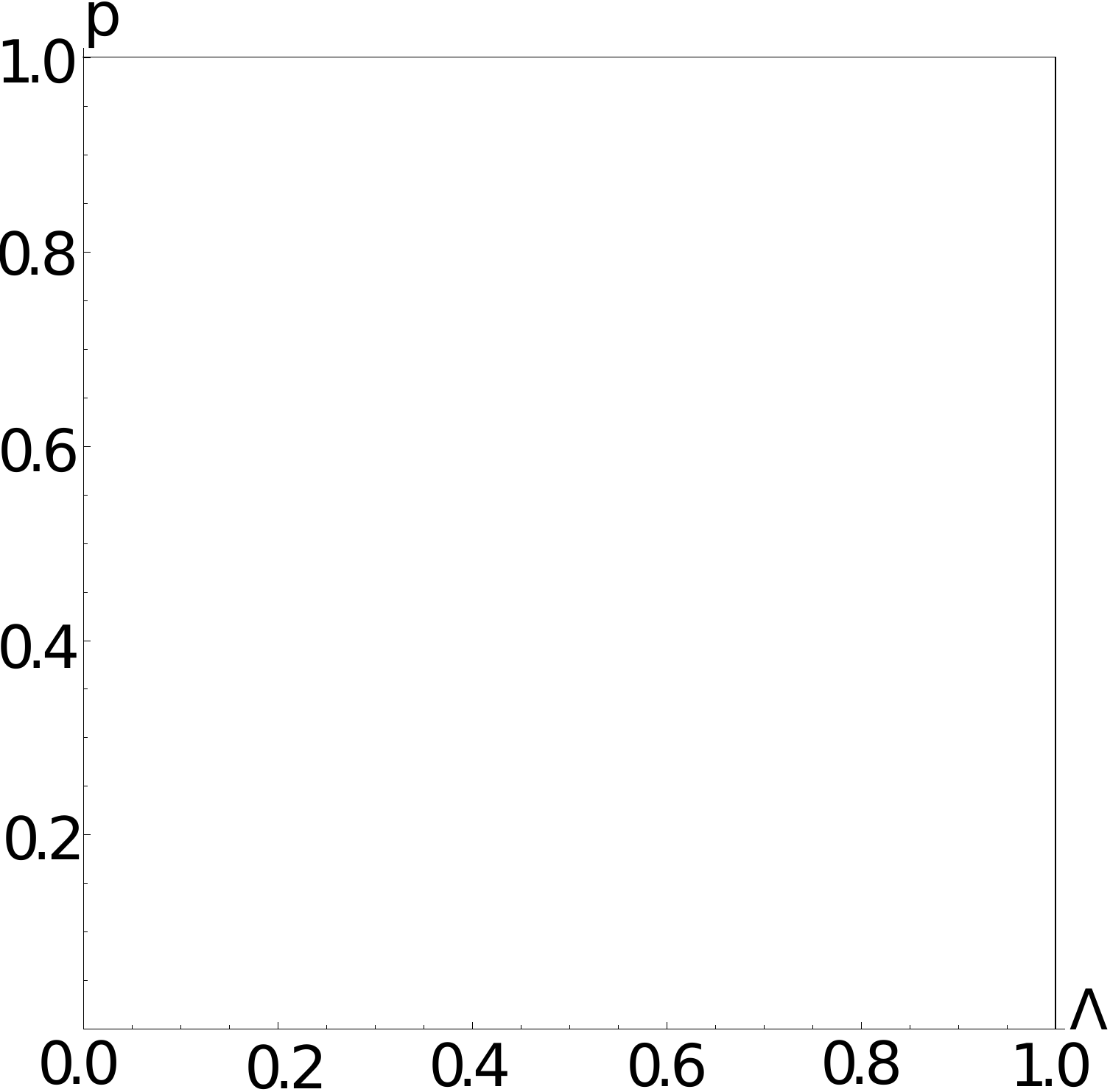}} 
   & \subfloat[r=0.9]{\includegraphics[width=2.5cm]{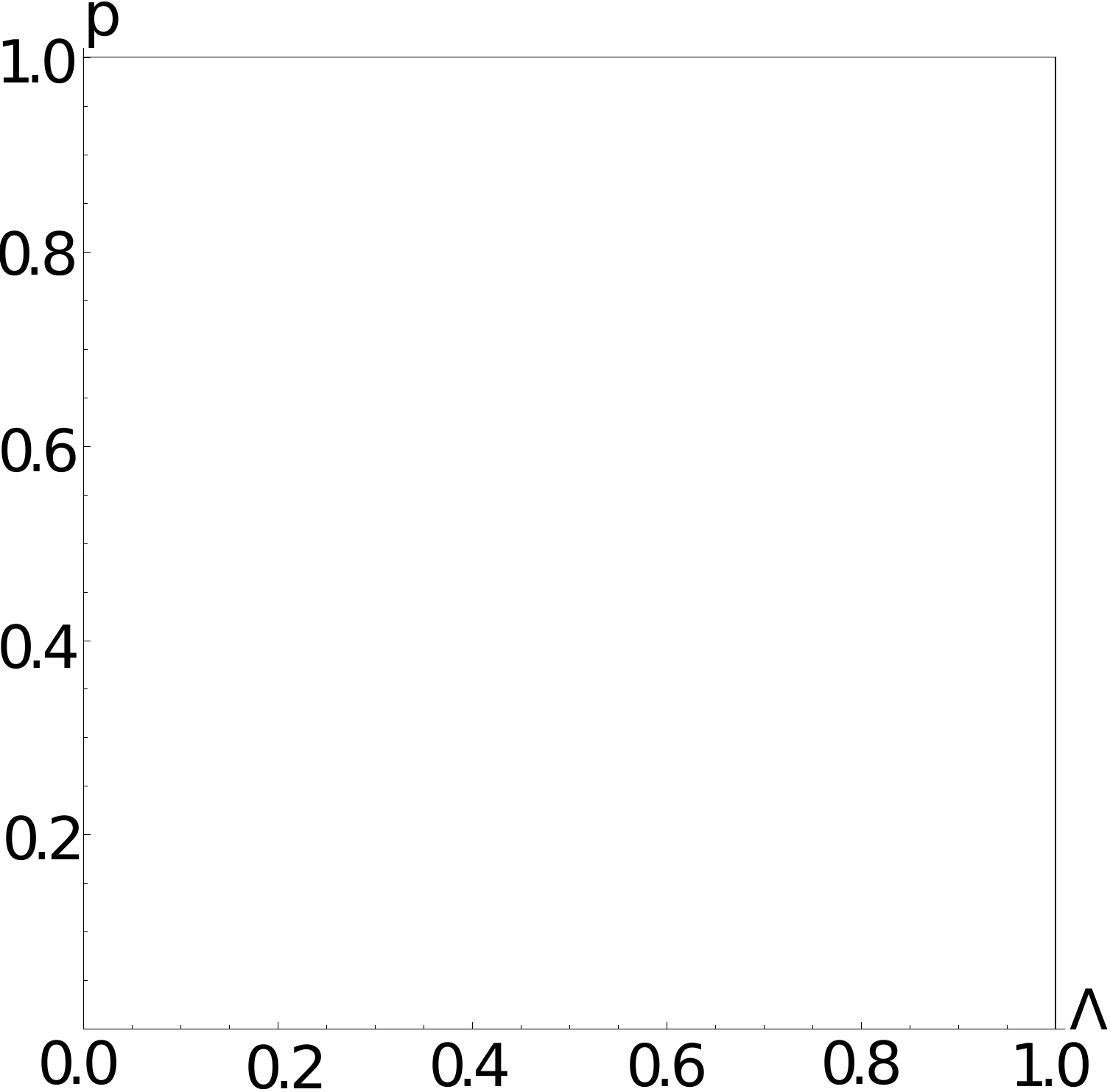}}
& \subfloat[r=0.7]{\includegraphics[width=2.5cm]{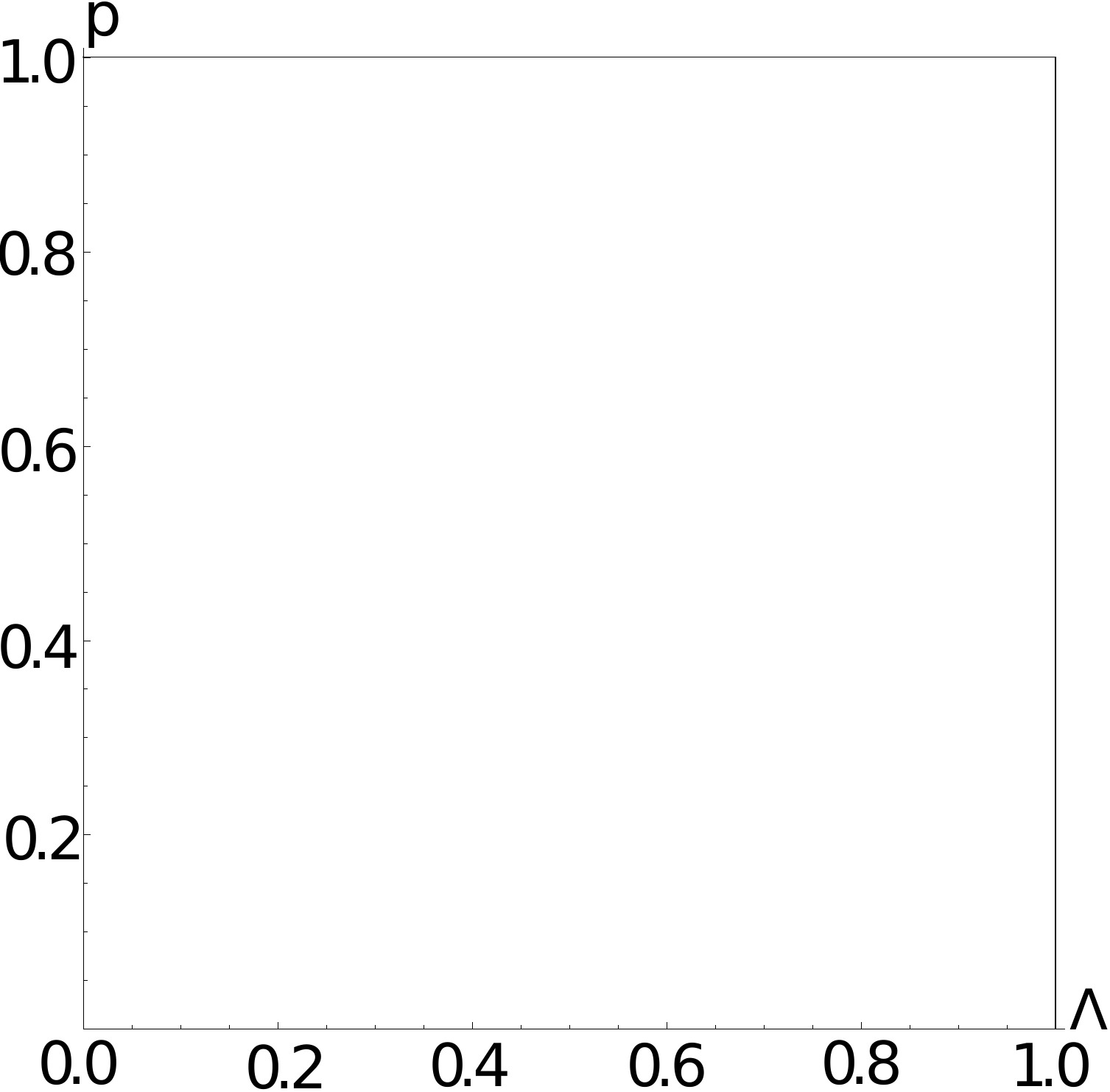}}\\

\subfloat[r=0.5]{\includegraphics[width=2.5cm]{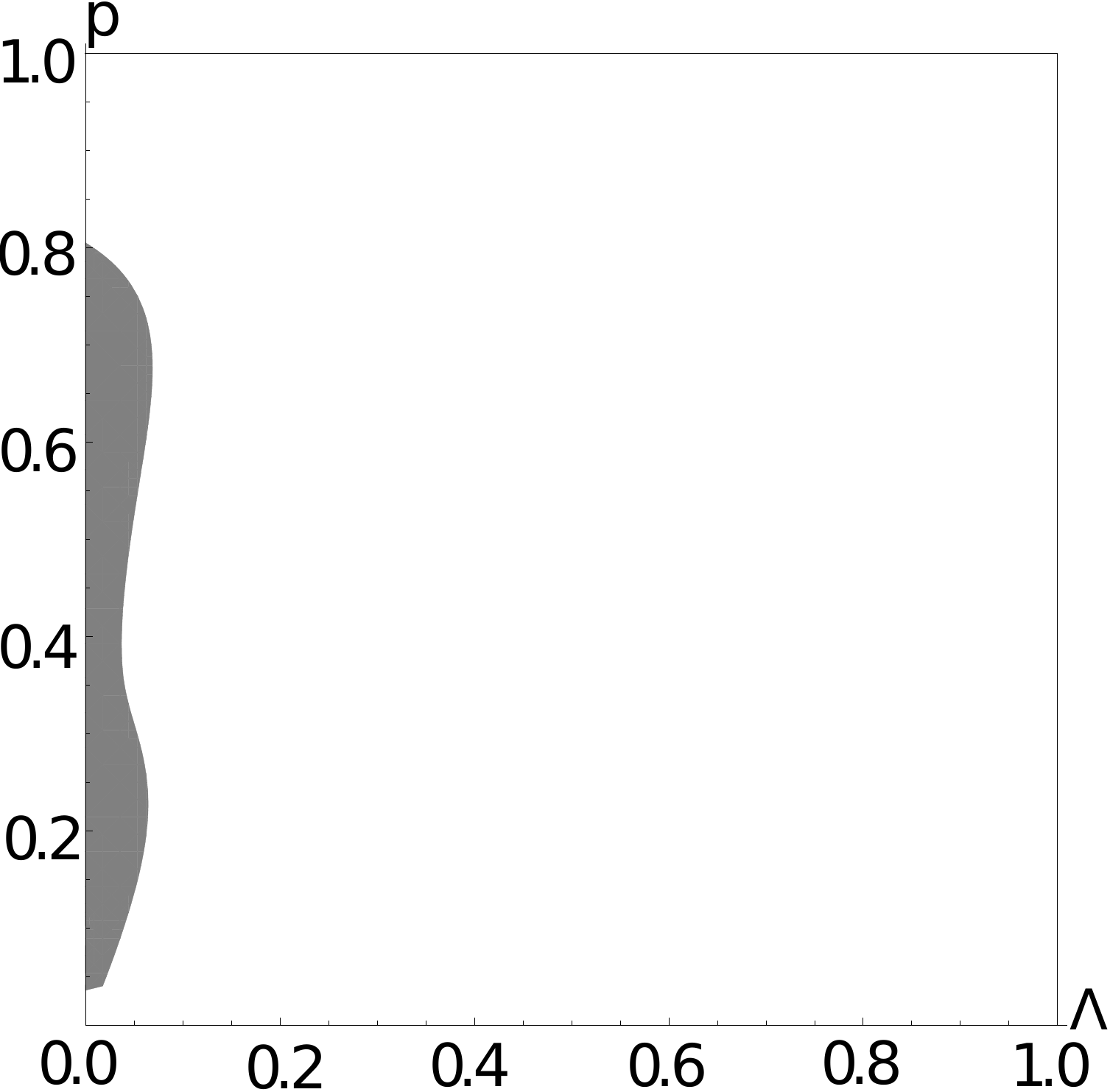}} 
   & \subfloat[r=0.3]{\includegraphics[width=2.5cm]{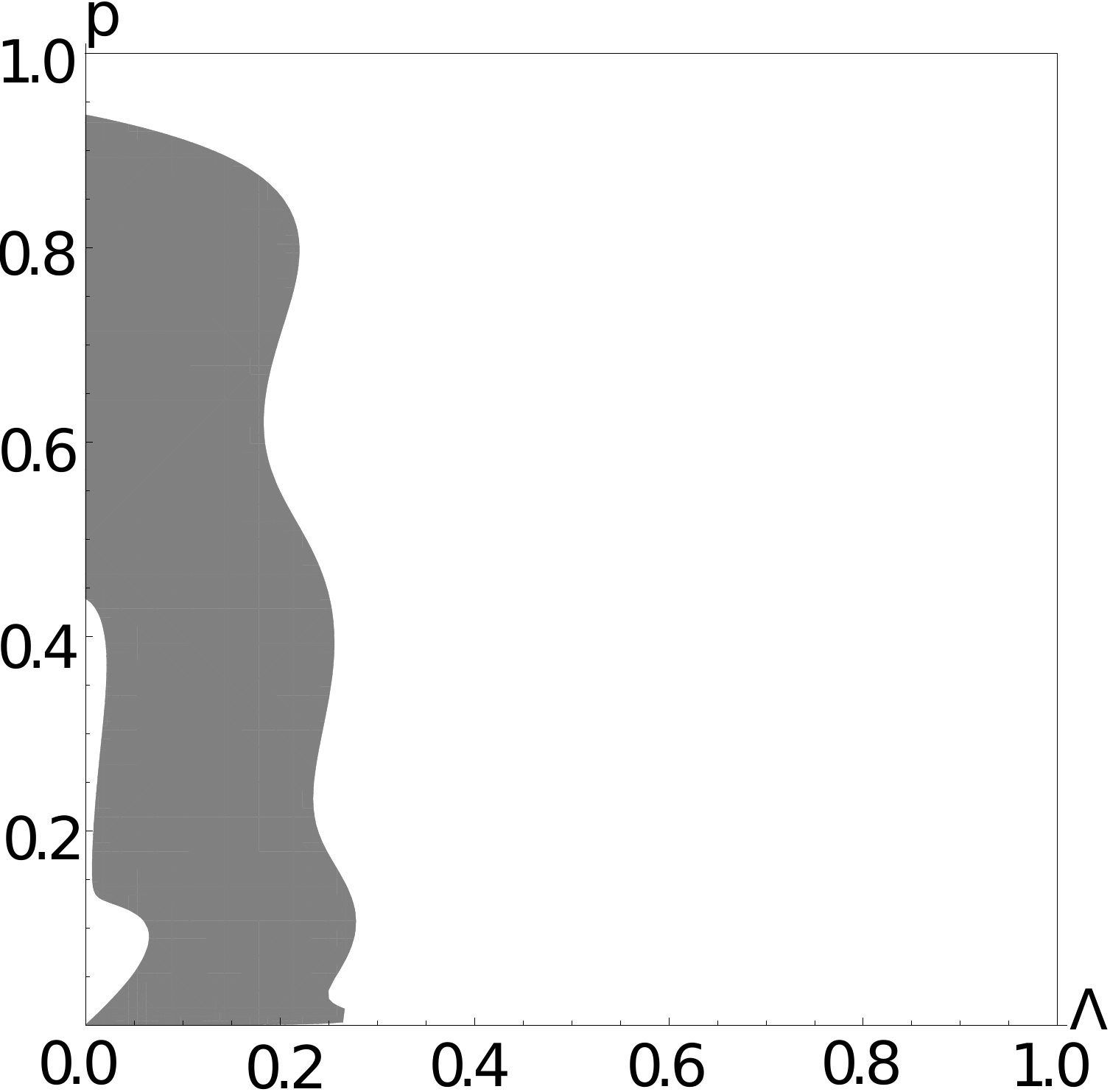}}
& \subfloat[r=0.1]{\includegraphics[width=2.5cm]{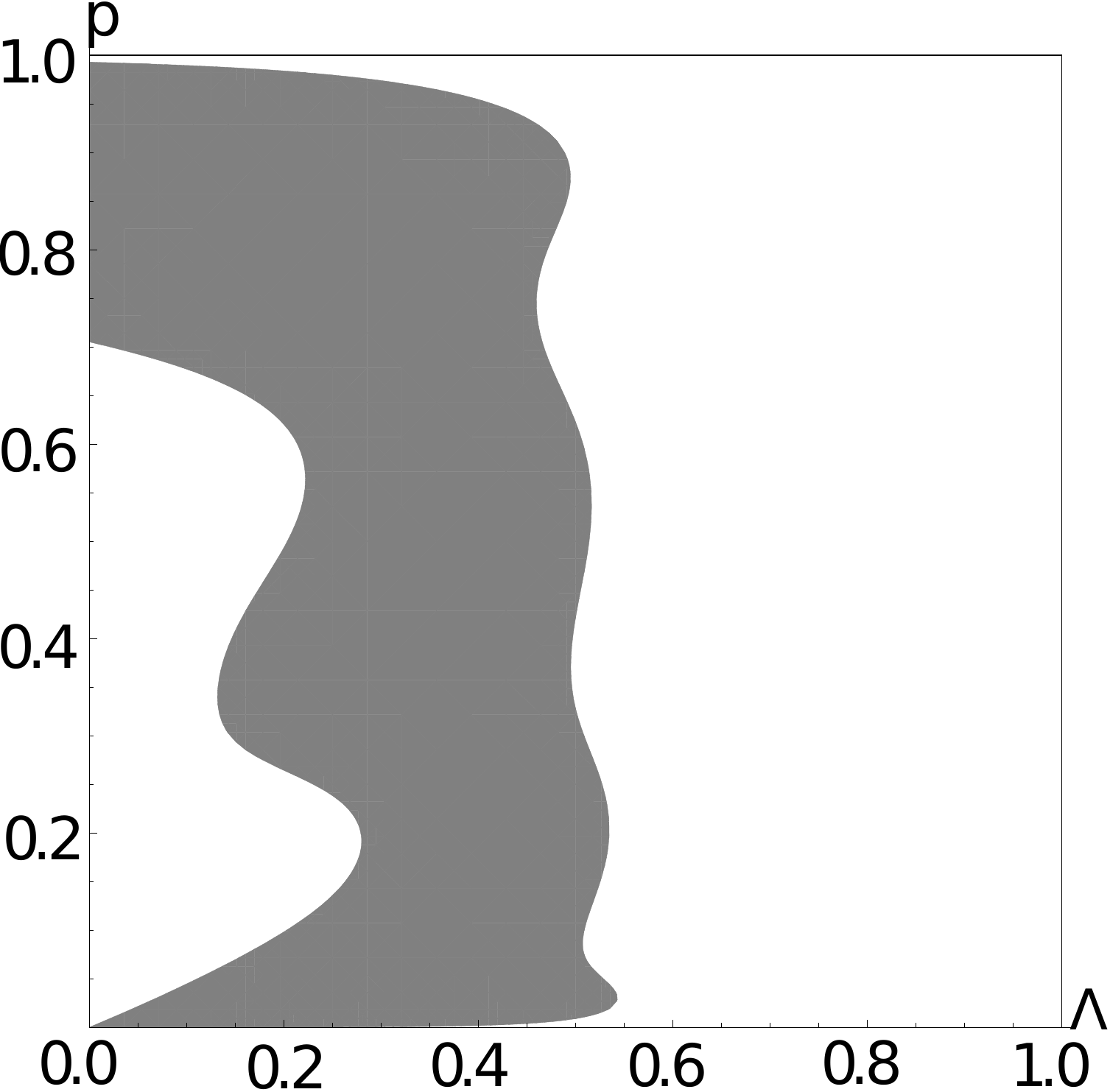}}\\

\subfloat[r=0.05]{\includegraphics[width=2.5cm]{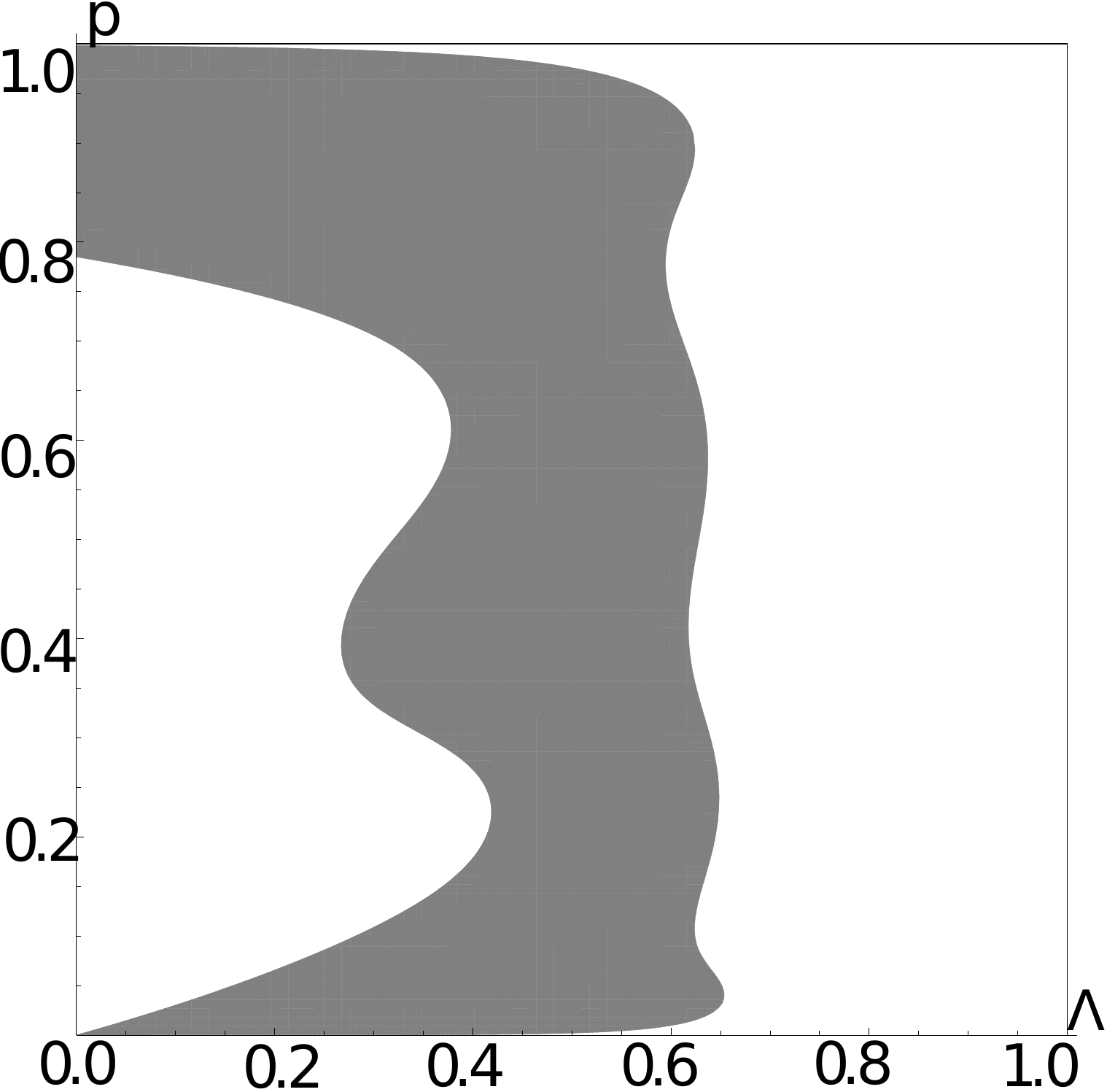}} 
   & \subfloat[r=0.01]{\includegraphics[width=2.5cm]{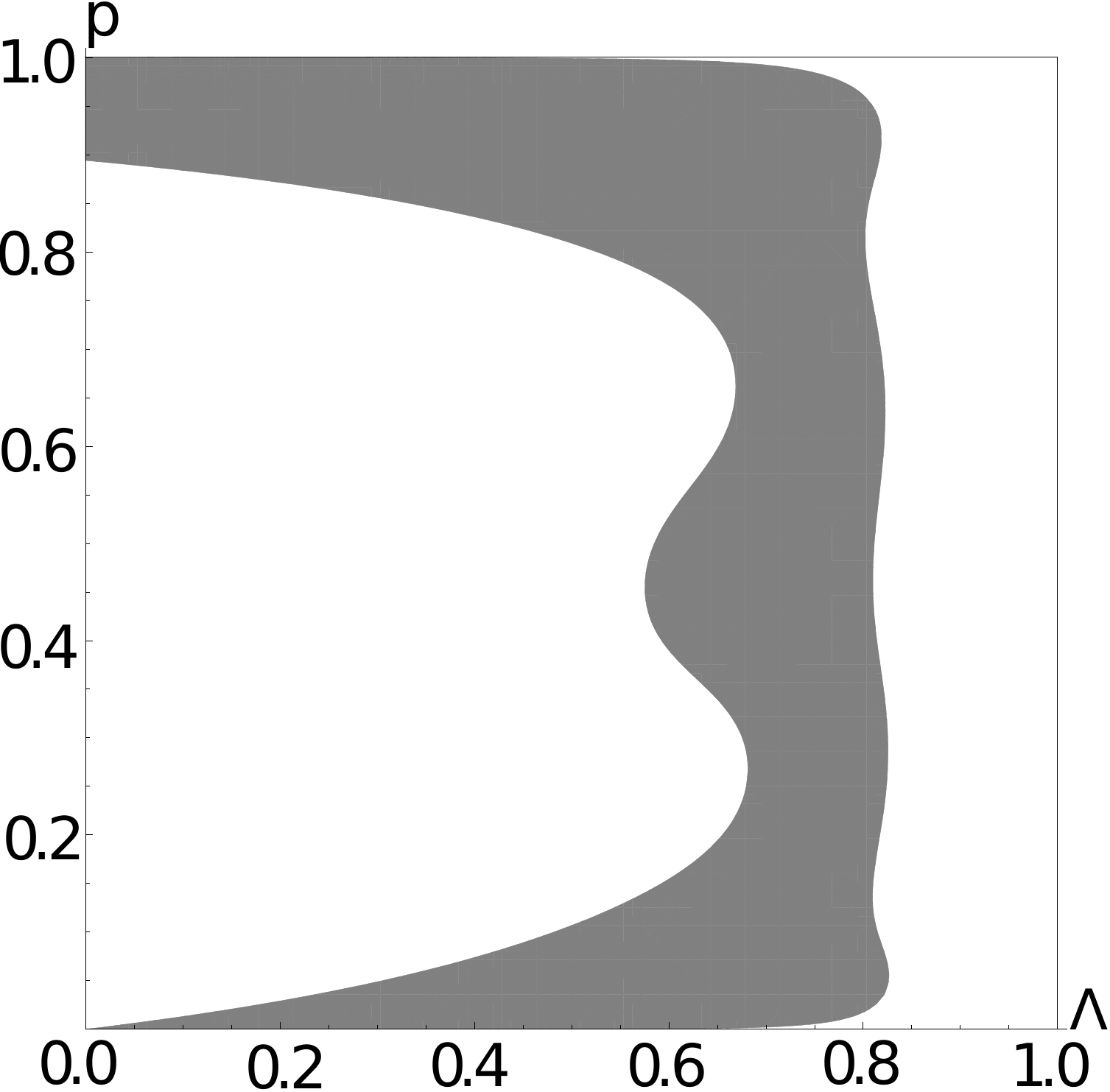}}
& \subfloat[r=0.001]{\includegraphics[width=2.5cm]{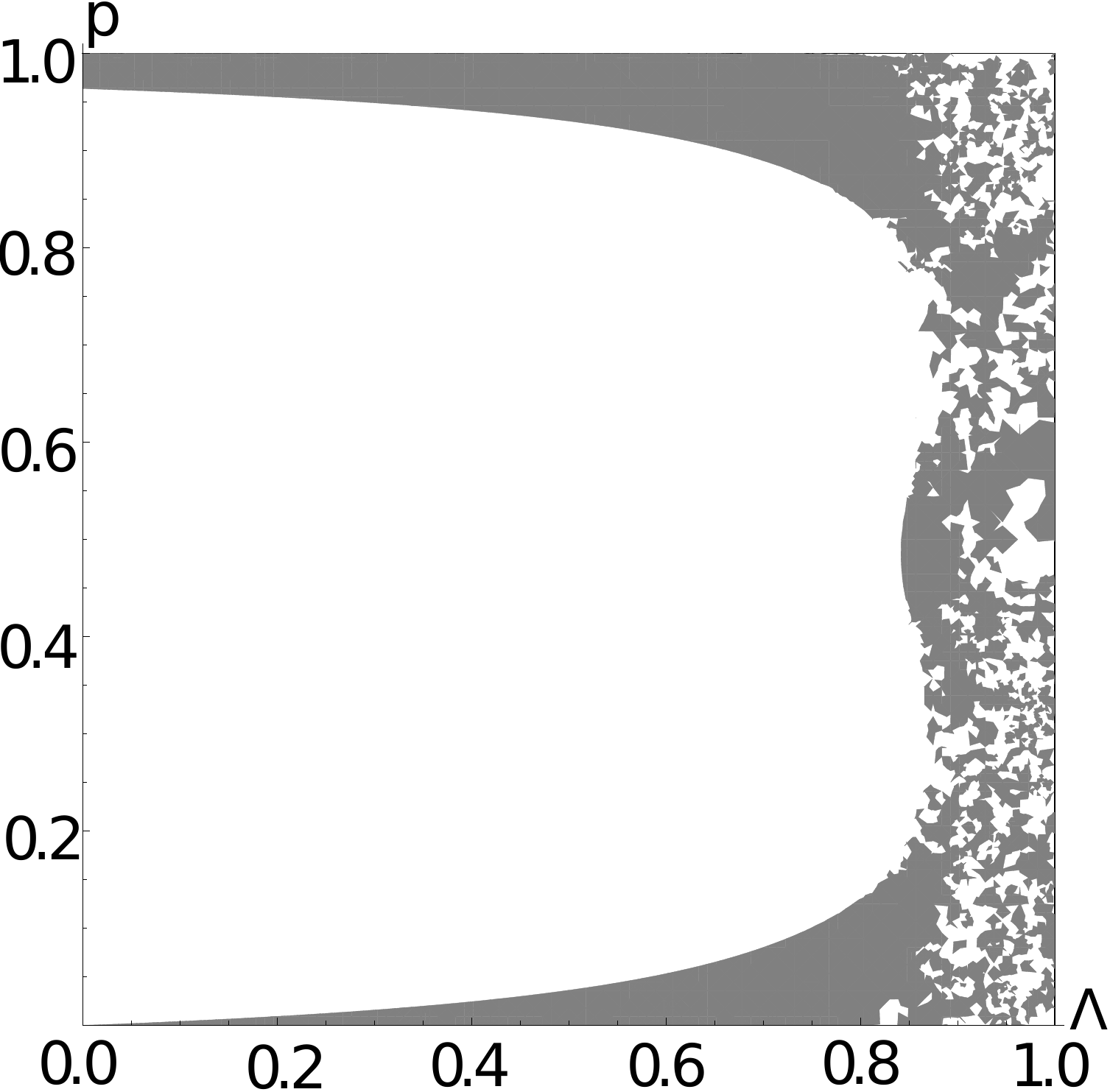}}\\
\end{tabular}

\caption{
Dark regions correspond to non-positivity of $\det Q$
 for the case when $Q$ is approximated by the $5 \times 5$ matrix built of double moments.
}
\label{fig55}
\end{figure}

\begin{figure}[]

\centering

\begin{tabular}{ccc}
\subfloat[r=0.99]{\includegraphics[width=2.5cm]{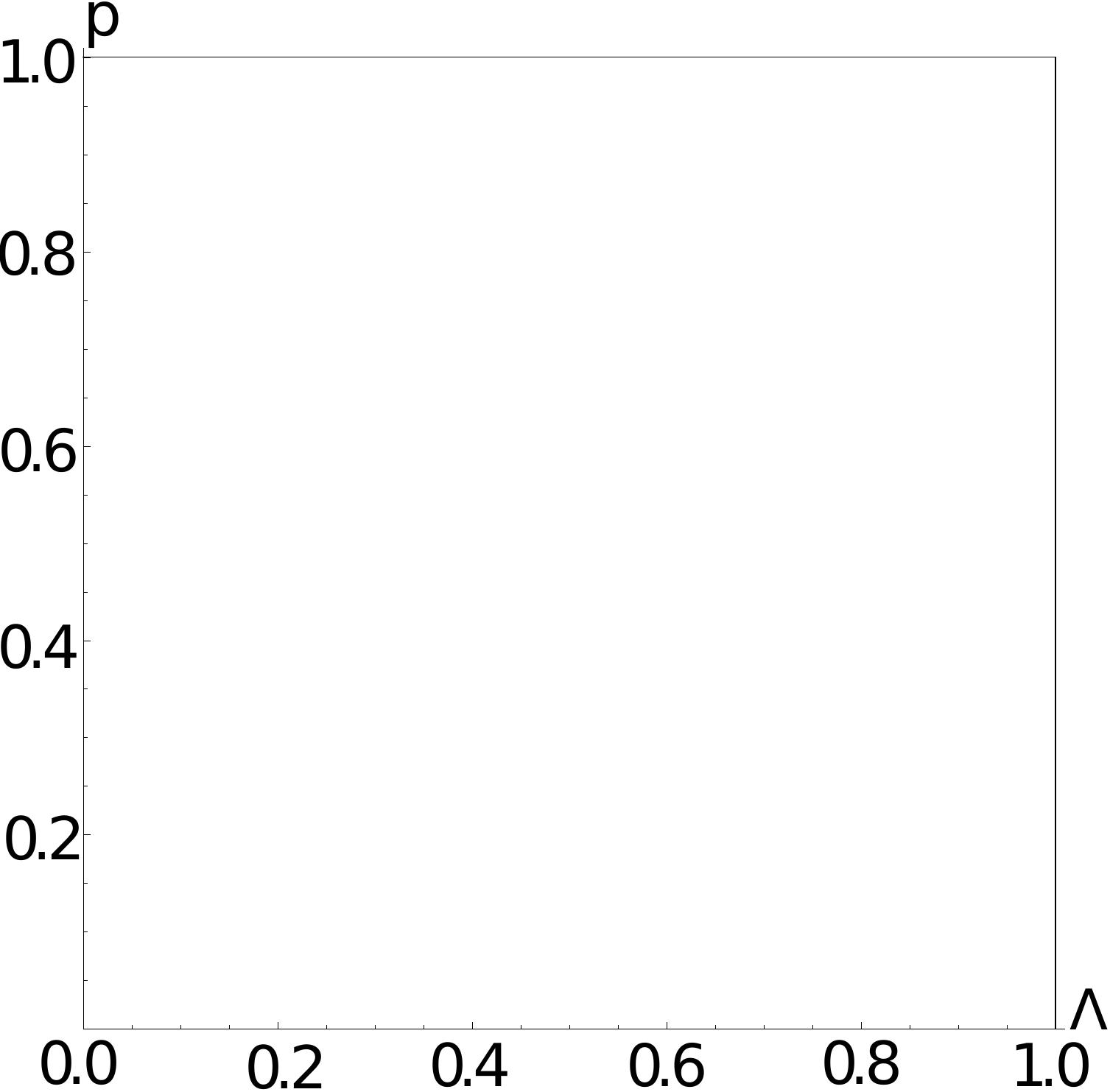}} 
   & \subfloat[r=0.9]{\includegraphics[width=2.5cm]{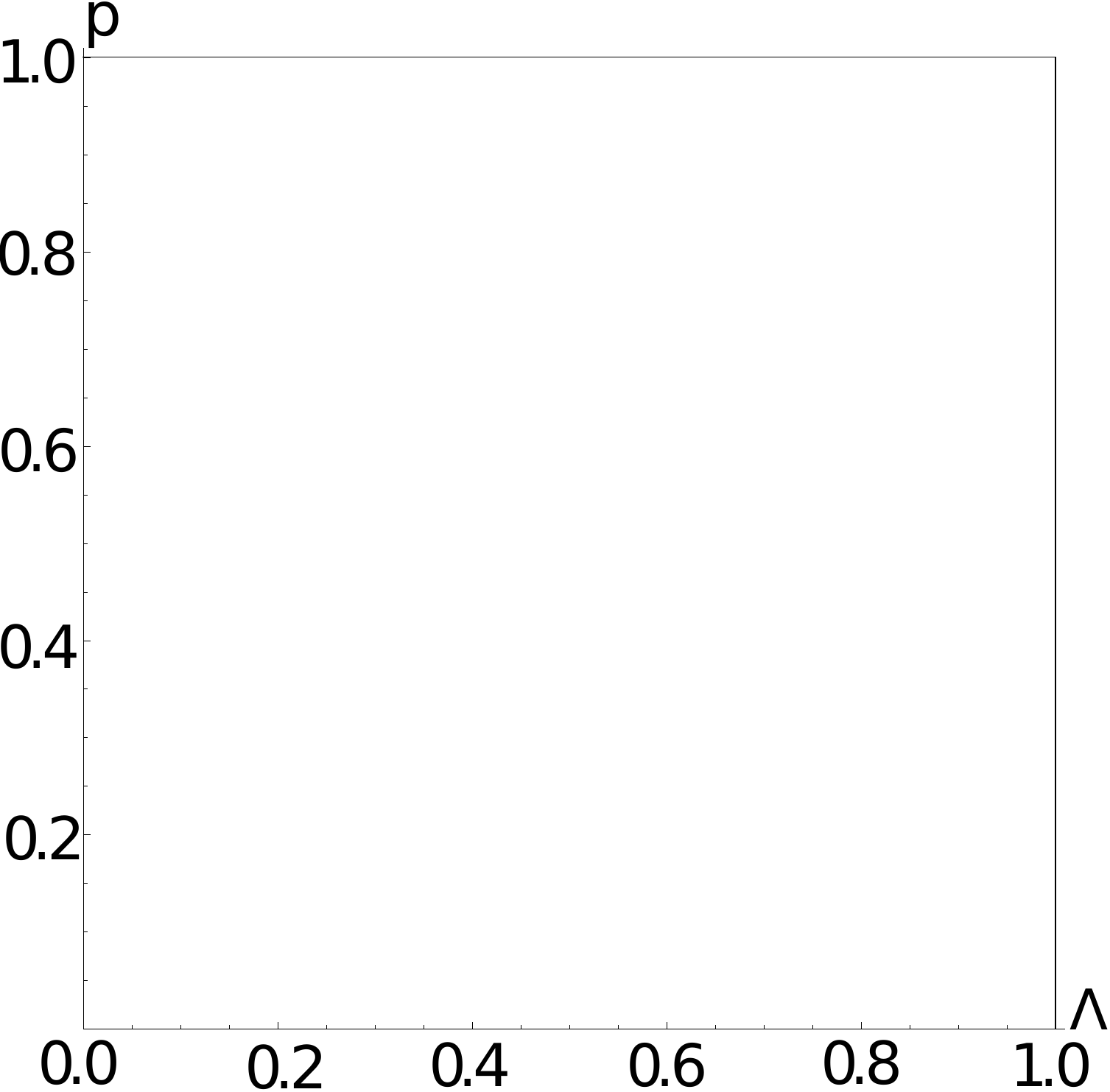}}
& \subfloat[r=0.7]{\includegraphics[width=2.5cm]{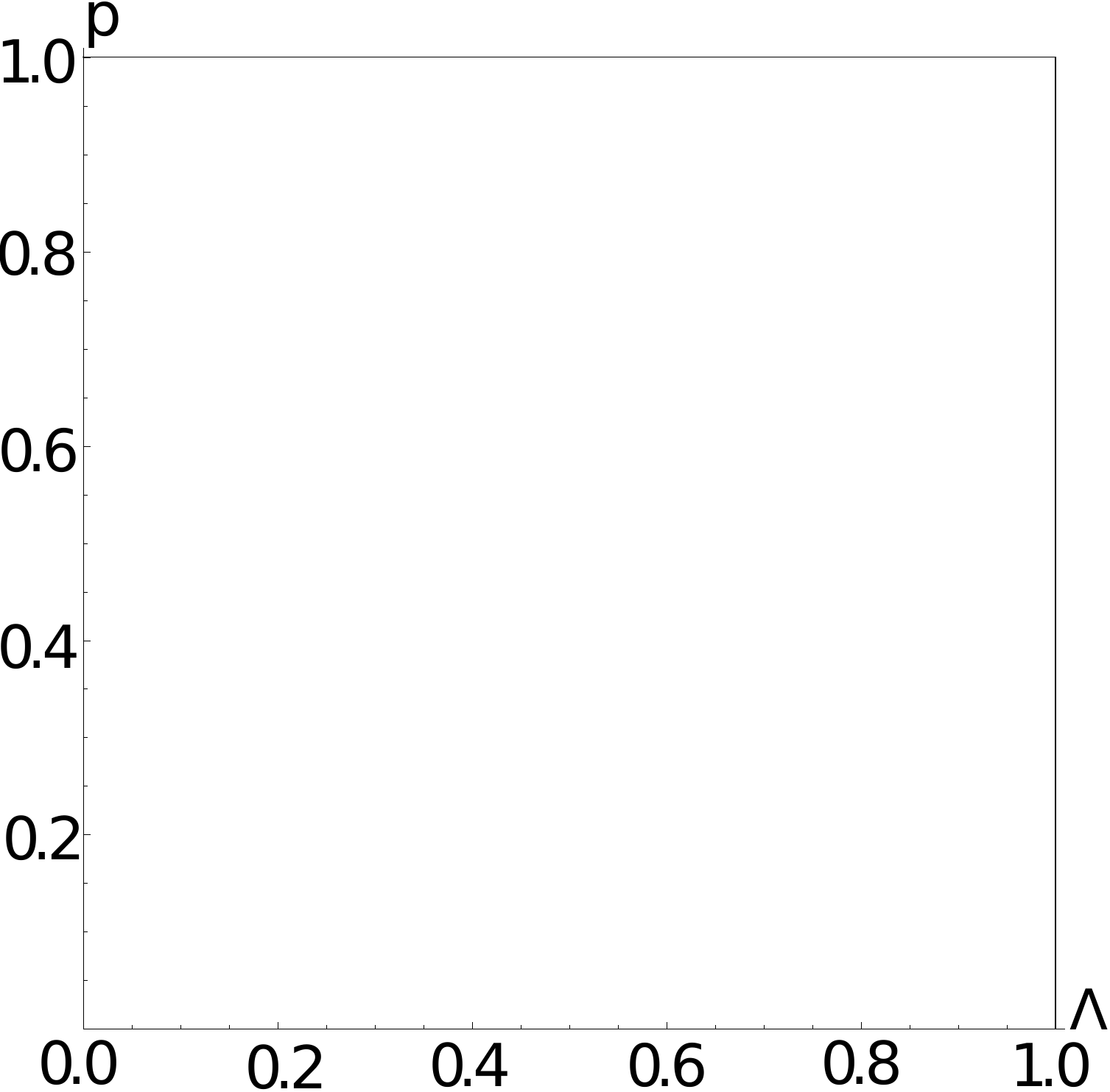}}\\

\subfloat[r=0.5]{\includegraphics[width=2.5cm]{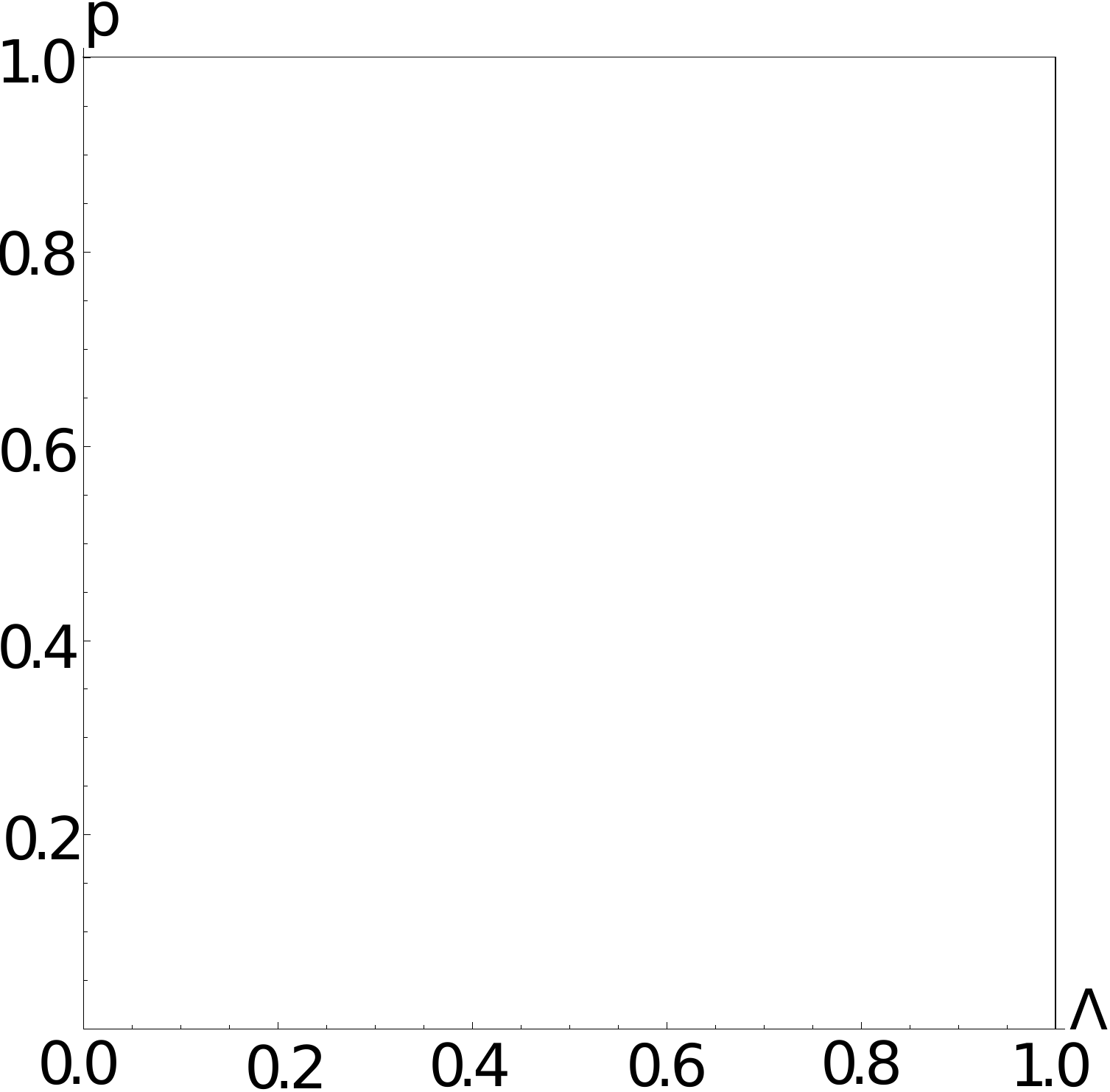}} 
   & \subfloat[r=0.3]{\includegraphics[width=2.5cm]{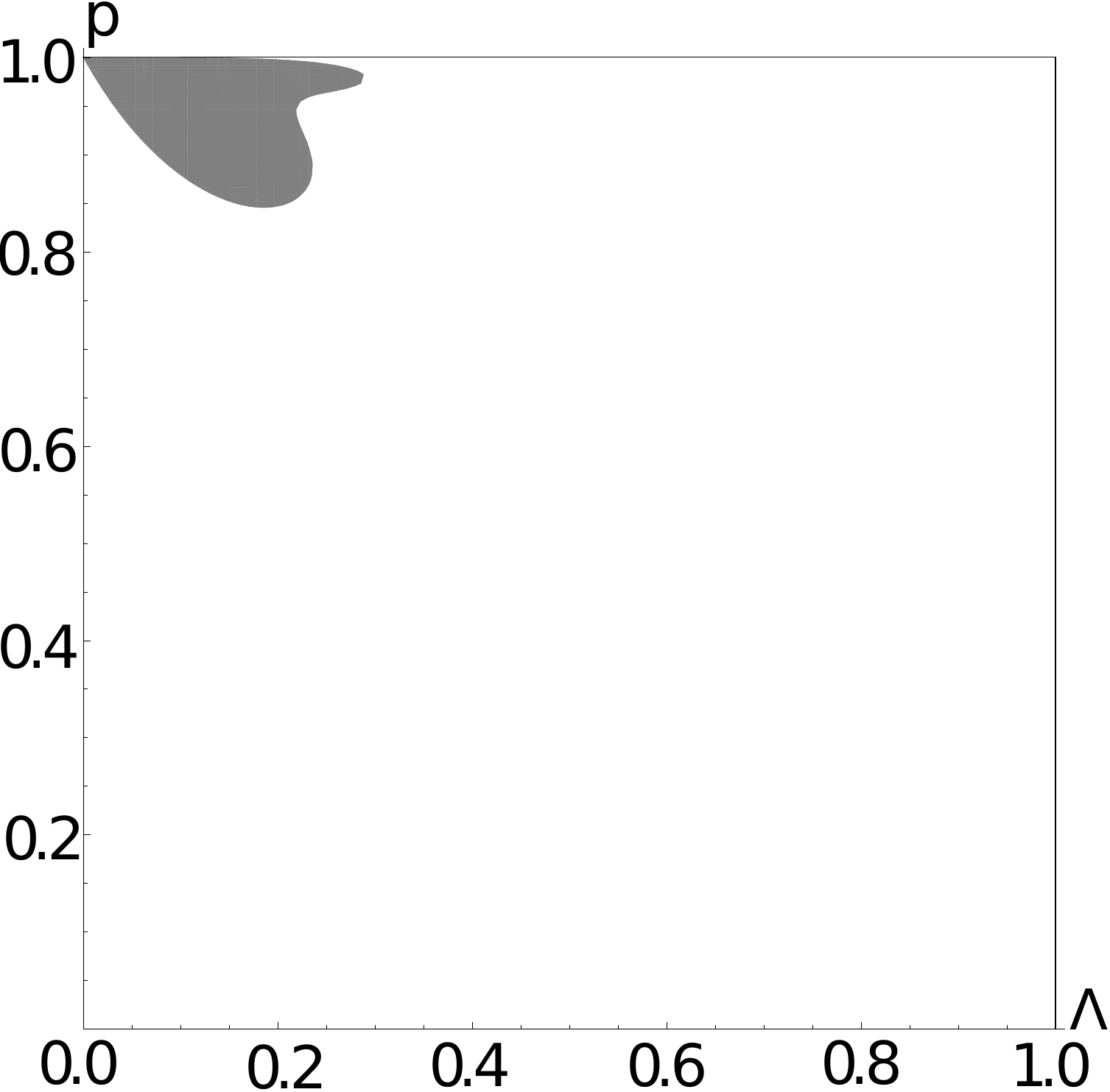}}
& \subfloat[r=0.1]{\includegraphics[width=2.5cm]{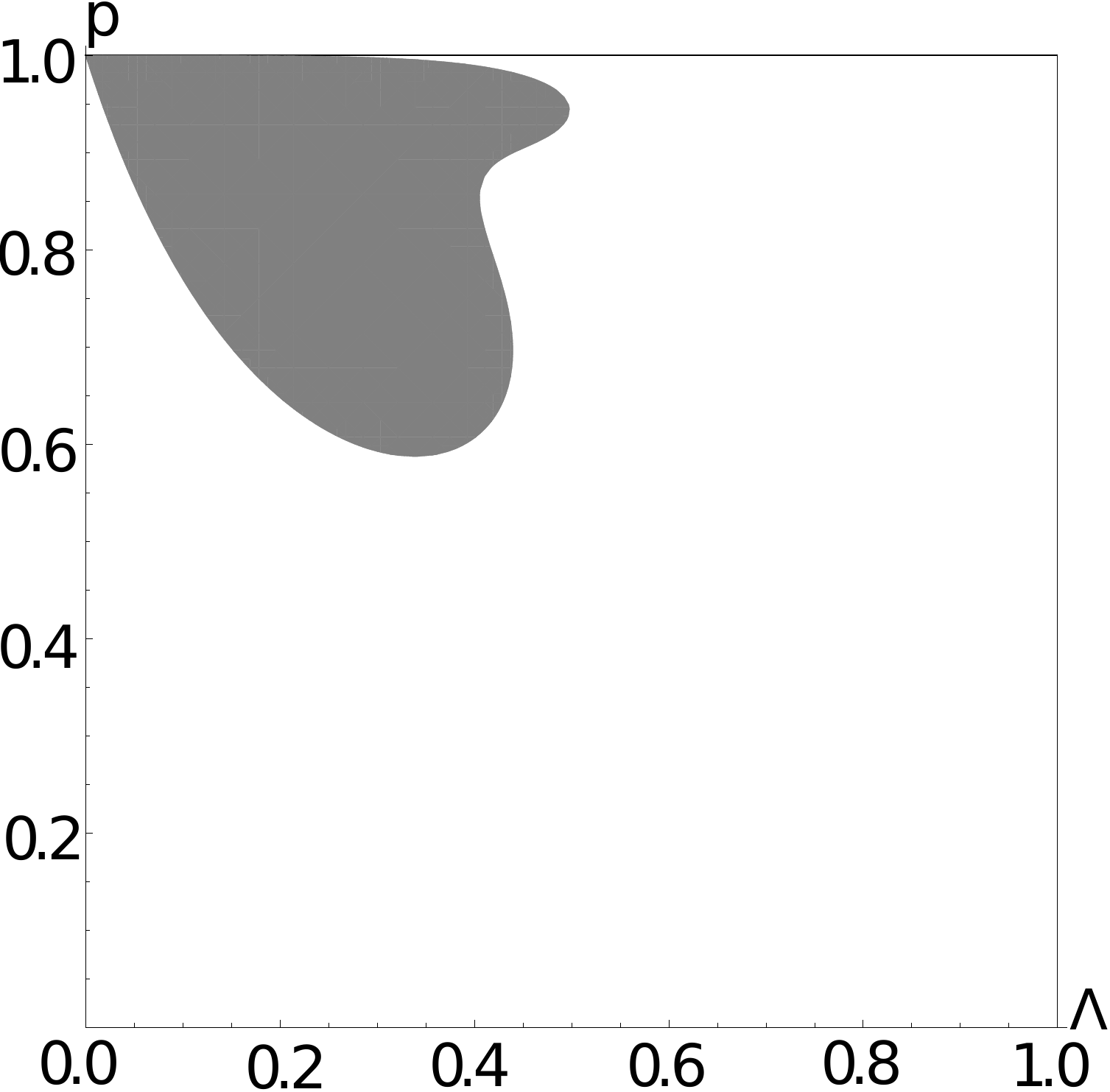}}\\

\subfloat[r=0.05]{\includegraphics[width=2.5cm]{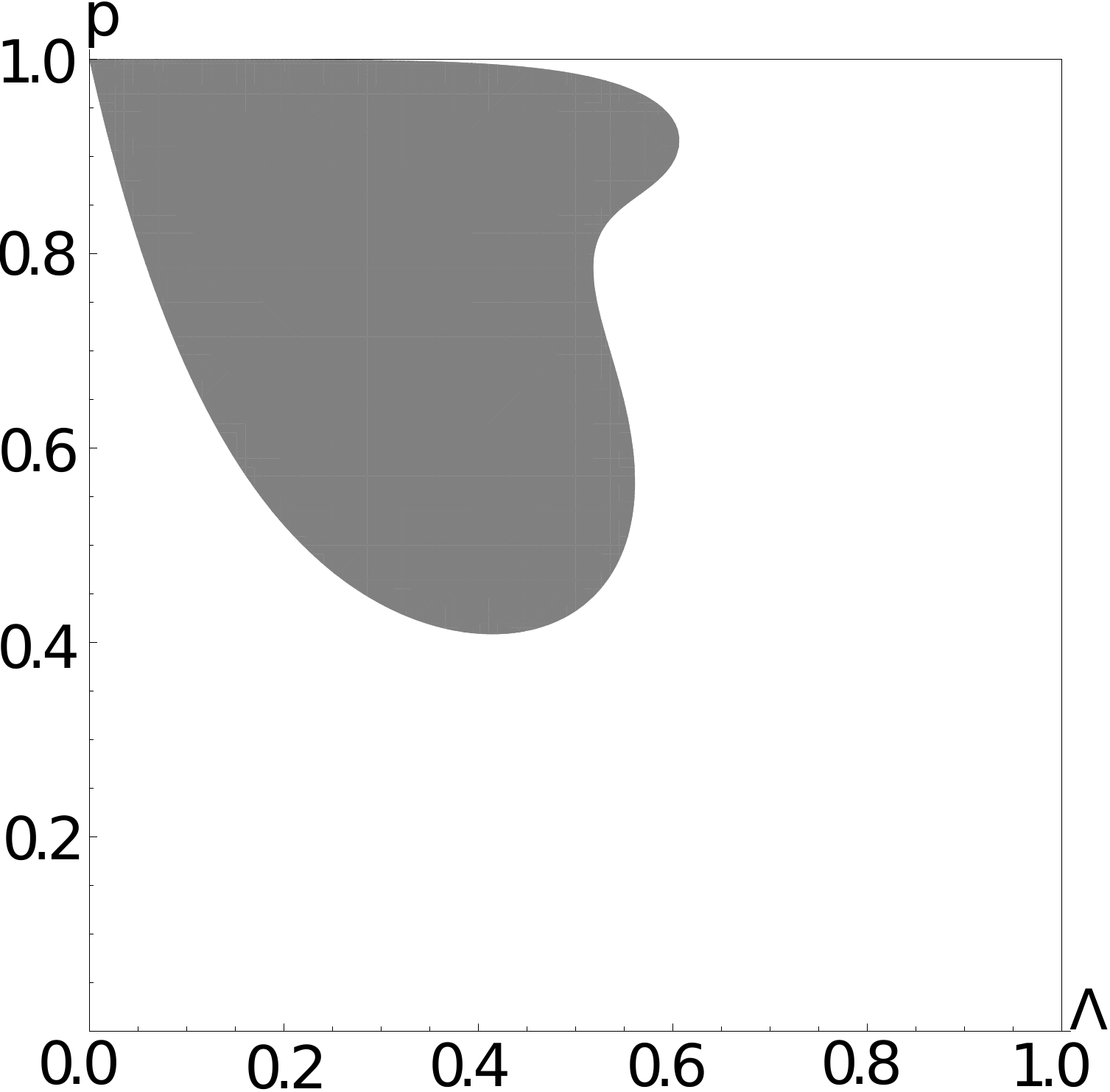}} 
   & \subfloat[r=0.01]{\includegraphics[width=2.5cm]{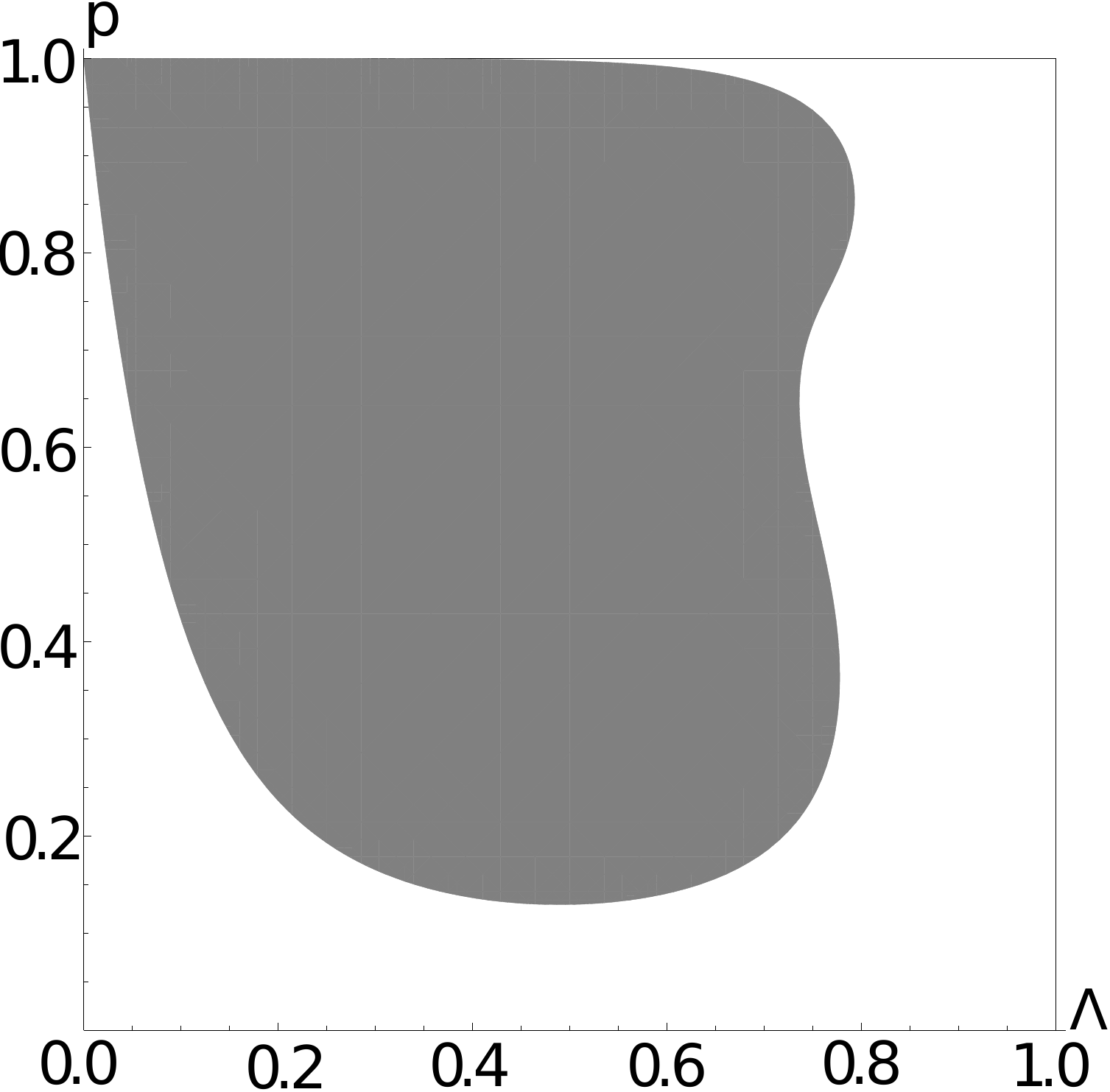}}
& \subfloat[r=0.001]{\includegraphics[width=2.5cm]{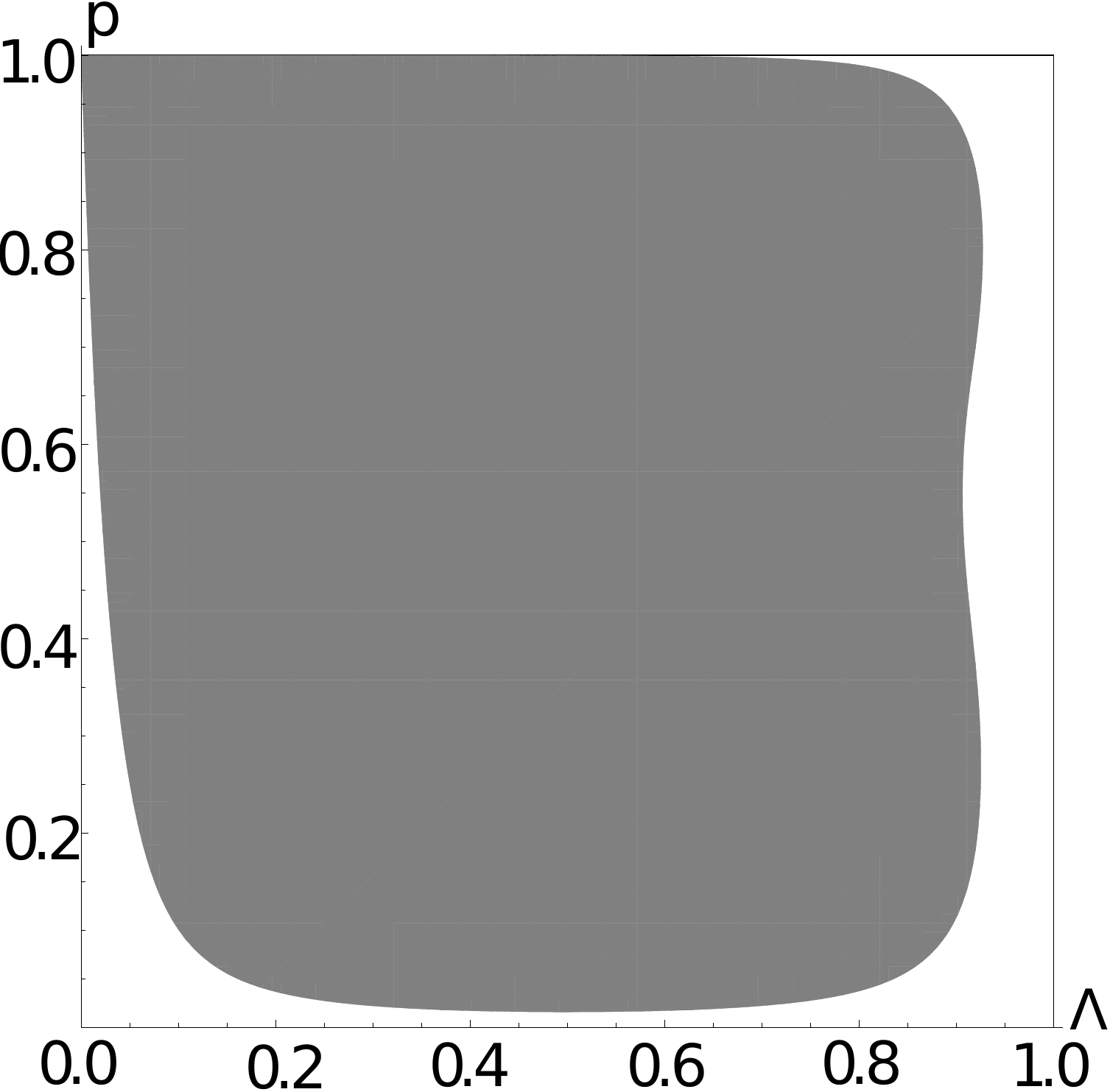}}\\
\end{tabular}

\caption{Dark regions correspond to non-positivity of $\det Q$
 for the case when $Q$ is approximated by the $3 \times 3$ matrix built of  dual double moments.}
\label{fig33d}
\end{figure}

\begin{figure}[]

\centering

\begin{tabular}{ccc}
\subfloat[r=0.99]{\includegraphics[width=2.5cm]{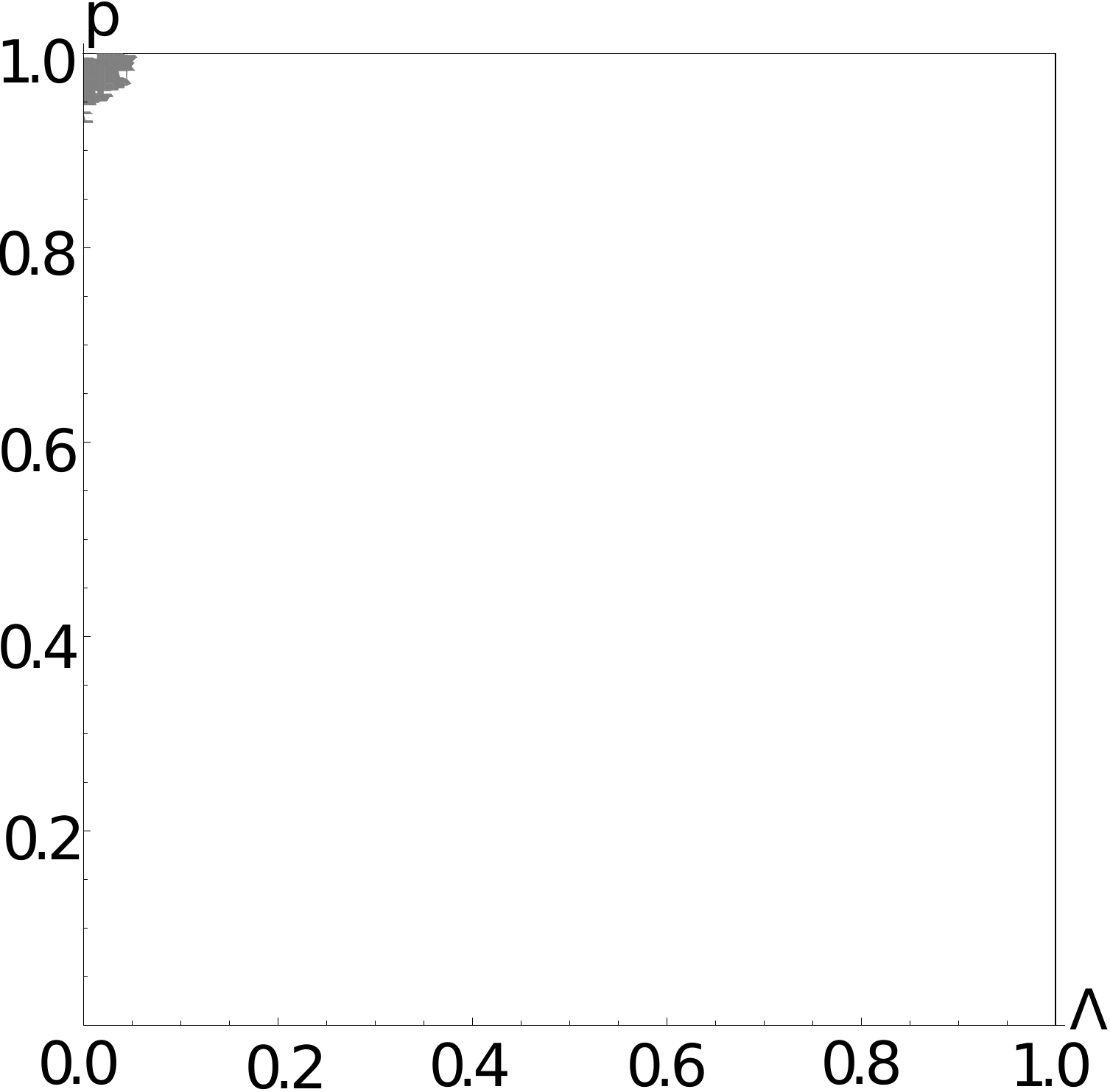}} 
   & \subfloat[r=0.9]{\includegraphics[width=2.5cm]{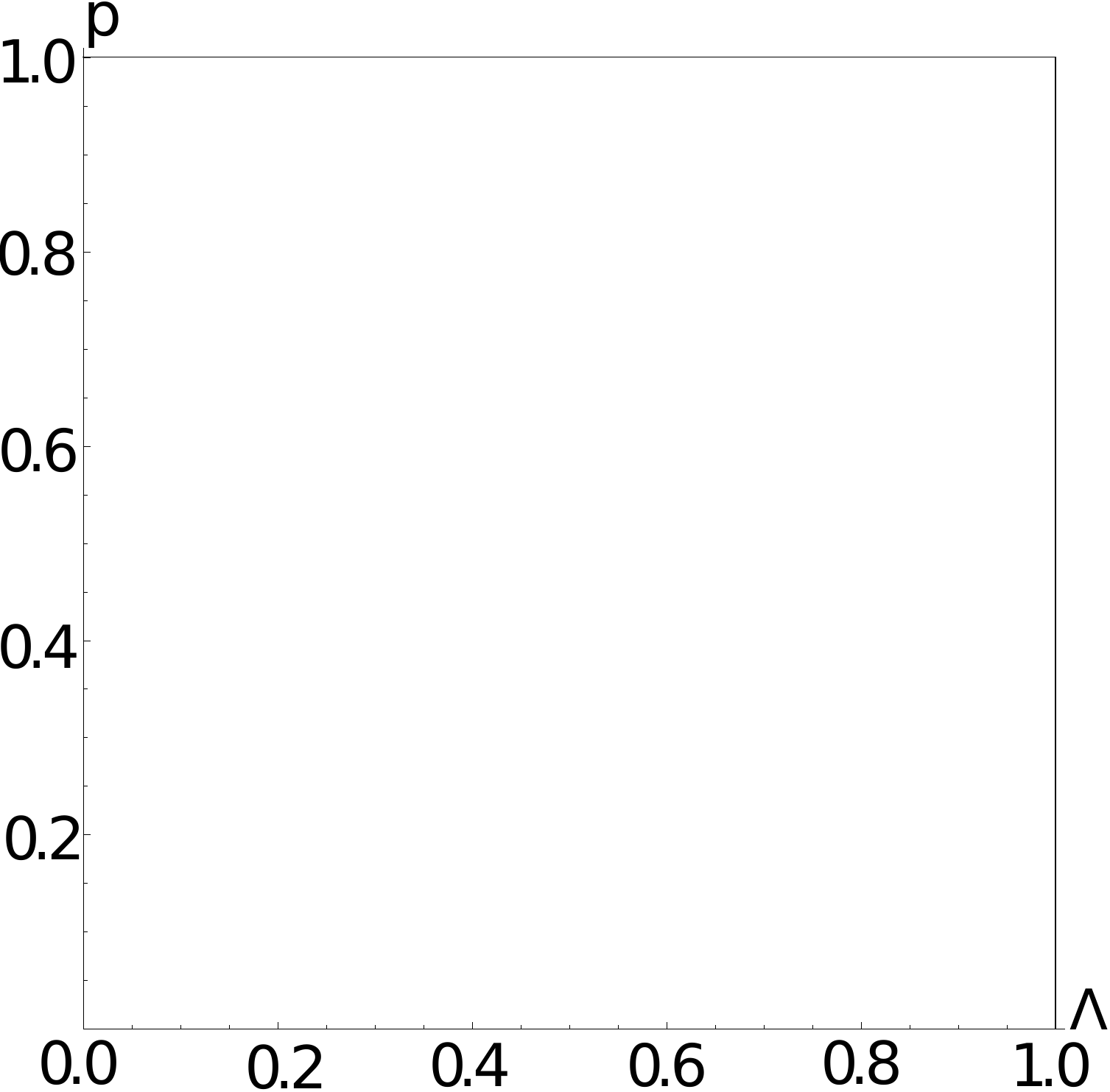}}
& \subfloat[r=0.7]{\includegraphics[width=2.5cm]{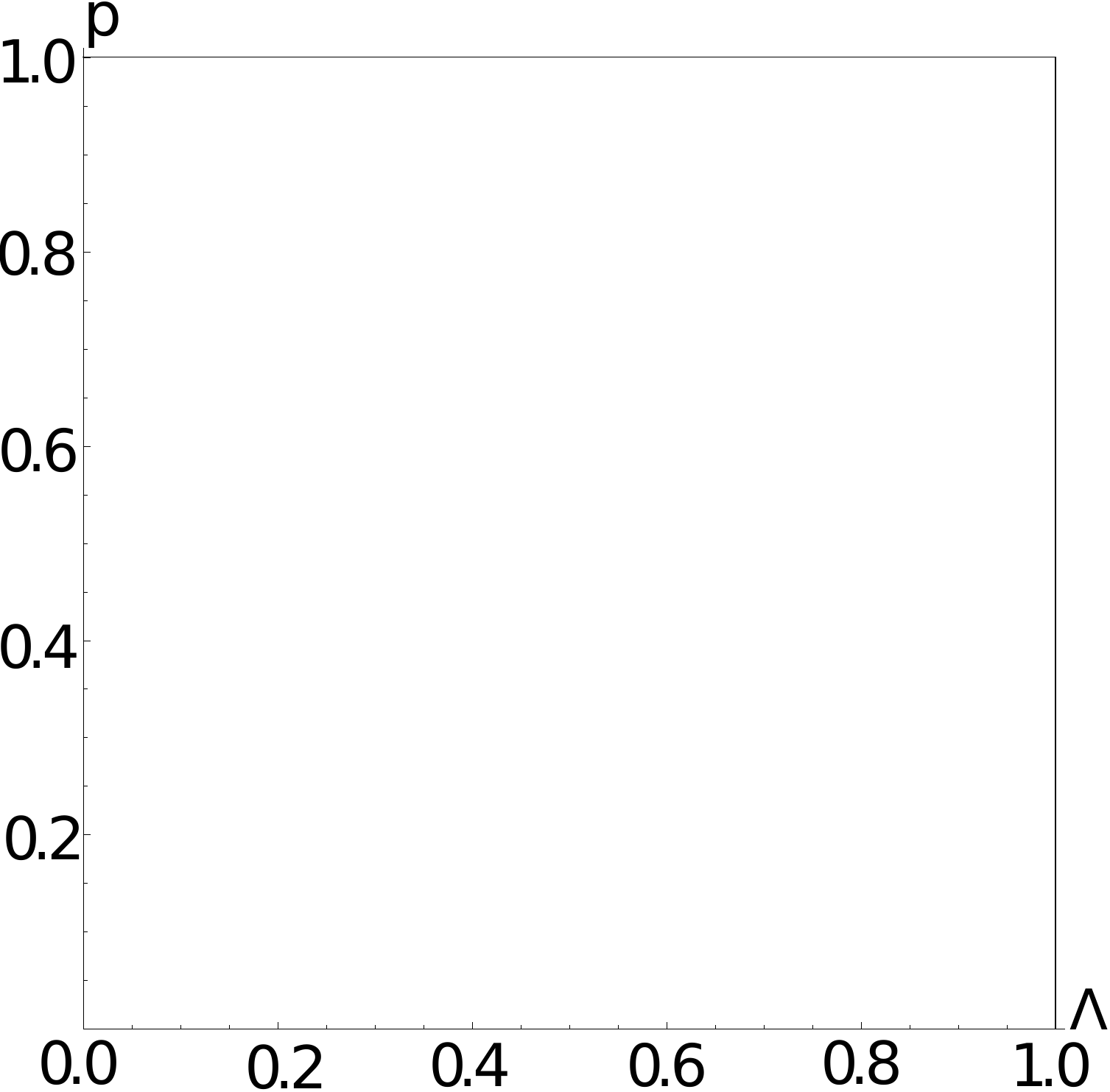}}\\

\subfloat[r=0.5]{\includegraphics[width=2.5cm]{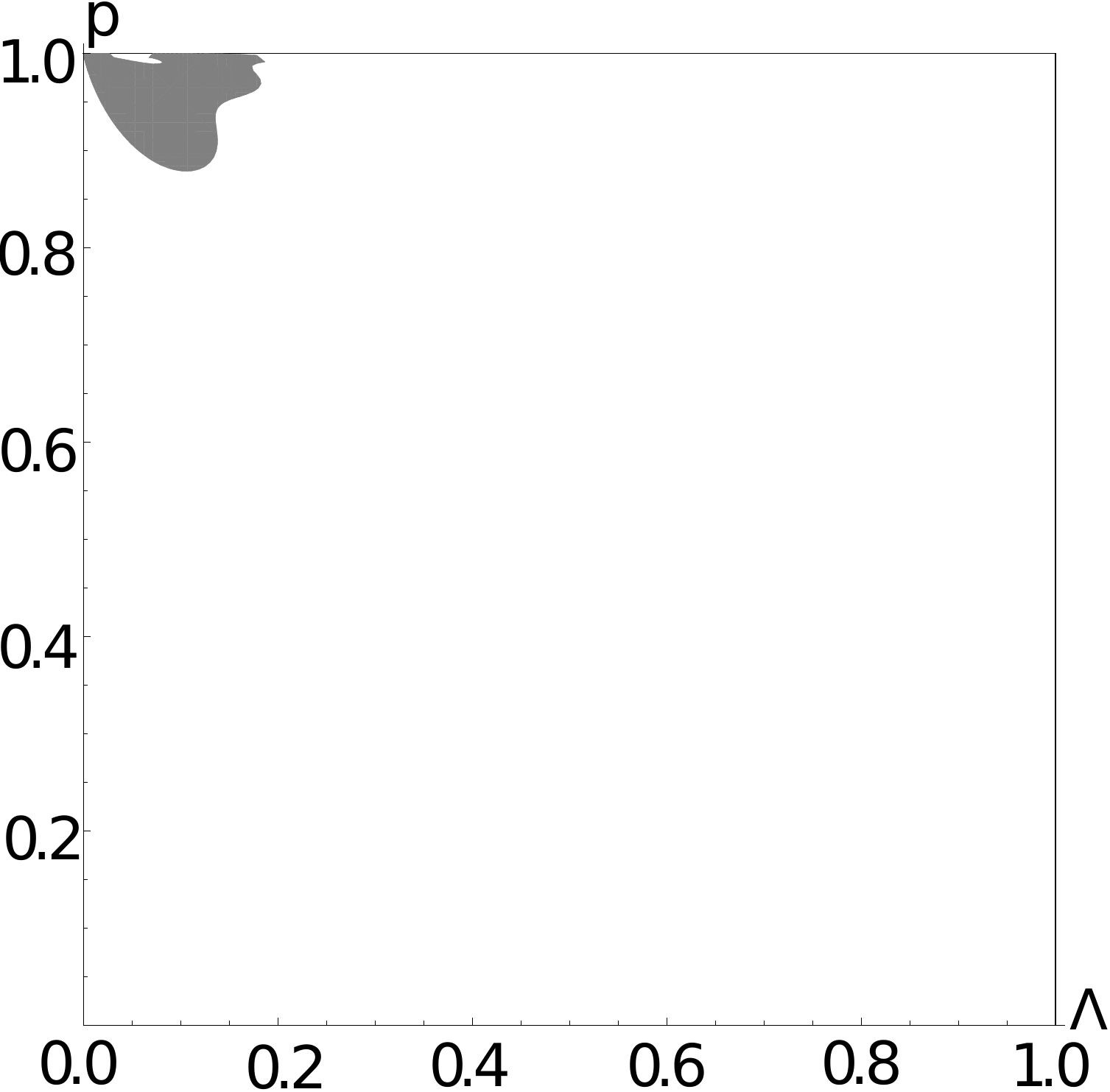}} 
   & \subfloat[r=0.3]{\includegraphics[width=2.5cm]{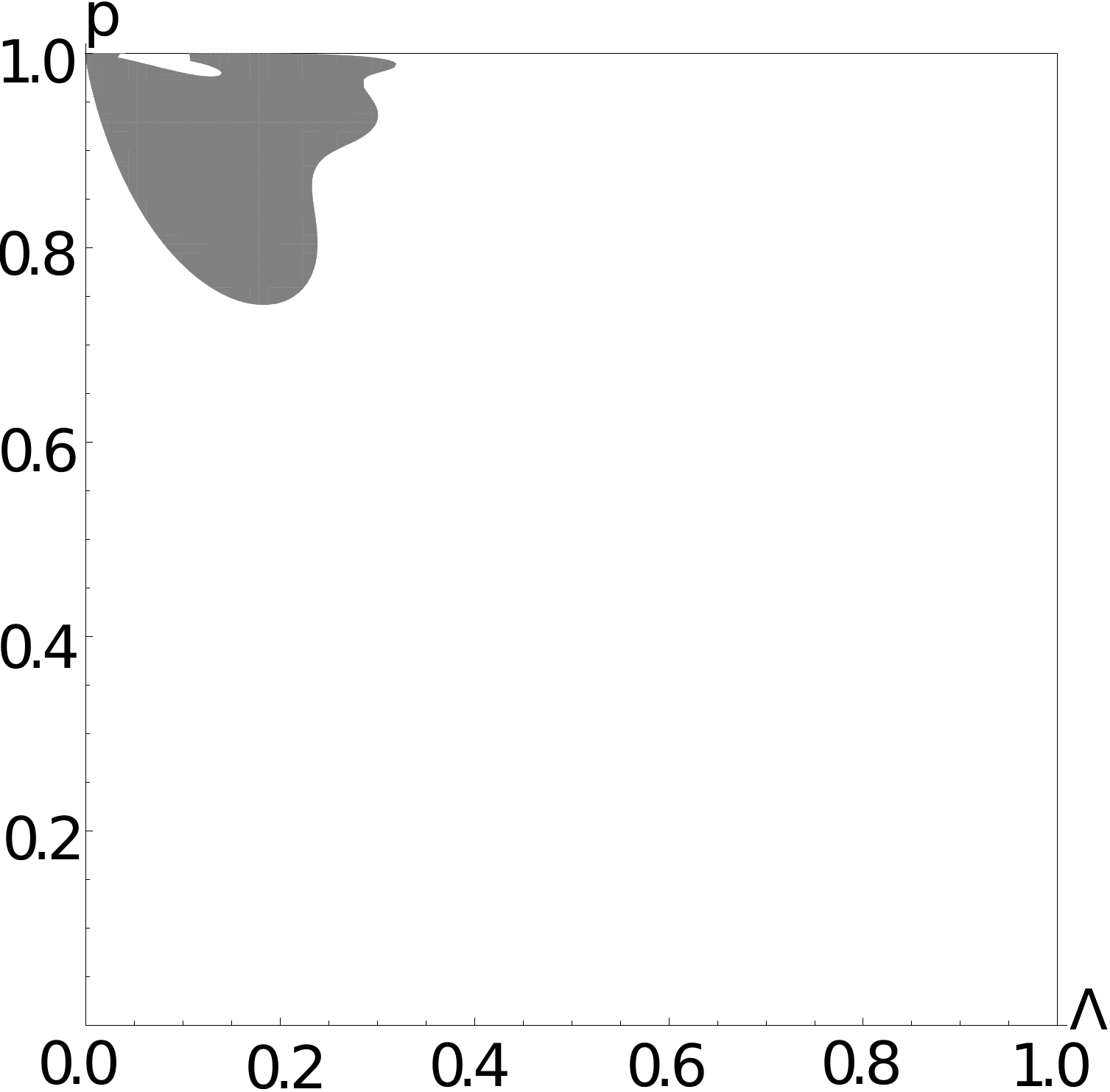}}
& \subfloat[r=0.1]{\includegraphics[width=2.5cm]{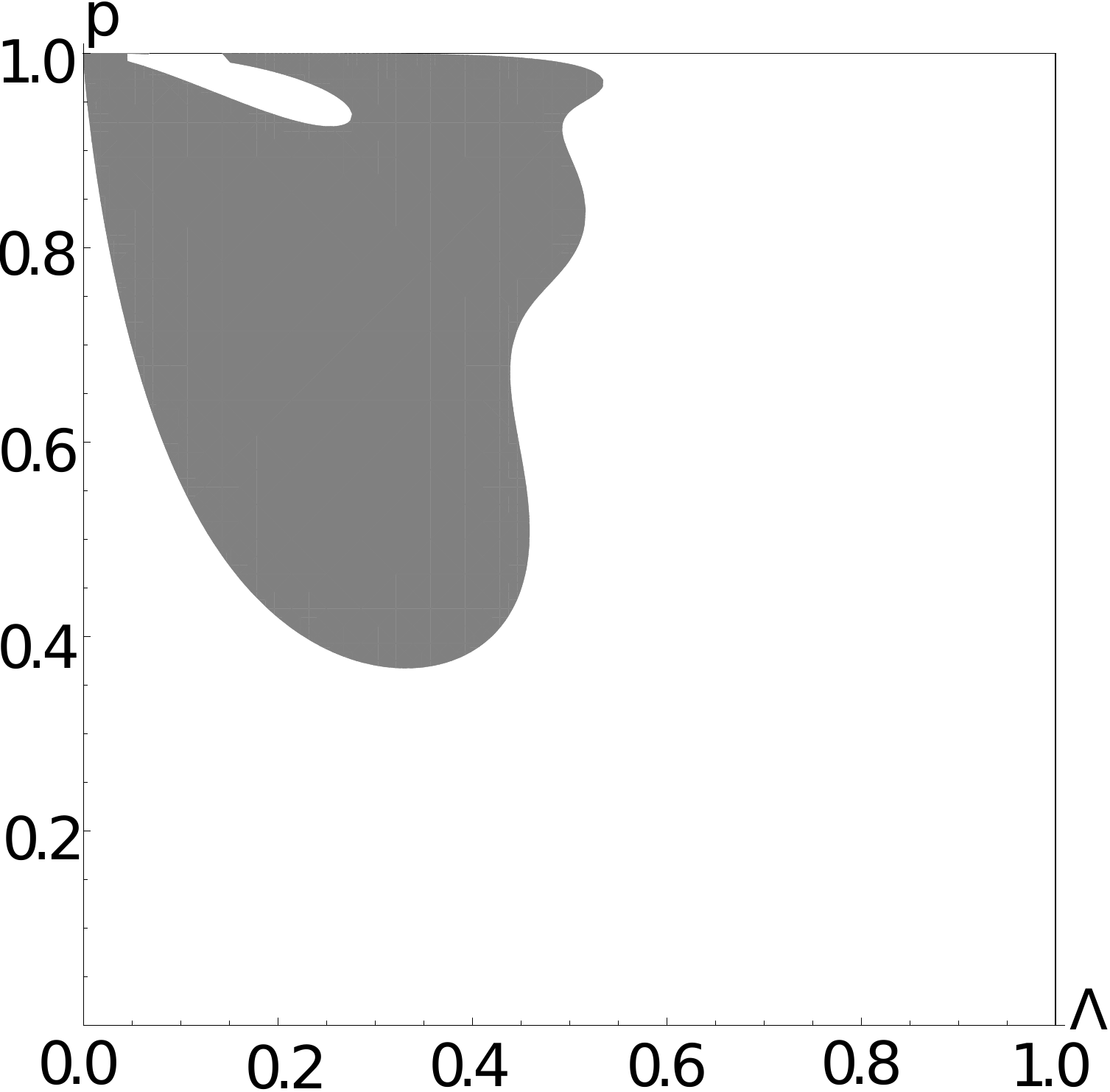}}\\

\subfloat[r=0.05]{\includegraphics[width=2.5cm]{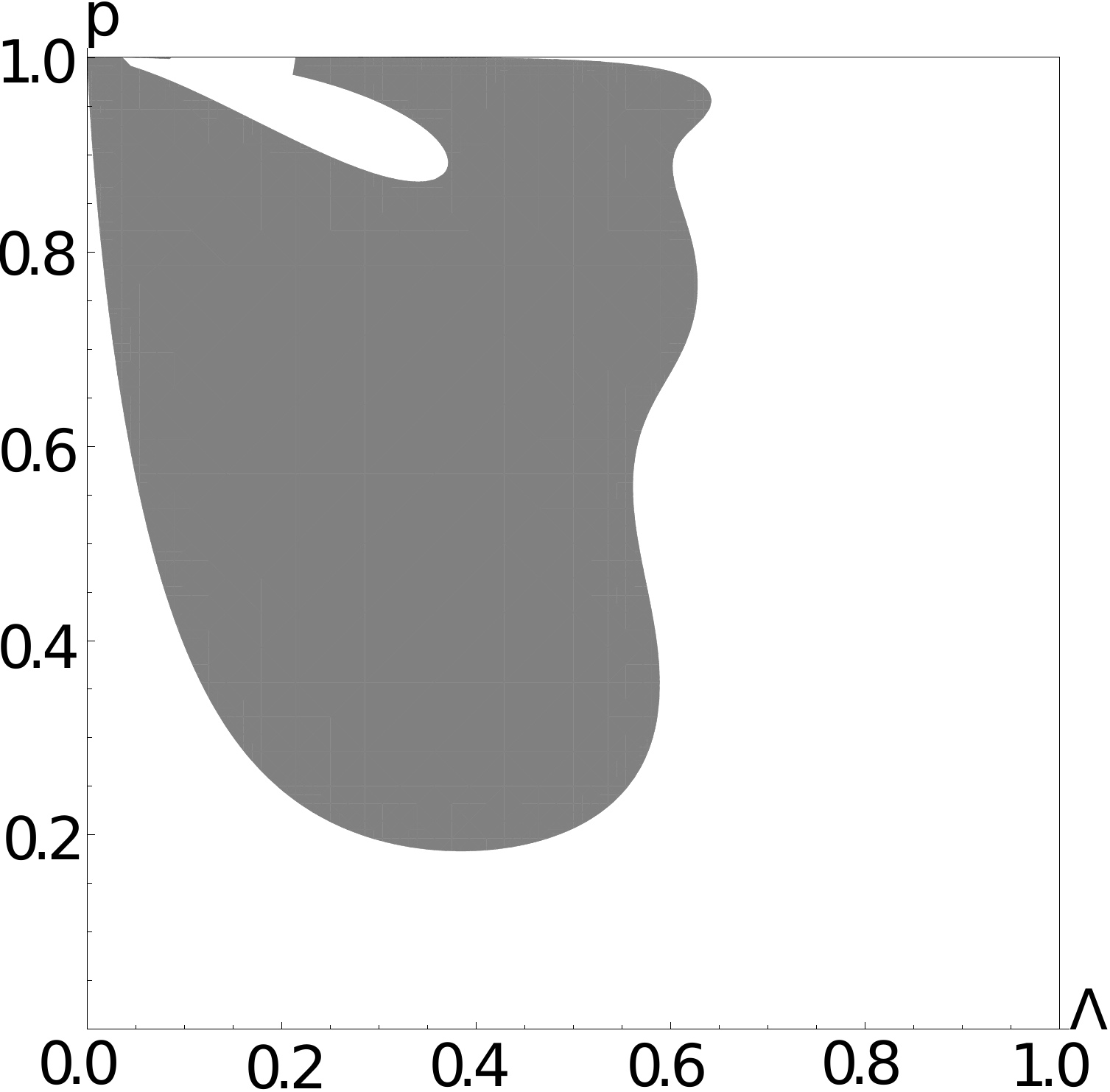}} 
   & \subfloat[r=0.01]{\includegraphics[width=2.5cm]{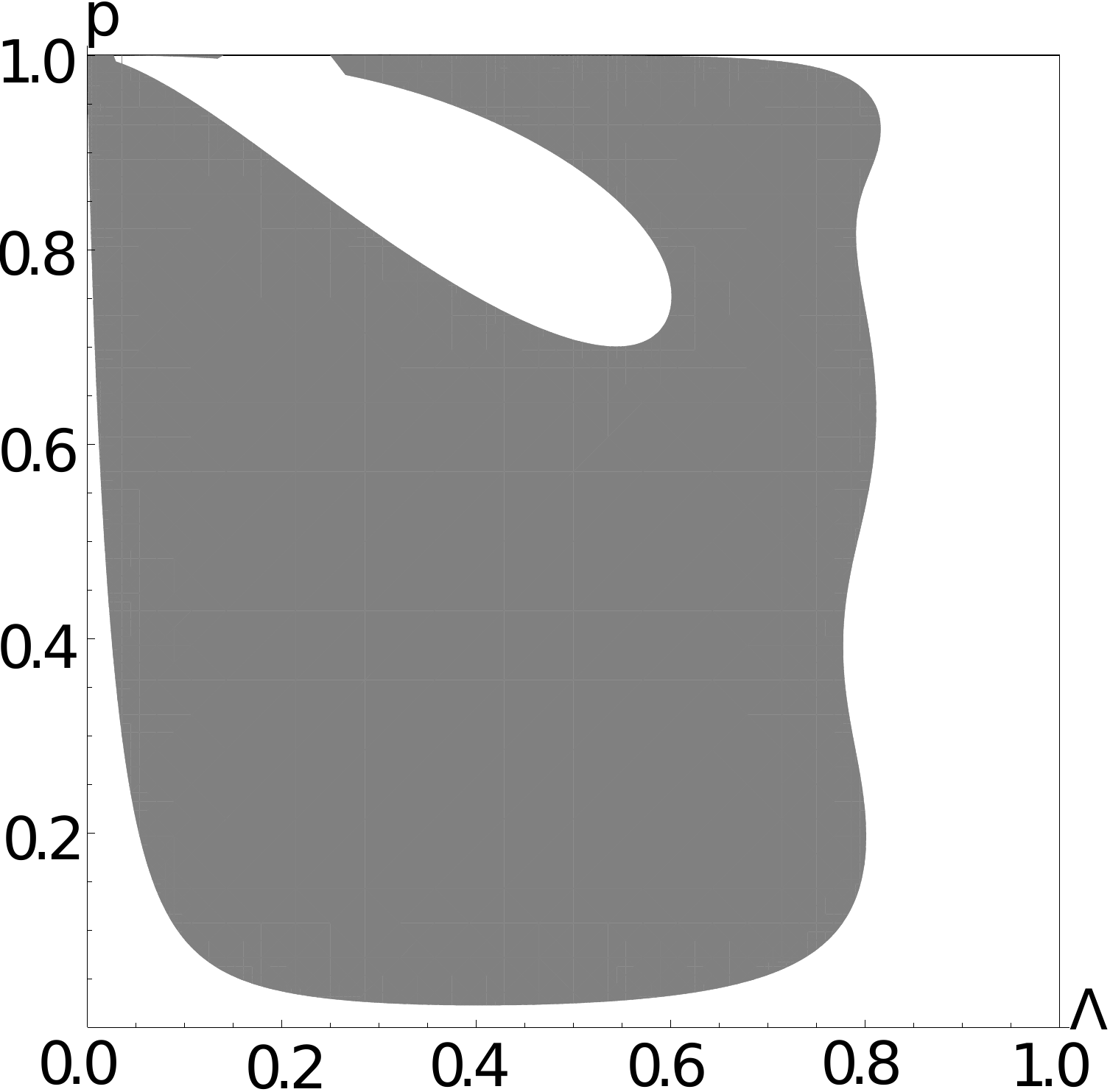}}
& \subfloat[r=0.001]{\includegraphics[width=2.5cm]{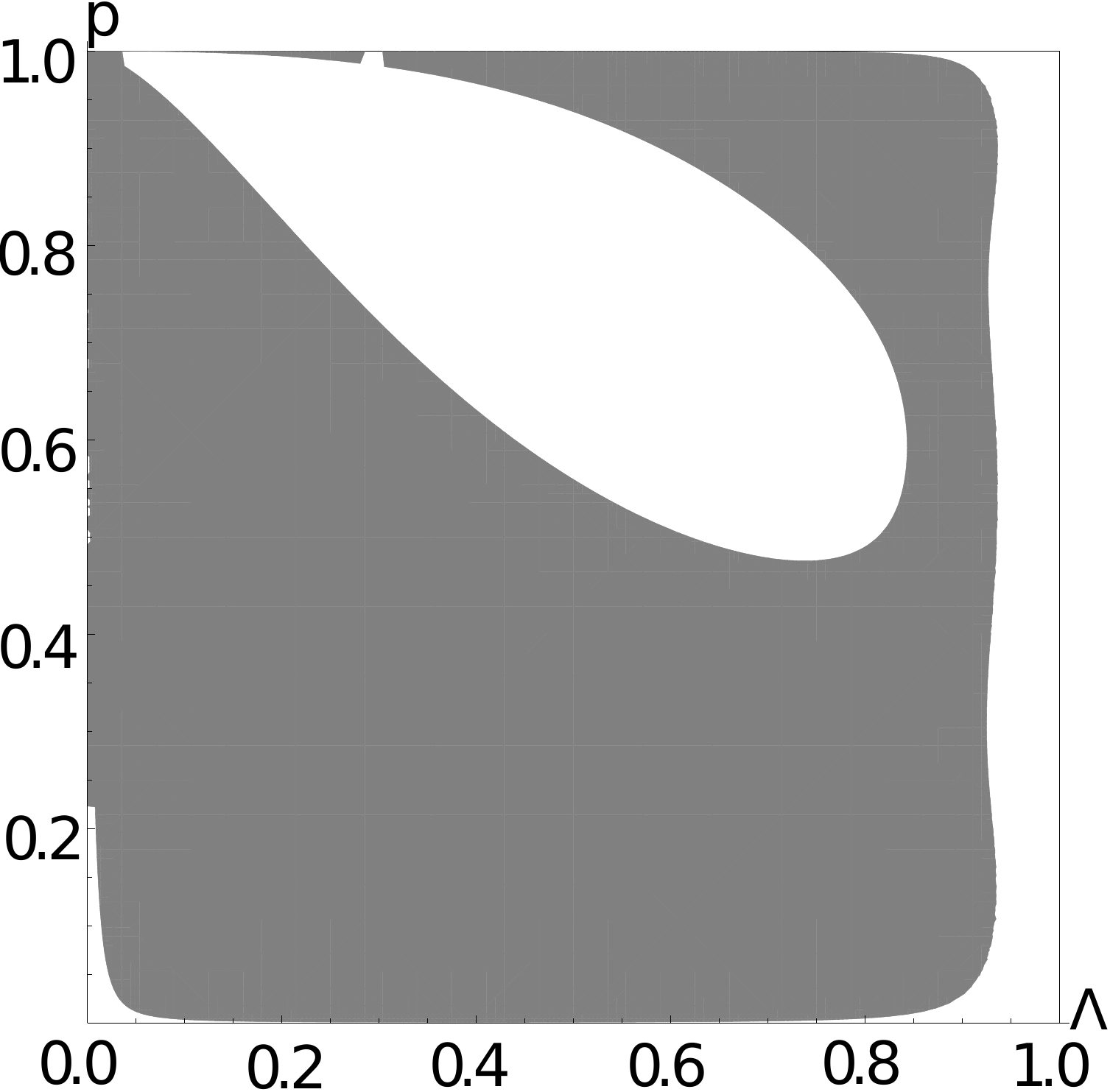}}\\
\end{tabular}

\caption{Dark regions correspond to non-positivity of $\det Q$
 for the case when $Q$ is approximated by the 4 by 4  matrix built of dual  double moments.}
\label{fig44d}
\end{figure}

\begin{figure}[]

\centering

\begin{tabular}{ccc}
\subfloat[r=0.99]{\includegraphics[width=2.5cm]{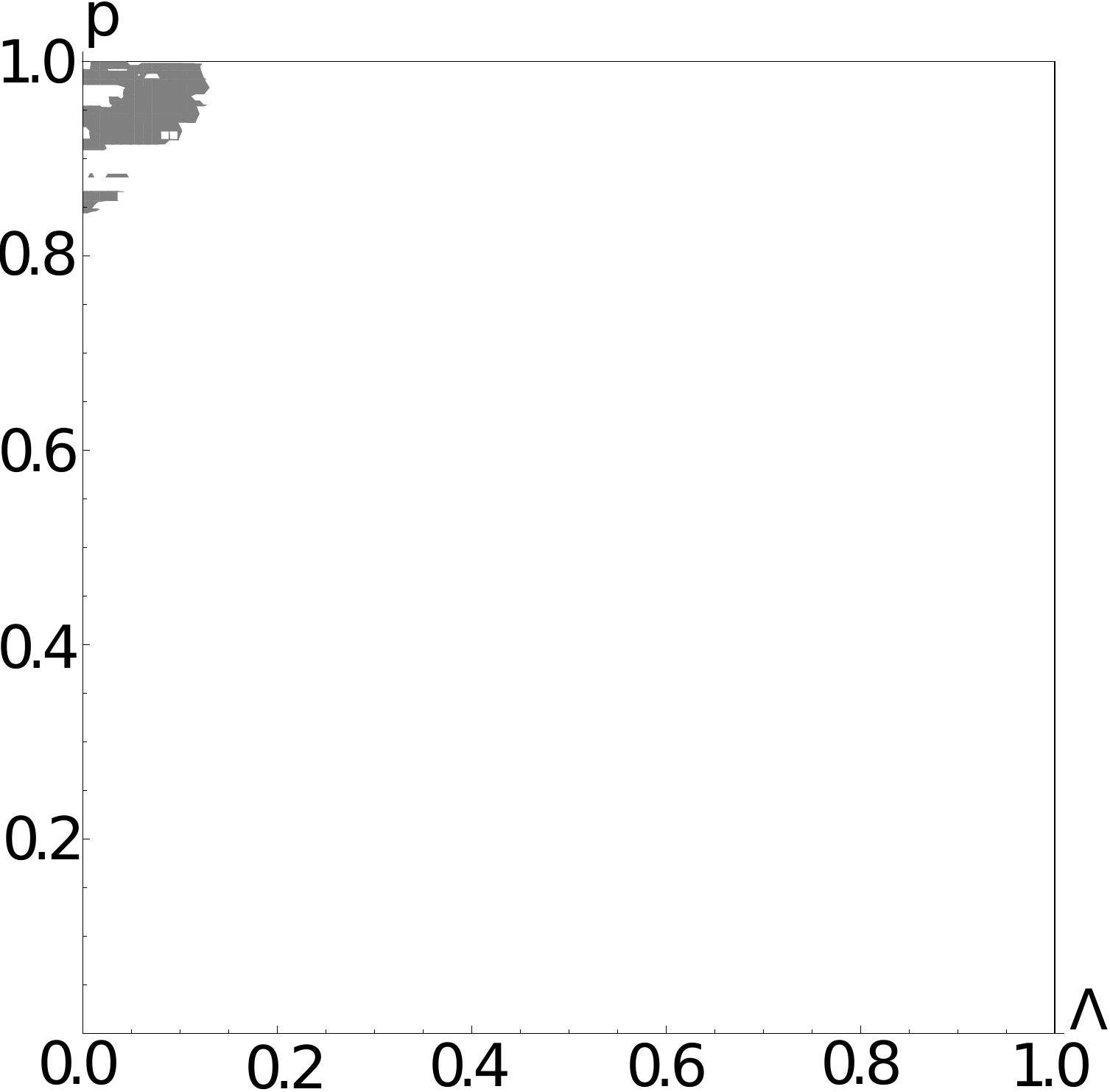}} 
   & \subfloat[r=0.9]{\includegraphics[width=2.5cm]{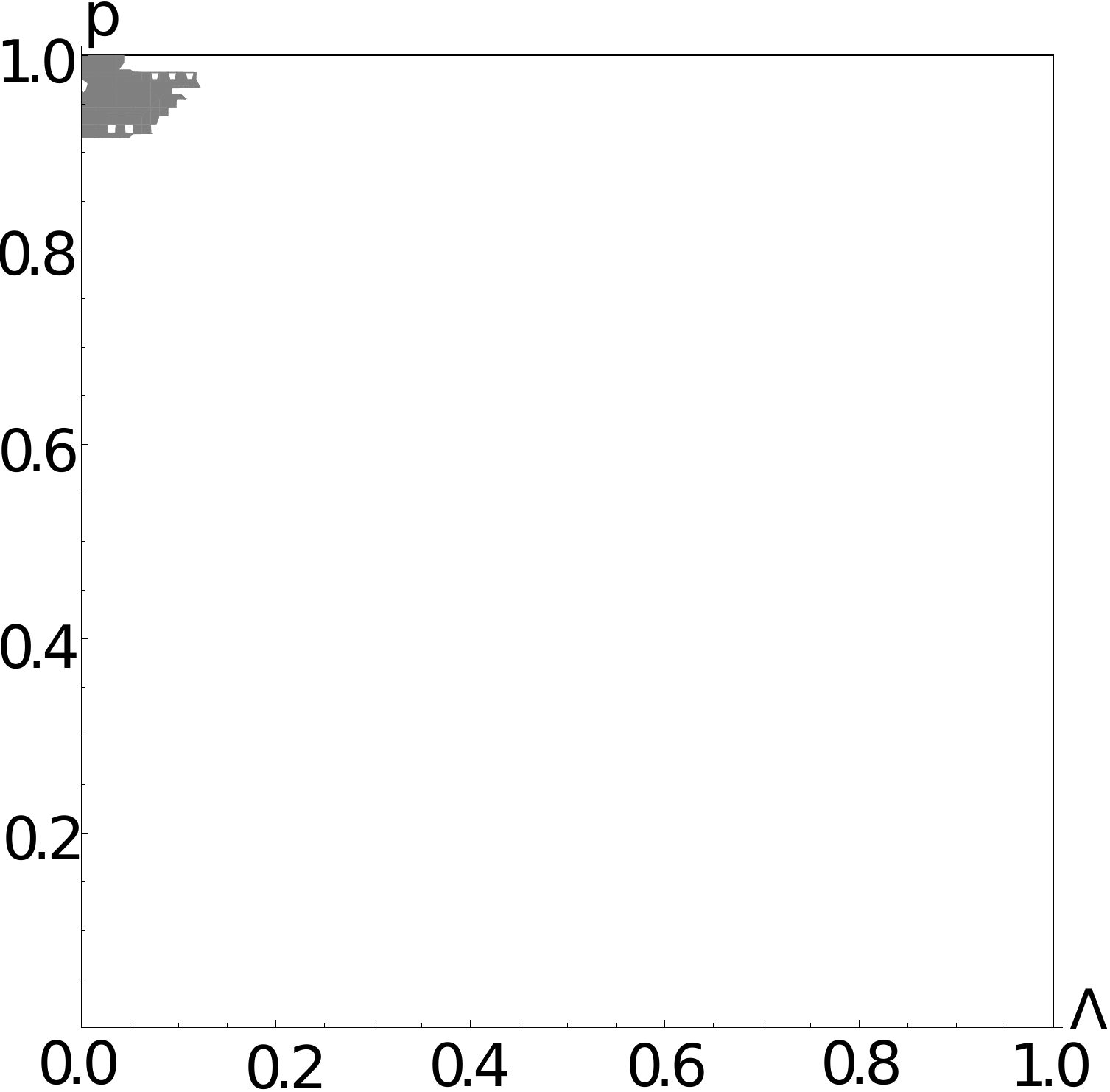}}
& \subfloat[r=0.7]{\includegraphics[width=2.5cm]{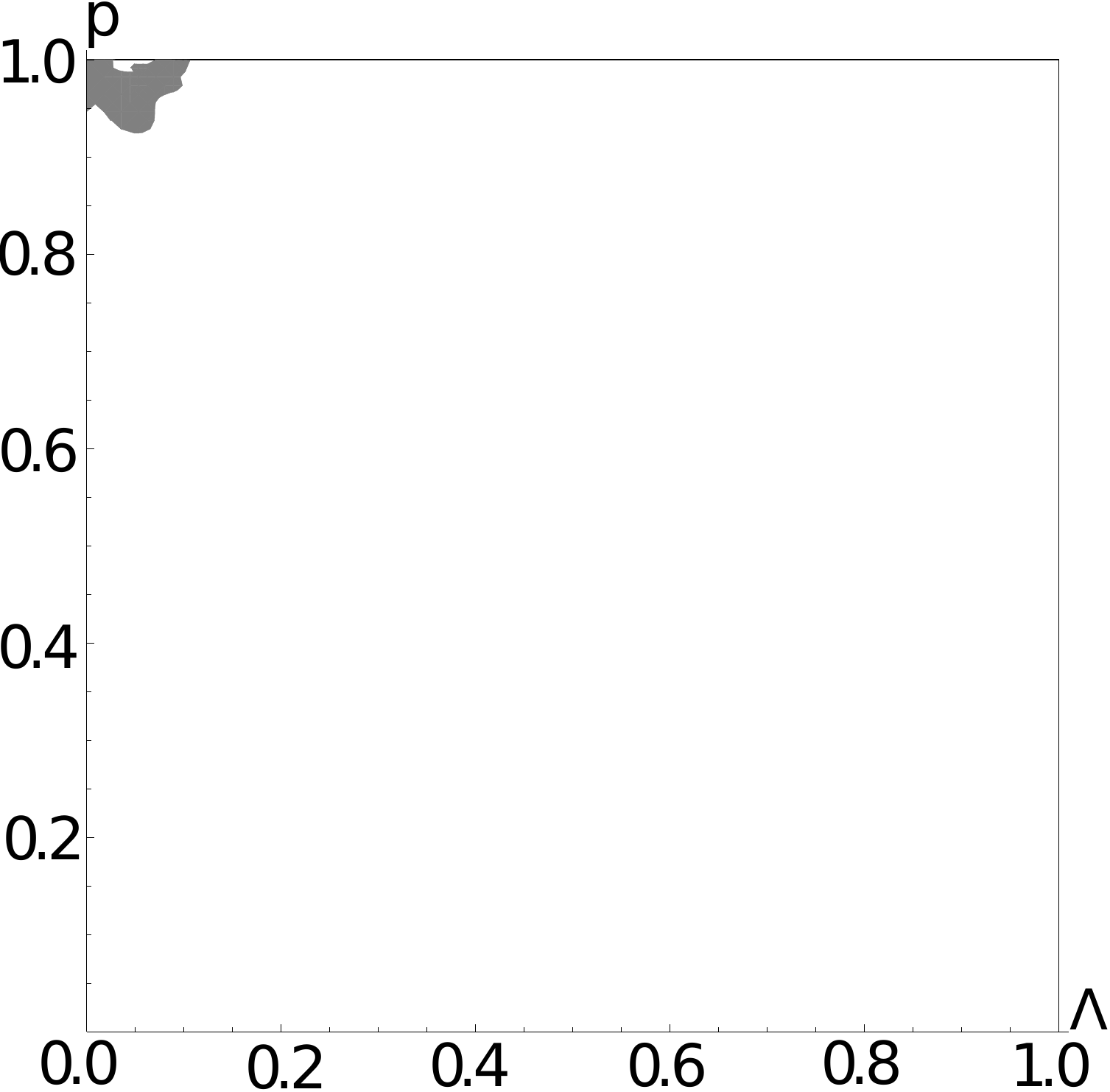}}\\

\subfloat[r=0.5]{\includegraphics[width=2.5cm]{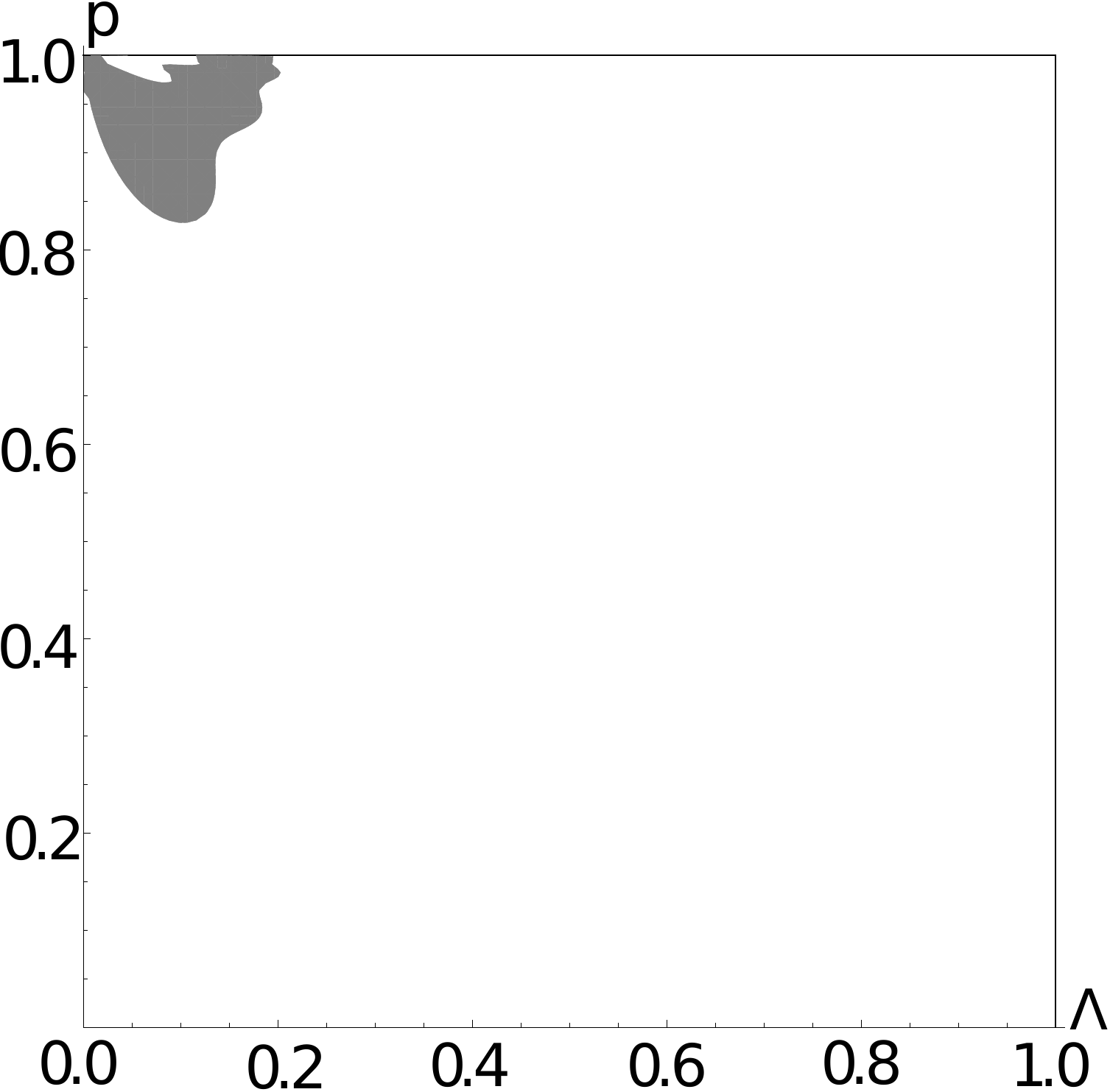}} 
   & \subfloat[r=0.3]{\includegraphics[width=2.5cm]{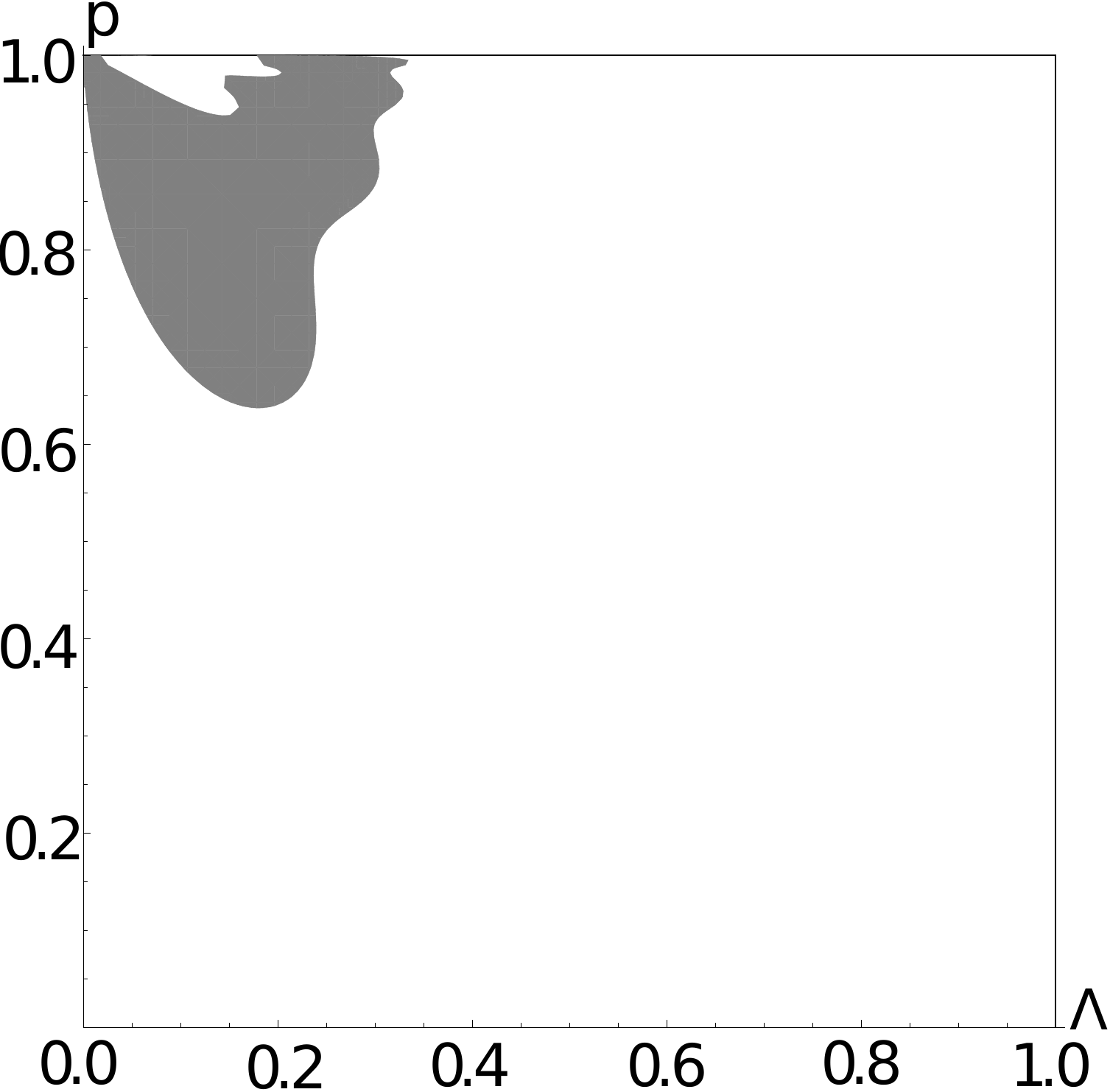}}
& \subfloat[r=0.1]{\includegraphics[width=2.5cm]{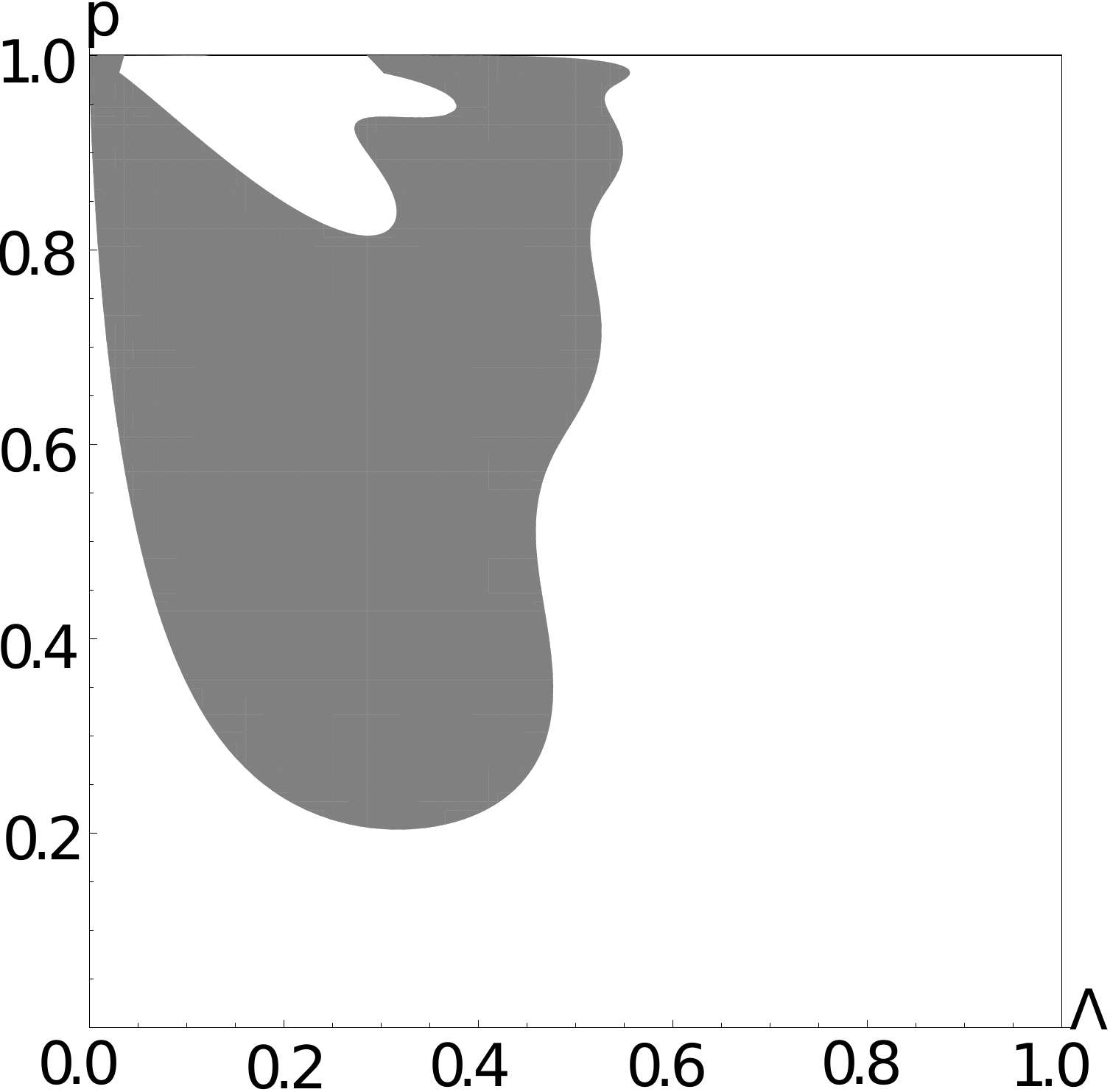}}\\

\subfloat[r=0.05]{\includegraphics[width=2.5cm]{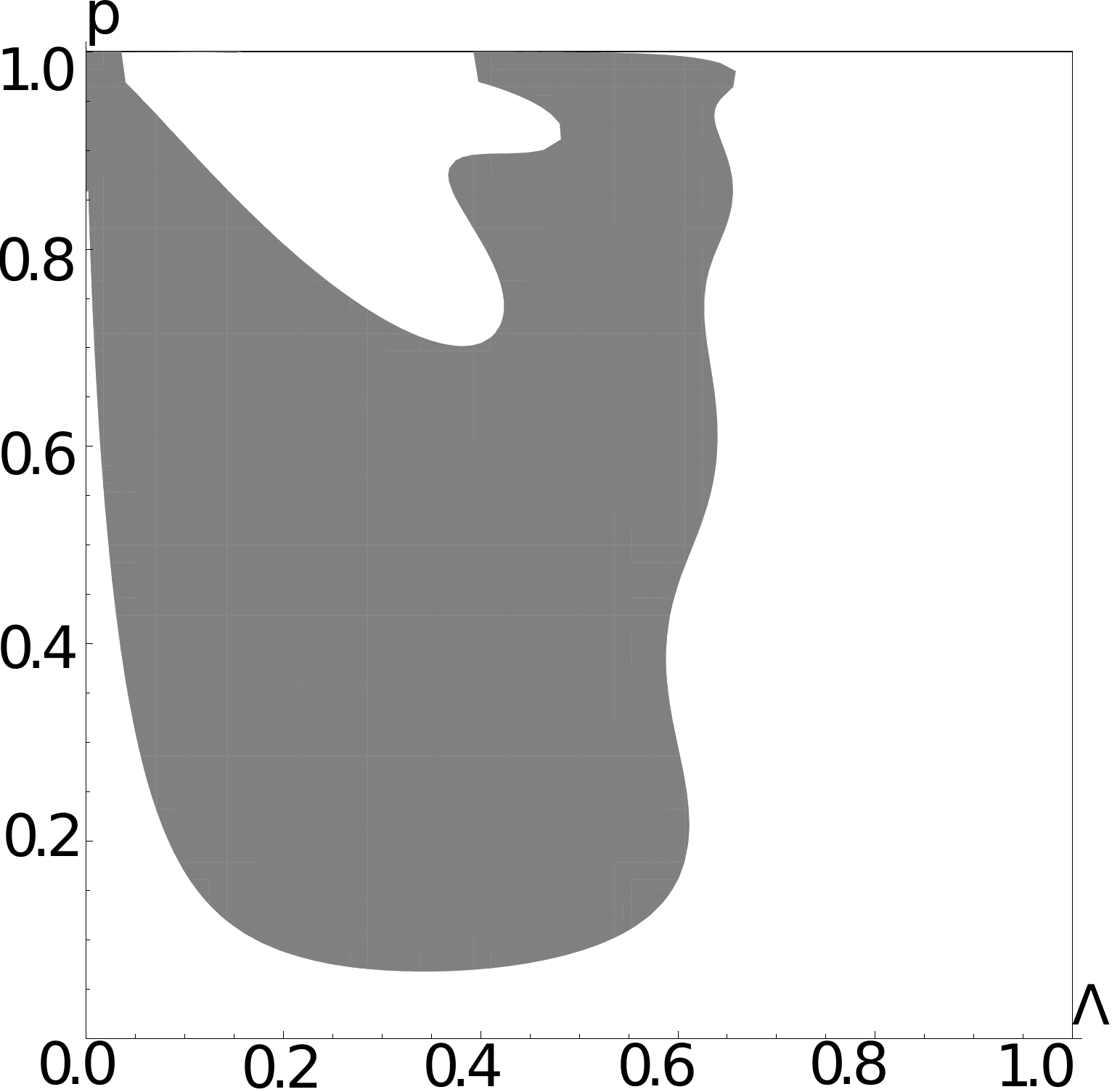}} 
   & \subfloat[r=0.01]{\includegraphics[width=2.5cm]{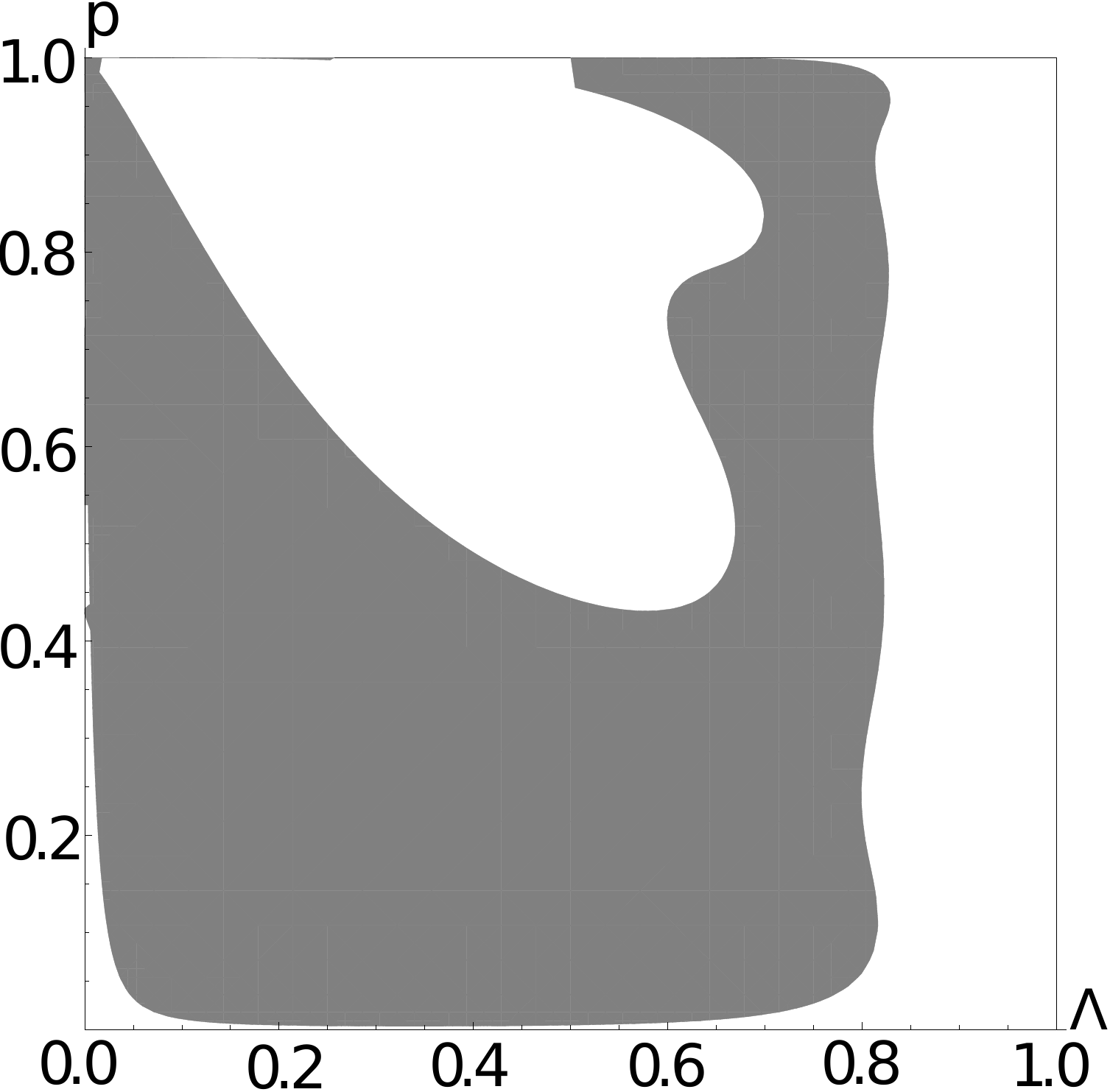}}
& \subfloat[r=0.001]{\includegraphics[width=2.5cm]{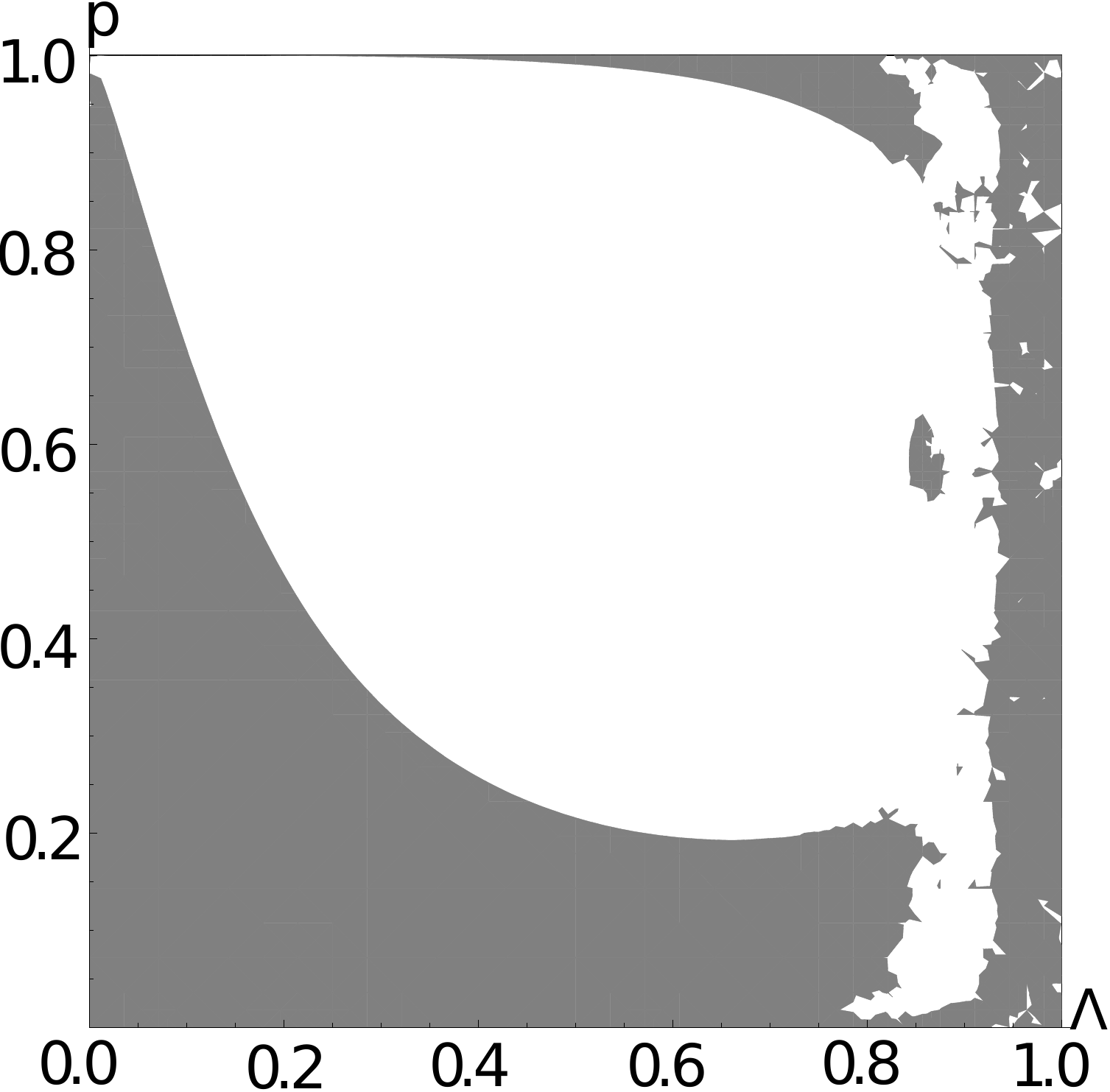}}\\
\end{tabular}

\caption{Dark regions correspond to non-positivity of $\det Q$
for the case when $Q$ is approximated by the 5 by 5 matrix built of dual double moments.}
\label{fig55d}
\end{figure}

The figure~\ref{fig33} shows the case when we estimate matrix $Q$ by 3 by 3 matrix, using statistical method based on double moments,  for several values of rectangularity parameter $r$. The shaded region  corresponds to the range of parameters  $\Lambda_s, p$ when the value of ${\rm det} Q$ is negative.  The smaller the values of $r$, the more pathological is the  behavior of ${\rm det} Q$, covering almost whole region of the  parameter space  in the case of extremely small $r=10^{-3}$.  But even in the case of "reasonable" $r=0.1$
one can notice large regions of wrong behavior of the entropic  term. 
The figure~\ref{fig44}  summarizes the repetition of  the above analysis for the same case, when  estimating  matrix $Q$ with the help of  4 by 4   matrix.  One can see an improvement, especially in the case of small $r$. Finally, the figure~\ref{fig55} shows the same case, when we estimate the matrix $Q$ with the help of 5 by 5 matrix. 
One can see the shrinkage of the shaded region for all values of $r$. This is expected  since the larger the  dimension of matrix $Q$, the better the convergence  toward the  "true" spectrum of the covariance matrix.  We would like to stress that in this case the estimator of matrix  $Q$  involves all double moments  up to $\alpha_{55}$.  All double moments are relatively complicated, e.g. $\alpha_{55}$ is composed of  42 terms involving various products of powers of single moments from $\alpha_1$   up to $\alpha_{10}$.  This clearly shows  that approaching the limiting distribution becomes less and less numerically tractable in the statistical method. 
The same problem holds when we apply  the dual statistical method.  The triple of figures~\ref{fig33d},\ref{fig44d} and~\ref{fig55d} summarize the  analysis for the same  values of the parameter $r$  as in the simple   statistical method. In general, we see the same tendency of improvement when the dimension of the estimator grows.  The "fractal-like" structures, visible for example in the  5 by 5 dual case for $r=10^{-4}$ are the artifact of numerical accuracy.  In general, we see that statistical  analysis works better comparing to the case of dual statistical analysis.  
 \section{Comment on real Wishart ensemble}

\begin{table*}[ht]
\footnotesize
\caption{Comparison between the real and the complex case} \label{tab:3}
\begin{tabular}{|l|l|l|l|l|l|l|l|l|l|l|}
\hline
\multicolumn{11}{|l|}{Explanation of the symbols:}\\
\multicolumn{11}{|l|}{A/S - the analytical/ statistical method}\\
\multicolumn{11}{|l|}{AD - the analytical dual method}\\ 
\multicolumn{11}{|l|}{3x3 - the size of the matrix Q used}\\
\multicolumn{11}{|l|}{RMSE - results from the article \cite{MIT}}\\
\multicolumn{11}{|l|}{q - subjective assessment of the quality of the estimation (1 - 4 stars)}\\
\multicolumn{11}{|l|}{n - number of matrices with all parameters estimated as positive real}\\
\multicolumn{11}{|l|}{$\langle \dots \rangle$ - arithmetical mean of the estimations}\\
\multicolumn{11}{|l|}{$\sigma( \dots )$ - standard deviation of the estimations}\\
\multicolumn{11}{|l|}{$\eta$ - a parameter measuring the quality of the estimation (less is better)}\\
\multicolumn{11}{|l|}{n.d. - no data}\\
\hline
method & q & n & $\langle \lambda_1 \rangle $ & $ \sigma (\lambda _1)$ & $\langle \lambda_2 \rangle $ & $ \sigma (\lambda _2)$ & $\langle p_1^S \rangle $ & $ \sigma (p_1^S)$ & $\eta$ & time [s]\\
\hline
\multicolumn{11}{|c|}{100 real matrices $80 \times 40$, $\Lambda_1=2$, $\Lambda_2=1$, $p_1=1/2$} \\
\hline
S 3x3  & ** & 100 & 2.1635 & 0.5167 & 0.7792 & 0.4445 & 0.5322 & 0.2844 & 0.6999 & 688.5\\
\hline
\multicolumn{11}{|c|}{100 real matrices $ 320 \times 160$, $\Lambda_1=2$, $\Lambda_2=1$, $p_1=1/2$} \\
\hline
S 3x3 & *** & 100 & 2.0690 & 0.1620 & 0.9986 & 0.1361 & 0.4713 & 0.1273 & 0.2409 & 364.1 \\
\hline
\multicolumn{11}{|c|}{100 real matrices $80 \times 82$, $\Lambda_1=2$, $\Lambda_2=1$, $p_1=1/2$} \\
\hline
S 3x3 & *** & 100 & 2.0512 & 0.2309 & 0.9634 & 0.1756 & 0.4971 & 0.1603 & 0.3185 & 506.6\\
\hline
\multicolumn{11}{|c|}{100 real matrices $320 \times 322$, $\Lambda_1=2$, $\Lambda_2=1$, $p_1=1/2$} \\
\hline
S 3x3 & ***& 100 & 2.0210 & 0.0557 & 1.0026 & 0.0486 & 0.4886 & 0.0489 & 0.0867 & 710.9\\
\hline
\multicolumn{11}{|c|}{1000 real matrices $80 \times 160$, $\Lambda_1=2$, $\Lambda_2=1$, $p_1=1/2$} \\
\hline
S 3x3 & *** & 1000 & 2.0320 & 0.1010 & 0.9961 & 0.0721 & 0.4869 & 0.0765 & 0.1402 & 10904.5  \\
\hline
\multicolumn{11}{|c|}{100 real matrices $320 \times 640$, $\Lambda_1=2$, $\Lambda_2=1$, $p_1=1/2$} \\
\hline
S 3x3 & ***& 100 & 2.0138 & 0.0274 & 1.0044 & 0.0187 & 0.4915 & 0.0202 & 0.0370 & 1365.5  \\
\hline
\hline
\multicolumn{11}{|c|}{100 complex matrices $80 \times 40$, $\Lambda_1=2$, $\Lambda_2=1$, $p_1=1/2$} \\
\hline
S 3x3  & ** & 100 & 2.0377 & 0.4955 & 0.6969 & 0.5212 & 0.5860 & 0.3163 & 0.7487 & 1076.9\\
\hline
\multicolumn{11}{|c|}{100 complex matrices $ 320 \times 160$, $\Lambda_1=2$, $\Lambda_2=1$, $p_1=1/2$} \\
\hline
S 3x3 & *** & 100 & 2.0117 & 0.1499 & 0.9654 & 0.1496 & 0.5105 & 0.1307 & 0.2425 & 387.7\\
\hline
\multicolumn{11}{|c|}{100 complex matrices $80 \times 82$, $\Lambda_1=2$, $\Lambda_2=1$, $p_1=1/2$} \\
\hline
S 3x3 & *** & 100 & 1.9461 & 0.2233 & 0.8630 & 0.2576 & 0.5834 & 0.1886 & 0.3738& 584.9\\
\hline
\multicolumn{11}{|c|}{100 complex matrices $320 \times 322$, $\Lambda_1=2$, $\Lambda_2=1$, $p_1=1/2$} \\
\hline
S 3x3 & *** & 100 & 2.0101 & 0.0540 & 1.0055 & 0.0453 & 0.4929 & 0.0473 & 0.0826 & 779.4\\
\hline
\multicolumn{11}{|c|}{100 complex matrices $80 \times 160$, $\Lambda_1=2$, $\Lambda_2=1$, $p_1=1/2$} \\
\hline
S 3x3 & *** & 100 & 2.0023 & 0.0924 & 0.9936 & 0.0783 & 0.5030 & 0.7990 & 0.1403 & 1188.3 \\
\hline
\multicolumn{11}{|c|}{100 complex matrices $320 \times 640$, $\Lambda_1=2$, $\Lambda_2=1$, $p_1=1/2$} \\
\hline
S 3x3 & *** & 100 & 2.0011 & 0.0220 & 1.0024 & 0.0180 & 0.4978 & 0.0192 & 0.0331 & 1385.6 \\
\hline
\end{tabular}
\end{table*}

Technically, complex Wishart ensemble is the easiest one from the point of view of the formal methods of random matrix theory, alike Gaussian Unitary Ensemble is the easiest among the triple of classical Dyson's ensembles.  However, in practice we encounter several situations when the measured data are strictly real, which leads to  the  question, to what extend the analysis and the comparison between both methods presented in this work is transferable to the real case. 
We start from the  analytical method. In this case, there is no difference between the Green's functions for the real and complex Wishart ensemble, since the value of the parameter $\beta$  can always be absorbed into the definition of the variance.  Similar statement holds for  the Green's functions generating dual moments. In the case of two-point functions situation is a bit more subtle. The first difference is of  similar origin to the one discussed for the Green's functions i.e. corresponds only to  the  redefinition of $Q$ by the value $2/\beta$. Note that this redefinition does not change the function $g_{\theta}$. The second difference is more fine. In the case of real Wishart ensemble, already  $1/N$ corrections are present, contrary to the complex  Wishart ensemble, where subleading corrections start at the order of $1/N^2$.  In this case the pdf of multidimensional Gaussian  $f(v_{\theta})$ develops additionally the non-zero mean-values $\mu_{\theta}$, which, unfortunately, are given only in terms of  some contour  integral, which makes the operational use of their representation  difficult. In the literature~\cite{MIT}, this problem was avoided in such a way that the non-zero means were neglected and the minimizer was based on central multidimensional Gaussian alike in the complex case.  The numerical accuracy  based on this approximation was quite satisfactory. We have performed similar studies  and we confirm the rationale of this approximation, see Table \ref{tab:3}. We believe that it is possible to get numerically tractable representation for the means $\mu_{\theta}$, but it is perhaps not worthy to  invest  a lot of work in order to achieve this goal, taking into account: first, how well the approximation of zero mean works; second, that the statistical method in general is much more complicated and time-consuming comparing to the analytical  one. 
For completeness,  we mention also the case of $\beta=4$, corresponding to quaternion-valued Wishart.  This is almost an academic case, since we are not aware of any statistical problem when quaternion-valued measurements appear.  On the other side,  there exists  a closely related to the quaternionic Wishart ensemble so-called chiral symplectic ensemble, which plays the role for certain lattice version of the Dirac operator in Quantum Chromodynamics.  In case the analysis of the moments of such operator would be needed, one should use  analytic method. Then, alike in the case of  the real Wishart, the effect of quaternion variable can be incorporated into the redefinition of the variance of the Gaussian, and all our formulae relating the moments still hold.

\section{Conclusions and prospects}
In this paper, we have compared two methods of eigen-inference, based on the analysis of one-point and two-point Green's functions, respectively. As far as we know, this is the first so extensive comparative analysis of this type of inference. We have also confirmed (both analytically and numerically) recent results based on two-point Green's functions performed by~\cite{RAO,MIT}.  Our analysis clearly points at the superiority of eigen-inference based on one-point Green's function. The procedure is of orders of magnitude faster comparing to the analysis based on two-point functions, and involves much less computer memory. It is also not restricted to the case of only two or maximum three distinct eigenvalues of the true covariance method. It works equally well for real and complex data points.  Second, we have observed a numerical instability of the statistical method in the case of very small values of the rectangularity parameter $r$.  This was a priori puzzling, since usually the smallness of $r$ improves the inference (in the case of the analysis based on one-point functions). We have identified the source of this puzzle, linking the failure of the method to the appearance of the spurious zero and negative modes in the {\it truncated} approximation of the double moments matrix $Q_{\theta}$. Third, we have performed analysis based on inverse single and double moments. In particular, we have  pointed out that in several cases the inverse single moments can be used to perform eigen-inference as well as the standard moments, whereas inverse double moments inherit the above-mentioned pathology in even more pronounced way.   
In this paper, we have not performed the comparison between the errors  of analytical method based on our conformal mapping (31)  and  the popular and powerful method of so-called G-estimators, proposed originally by Girko~\cite{GIRKO} and widely used e.g. by Mestre et al~\cite{MESTRE}.  Examples are so-called Generalized Likehood Ratio Test (GLRT)   or Frobenius test.   From preliminary numerical studies done by us, we got relatively similar results for the eigen-inference, with slight advantage of the G-estimators method.  It is not puzzling, since the conformal mapping we use~\cite{JUREKZDZICH,OURFIN}  is closely related  G-estimators. 
Simple  comparison is however not easy,    since G-estimator method~\cite{MESTRE} requires the knowledge of probabilities $p_i$ and infers the values on the unknown eigenvalues only, whereas analytic method infers  both  sets  of values of unknown probabilities  and spectrum. 
G-method gives good results in the case when one of the eigenvalues is "spiked" with a very low (known) probability $p_i \sim 1/N$.  
Analytic method assumes that all probabilities are of  the same order, so again a direct comparison is not justified. Taking into account the importance of the "spiked" events, we plan to extend   our analysis in the future for the case of unusual $N$ scaling of both probabilities and eigenvalues, and to present the results of such  analysis in future publication. 

\section{Appendices}
\subsection{Conformal mapping}
Let us consider the case when the true covariance matrix is given by the unknown, multidimensional, complex correlated Gaussian distribution 
\be
P(X)&=& (\pi)^{-NT} (\det B)^{-T} (\det A)^{-N}  \nonumber \\ 
 && e^{-  \sum_{i,j=1}^N \sum_{a,b=1}^T X_{ia}[B^{-1}]_{ij} X_{jb}^{*} [A^{-1}]_{ba}    }
\ee
where the true matrices $A$ and $B$ are unknown. Standard procedure relies on approximating them by the Pearson estimators,  built from  empirical data $b=\frac{1}{T} XX^{\dagger}$
and $c=\frac{1}{N} X^{\dagger}X$.  Introducing functional inverse of the generating function $M(z)$, i.e. the function $N(z)$ such that $N[M(z)]=M[N(z)]=z$, and using the theory of free random variables~\cite{VOICULESCU} (valid in the limit when dimensions $N, T$ tend to infinity  while the ratio is $r=N/T$  is kept fixed), one can reduce the problem of inference to surprisingly simple relation~\cite{OURFIN}
\be
N_c(z)=rzN_A(rz)N_B(z)
\label{inference}
\ee
or,  equivalently, after substitution $z \rightarrow M_c(z)$, 
\be
z=rM_c(z) N_A(rM_c(z))N_B(M_c(z))
\label{explicit}
\ee
The formula~(\ref{conformal}),  representing  conformal mapping between the $z$ complex plane and $Z$ complex plane,  originally derived by  diagrammatical methods~\cite{JUREKZDZICH}, is a special case  of last  relation, corresponding to  the case when 
$A={\bf 1}_T$, $c=S$ and $B=\Sigma$, since in this case  $N_A(z)=1+1/z$.  Similar mapping appears also as the heart of the G-estimator method. 
\subsection{Tables of double and double dual  moments of $\Sigma$ rephrased in terms of moments of $S$} 
\noindent
Single and dual moments can be easily calculated using the conformal mappings. The authors are willing to provide appropriate symbolic codes.
Calculation of double moments is more involved,  so we 
 list the Table of double  moments up to $\alpha_{55}$. 
 The values of all coefficients agree with~\cite{RAO,MIT}. 
 For completeness, we list also the double dial moments  up to $\ta_{55}$. As far as we know, this result was never published.  Alike  in the case of single moments, we are willing to provide appropriate symbolic codes for double moments as well.

\subsection{Pad\'{e} approximants}
Pad\'{e} approximant  of function $G(x)$  of order $[m/n]$, denoted  usually as $[m/n]_G(x)$, represents  the "optimal" approximation of the unknown function $G(x)$ by the ratio of two polynomials $A(x)$  and $B(x)$, of orders $m$ and $n$, correspondingly. By construction, Pad\'{e} approximant  agrees with $G(x)$ to the highest possible order, i.e. to $n+m$ term in Taylor expansion of $G(x)$.  It is perfectly suited for numerical analysis, since it works even in the case when the convergence of the Taylor series is difficult to hold. 
On may therefore say, that the Pad\'{e} approximate is the optimal deterministic G-estimator.
In our case, after simple change of variables in the Green's function $G(z)$, by the nature of the resolvent we have an approximant 
 $G(z) \approx  [K_{max}-1, K_{max}]_G(x=1/z)$.  Since fast Pad\'{e}  algorithms are incorporated into  several  standard numerical packages,  "Pad\'{e}ization" of the calculation speeds up considerably the eigen-inference. 

\section{Acknowledgments}
 The authors appreciate discussions with Roland Speicher and Xavier Mestre. 
MAN is   supported by the Grant DEC-2011/02/A/ST1/00119 of the National Centre of Science.
JJ is supported by the       Grant  DEC-2012/06/A/ST2/00389  of the National Centre of Science.



\begin{table*}[ht]
\caption{Double moments.}
\centering
\begin{tabularx}{17.8cm}{l l}
\hline
$\alpha_{1,1}=$&$-\alpha_1{}^2+\alpha_2$\\
\hline
$\alpha_{1.2}=$&$\begin{array}{l} 2 (\alpha _1^3-  2 \alpha _2 \alpha _1+\alpha _3)   \end{array} $             \\
\hline
$\alpha_{1.3}=$&$\begin{array}{l} -3 (\alpha _1^4-3 \alpha _2 \alpha _1^2+2 \alpha _3 \alpha _1+\alpha _2^2-\alpha _4)   \end{array} $             \\
\hline
$\alpha_{1.4}=$&$\begin{array}{l} 4 (\alpha _1^5-4 \alpha _2 \alpha _1^3+3 \alpha _3 \alpha _1^2+(3 \alpha _2^2-2 \alpha _4) \alpha _1-2 \alpha _2 \alpha _3+\alpha _5)  \end{array} $             \\
\hline
$\alpha_{1.5}=$&$\begin{array}{l}  -5 (\alpha _1^6-5 \alpha _2 \alpha _1^4+4 \alpha _3 \alpha _1^3+(6 \alpha _2^2-3 \alpha _4) \alpha _1^2+(2 \alpha _5-6 \alpha _2 \alpha _3) \alpha _1-\alpha _2^3+\alpha _3^2+2 \alpha _2 \alpha _4-\alpha _6)  \end{array} $             \\
\hline
$\alpha_{2,2}=$&$\begin{array}{l} -6 \alpha _1^4+16 \alpha _2 \alpha _1^2-8 \alpha _3 \alpha _1-6 \alpha _2^2+4 \alpha _4   \end{array} $             \\
\hline
$\alpha_{2.3}=$&$\begin{array}{l} 6 (2 \alpha _1^5-7 \alpha _2 \alpha _1^3+4 \alpha _3 \alpha _1^2+(5 \alpha _2^2-2 \alpha _4) \alpha _1-3 \alpha _2 \alpha _3+\alpha _5)  \end{array} $             \\
\hline
$\alpha_{2.4}=$&$\begin{array}{l}  -4 (5 \alpha _1^6-22 \alpha _2 \alpha _1^4+14 \alpha _3 \alpha _1^3+8 (3 \alpha _2^2-\alpha _4) \alpha _1^2+4 (\alpha _5-5 \alpha _2 \alpha _3) \alpha _1-4 \alpha _2^3+3 \alpha _3^2+6 \alpha _2 \alpha _4-2 \alpha _6)  \end{array} $             \\
\hline
$\alpha_{2.5}=$&$\begin{array}{l} 10 (3 \alpha _1^7-16 \alpha _2 \alpha _1^5+11 \alpha _3 \alpha _1^4+(24 \alpha _2^2-7 \alpha _4) \alpha _1^3+4 (\alpha _5-6 \alpha _2 \alpha _3) \alpha _1^2+(-9 \alpha _2^3+10 \alpha _4 \alpha _2+5 \alpha _3^2-2 \alpha _6) \alpha _1+6 \alpha _2^2 \alpha _3 \\ -3 \alpha _3 \alpha _4-3 \alpha _2 \alpha _5+\alpha _7)   \end{array} $             \\
\hline
$\alpha_{3.3}=$&$\begin{array}{l} -3 (10 \alpha _1^6-42 \alpha _2 \alpha _1^4+24 \alpha _3 \alpha _1^3+3 (15 \alpha _2^2-4 \alpha _4) \alpha _1^2+6 (\alpha _5-6 \alpha _2 \alpha _3) \alpha _1-7 \alpha _2^3+6 \alpha _3^2+9 \alpha _2 \alpha _4-3 \alpha _6)   \end{array} $             \\
\hline
$\alpha_{3,4}=$&$\begin{array}{l}12 (5 \alpha _1^7-25 \alpha _2 \alpha _1^5+15 \alpha _3 \alpha _1^4+4 (9 \alpha _2^2-2 \alpha _4) \alpha _1^3+(4 \alpha _5-33 \alpha _2 \alpha _3) \alpha _1^2+(-13 \alpha _2^3+12 \alpha _4 \alpha _2+7 \alpha _3^2-2 \alpha _6) \alpha _1+8 \alpha _2^2 \alpha _3 \\ -4 \alpha _3 \alpha _4-3 \alpha _2 \alpha _5+\alpha _7)   \end{array} $             \\
\hline
$\alpha_{3.5}=$&$\begin{array}{l} -15 (7 \alpha _1^8-41 \alpha _2 \alpha _1^6+26 \alpha _3 \alpha _1^5+15 (5 \alpha _2^2-\alpha _4) \alpha _1^4+(8 \alpha _5-76 \alpha _2 \alpha _3) \alpha _1^3+(-44 \alpha _2^3+33 \alpha _4 \alpha _2+18 \alpha _3^2-4 \alpha _6) \alpha _1^2 \\ +2 (21 \alpha _3 \alpha _2^2-6 \alpha _5 \alpha _2-7 \alpha _3 \alpha _4+\alpha _7) \alpha _1+4 \alpha _2^4+2 \alpha _4^2-8 \alpha _2^2 \alpha _4+4 \alpha _3 \alpha _5+\alpha _2 (3 \alpha _6-9 \alpha _3^2)-\alpha _8)  \end{array} $             \\
\hline
$\alpha_{4.4}=$&$\begin{array}{l}  -4 (35 \alpha _1^8-200 \alpha _2 \alpha _1^6+120 \alpha _3 \alpha _1^5+8 (45 \alpha _2^2-8 \alpha _4) \alpha _1^4+32 (\alpha _5-11 \alpha _2 \alpha _3) \alpha _1^3-4 (52 \alpha _2^3-36 \alpha _4 \alpha _2-21 \alpha _3^2+4 \alpha _6) \alpha _1^2 \\ +8 (24 \alpha _3 \alpha _2^2-6 \alpha _5 \alpha _2-8 \alpha _3 \alpha _4+\alpha _7) \alpha _1+19 \alpha _2^4-40 \alpha _2 \alpha _3^2+10 \alpha _4^2-36 \alpha _2^2 \alpha _4+16 \alpha _3 \alpha _5+12 \alpha _2 \alpha _6-4 \alpha _8)  \end{array} $             \\
\hline
$\alpha_{4.5}=$&$\begin{array}{l} 20 (14 \alpha _1^9-91 \alpha _2 \alpha _1^7+56 \alpha _3 \alpha _1^6+(198 \alpha _2^2-31 \alpha _4) \alpha _1^5+(16 \alpha _5-205 \alpha _2 \alpha _3) \alpha _1^4+(-160 \alpha _2^3+92 \alpha _4 \alpha _2+52 \alpha _3^2-8 \alpha _6) \alpha _1^3 \\ +(180 \alpha _3 \alpha _2^2-36 \alpha _5 \alpha _2-45 \alpha _3 \alpha _4+4 \alpha _7) \alpha _1^2+(35 \alpha _2^4-51 \alpha _4 \alpha _2^2+(12 \alpha _6-57 \alpha _3^2) \alpha _2+9 \alpha _4^2+16 \alpha _3 \alpha _5-2 \alpha _8) \alpha _1+4 \alpha _3^3 \\ -22 \alpha _2^3 \alpha _3+9 \alpha _2^2 \alpha _5-5 \alpha _4 \alpha _5-4 \alpha _3 \alpha _6+\alpha _2 (22 \alpha _3 \alpha _4-3 \alpha _7)+\alpha _9)   \end{array} $             \\
\hline
$\alpha_{5.5}=$&$\begin{array}{l} -5 (126 \alpha _1^{10}-910 \alpha _2 \alpha _1^8+560 \alpha _3 \alpha _1^7+10 (231 \alpha _2^2-31 \alpha _4) \alpha _1^6-20 (123 \alpha _2 \alpha _3-8 \alpha _5) \alpha _1^5-10 (240 \alpha _2^3-115 \alpha _4 \alpha _2-65 \alpha _3^2 \\ +8 \alpha _6) \alpha _1^4+40 (75 \alpha _3 \alpha _2^2-12 \alpha _5 \alpha _2-15 \alpha _3 \alpha _4+\alpha _7) \alpha _1^3+5 (175 \alpha _2^4-204 \alpha _4 \alpha _2^2+(36 \alpha _6-228 \alpha _3^2) \alpha _2+27 \alpha _4^2+48 \alpha _3 \alpha _5 \\ -4 \alpha _8) \alpha _1^2-10 (88 \alpha _3 \alpha _2^3-27 \alpha _5 \alpha _2^2+(6 \alpha _7-66 \alpha _3 \alpha _4) \alpha _2-12 \alpha _3^3+10 \alpha _4 \alpha _5+8 \alpha _3 \alpha _6-\alpha _9) \alpha _1-51 \alpha _2^5+125 \alpha _2^3 \alpha _4 \\ +15 \alpha _2^2 (14 \alpha _3^2-3 \alpha _6)-5 \alpha _2 (13 \alpha _4^2+24 \alpha _3 \alpha _5-3 \alpha _8)-5 (14 \alpha _4 \alpha _3^2-4 \alpha _7 \alpha _3-3 \alpha _5^2-5 \alpha _4 \alpha _6+\alpha _{10}))  \end{array} $             \\
\hline
\end{tabularx}
\end{table*}

\begin{table*}[]
\caption{Double dual moments.}
\centering
\begin{tabularx}{17.8cm}{l l}
\hline
$\ta_{1,1}=$&$(\ta_2 \ta_4-\ta_3^2)/\ta_2^2 $\\
\hline
$\ta_{1.2}=$&$\begin{array}{l} 2 (\ta_3^3-2 \ta_2 \ta_4 \ta_3+\ta_2^2 \ta_5)/\ta_2^3   \end{array} $             \\
\hline
$\ta_{1.3}=$&$\begin{array}{l} -3 (\ta_3^4-3 \ta_2 \ta_4 \ta_3^2+2 \ta_2^2 \ta_5 \ta_3  +\ta_2^2 (\ta_4^2-\ta_2 \ta_6))/\ta_2^4   \end{array} $             \\
\hline
$\ta_{1.4}=$&$\begin{array}{l} 4 (\ta_3^5-4 \ta_2 \ta_4 \ta_3^3+3 \ta_2^2 \ta_5 \ta_3^2+\ta_2^2 (3 \ta_4^2-2 \ta_2 \ta_6) \ta_3+\ta_2^3 (\ta_2 \ta_7-2 \ta_4 \ta_5))/\ta_2^5  \end{array} $             \\
\hline
$\ta_{1.5}=$&$\begin{array}{l} - 5 (\ta_3^6-5 \ta_2 \ta_4 \ta_3^4+4 \ta_2^2 \ta_5 \ta_3^3-3 \ta_2^2 (\ta_2 \ta_6-2 \ta_4^2) \ta_3^2  +2 \ta_2^3 (\ta_2 \ta_7-3 \ta_4 \ta_5) \ta_3-\ta_2^3 (\ta_4^3-2 \ta_2 \ta_6 \ta_4 \\ +\ta_2 (\ta_2 \ta_8-\ta_5^2)))/\ta_2^6  \end{array} $             \\
\hline
$\ta_{2,2}=$&$\begin{array}{l} 2 (-3 \ta_3^4+8 \ta_2 \ta_4 \ta_3^2-4 \ta_2^2 \ta_5 \ta_3+\ta_2^2 (2 \ta_2 \ta_6-3 \ta_4^2))/\ta_2^4   \end{array} $             \\
\hline
$\ta_{2.3}=$&$\begin{array}{l} 6 (2 \ta_3^5-7 \ta_2 \ta_4 \ta_3^3+4 \ta_2^2 \ta_5 \ta_3^2+\ta_2^2 (5 \ta_4^2-2 \ta_2 \ta_6) \ta_3  +\ta_2^3 (\ta_2 \ta_7-3 \ta_4 \ta_5))/\ta_2^5  \end{array} $             \\
\hline
$\ta_{2.4}=$&$\begin{array}{l} -4 (5 \ta_3^6-22 \ta_2 \ta_4 \ta_3^4+14 \ta_2^2 \ta_5 \ta_3^3-8 \ta_2^2 (\ta_2 \ta_6-3 \ta_4^2) \ta_3^2+4 \ta_2^3 (\ta_2 \ta_7-5 \ta_4 \ta_5) \ta_3+\ta_2^3 (-4 \ta_4^3+6 \ta_2 \ta_6 \ta_4 \\ +\ta_2 (3 \ta_5^2-2 \ta_2 \ta_8)))/\ta_2^6   \end{array} $             \\
\hline
$\ta_{2.5}=$&$\begin{array}{l} 10 (3 \ta_3^7-16 \ta_2 \ta_4 \ta_3^5+11 \ta_2^2 \ta_5 \ta_3^4+\ta_2^2 (24 \ta_4^2-7 \ta_2 \ta_6) \ta_3^3+4 \ta_2^3 (\ta_2 \ta_7-6 \ta_4 \ta_5) \ta_3^2+\ta_2^3 (-9 \ta_4^3+10 \ta_2 \ta_6 \ta_4 \\ + \ta_2 (5 \ta_5^2-2 \ta_2 \ta_8)) \ta_3+\ta_2^4 (\ta_5 (6 \ta_4^2-3 \ta_2 \ta_6)  +\ta_2 (\ta_2 \ta_9-3 \ta_4 \ta_7)))/\ta_2^7   \end{array} $             \\
\hline
$\ta_{3.3}=$&$\begin{array}{l} -3 (10 \ta_3^6-42 \ta_2 \ta_4 \ta_3^4+24 \ta_2^2 \ta_5 \ta_3^3+3 \ta_2^2 (15 \ta_4^2-4 \ta_2 \ta_6) \ta_3^2+6 \ta_2^3 (\ta_2 \ta_7-6 \ta_4 \ta_5) \ta_3+\ta_2^3 (-7 \ta_4^3+9 \ta_2 \ta_6 \ta_4 \\ +6 \ta_2 \ta_5^2-3 \ta_2^2 \ta_8))/\ta_2^6   \end{array} $             \\
\hline
$\ta_{3,4}=$&$\begin{array}{l} 12 (5 \ta_3^7-25 \ta_2 \ta_4 \ta_3^5+15 \ta_2^2 \ta_5 \ta_3^4+4 \ta_2^2 (9 \ta_4^2-2 \ta_2 \ta_6) \ta_3^3+\ta_2^3 (4 \ta_2 \ta_7-33 \ta_4 \ta_5) \ta_3^2+\ta_2^3 (-13 \ta_4^3+12 \ta_2 \ta_6 \ta_4 \\ +\ta_2 (7 \ta_5^2-2 \ta_2 \ta_8)) \ta_3+\ta_2^4 (\ta_5 (8 \ta_4^2-4 \ta_2 \ta_6)+\ta_2 (\ta_2 \ta_9-3 \ta_4 \ta_7)))/\ta_2^7  \end{array} $             \\
\hline
$\ta_{3.5}=$&$\begin{array}{l} -15 (7 \ta_3^8-41 \ta_2 \ta_4 \ta_3^6+26 \ta_2^2 \ta_5 \ta_3^5-15 \ta_2^2 (\ta_2 \ta_6-5 \ta_4^2) \ta_3^4+(8 \ta_2^4 \ta_7-76 \ta_2^3 \ta_4 \ta_5) \ta_3^3+\ta_2^3 (-44 \ta_4^3+33 \ta_2 \ta_6 \ta_4 \\ +2 \ta_2 (9 \ta_5^2-2 \ta_2 \ta_8)) \ta_3^2+2 \ta_2^4 (7 \ta_5 (3 \ta_4^2-\ta_2 \ta_6)+\ta_2 (\ta_2 \ta_9-6 \ta_4 \ta_7)) \ta_3+\ta_2^4 (4 \ta_4^4-8 \ta_2 \ta_6 \ta_4^2+3 \ta_2 (\ta_2 \ta_8 \\ -3 \ta_5^2) \ta_4+\ta_2^2 (2 \ta_6^2+4 \ta_5 \ta_7-\ta_2 \ta_{10})))/\ta_2^8   \end{array} $             \\
\hline
$\ta_{4.4}=$&$\begin{array}{l}  -4 (35 \ta_3^8-200 \ta_2 \ta_4 \ta_3^6+120 \ta_2^2 \ta_5 \ta_3^5+8 \ta_2^2 (45 \ta_4^2-8 \ta_2 \ta_6) \ta_3^4+32 \ta_2^3 (\ta_2 \ta_7-11 \ta_4 \ta_5) \ta_3^3-4 \ta_2^3 (52 \ta_4^3-36 \ta_2 \ta_6 \ta_4 \\ +\ta_2 (4 \ta_2 \ta_8-21 \ta_5^2)) \ta_3^2+8 \ta_2^4 (8 \ta_5 (3 \ta_4^2-\ta_2 \ta_6)+\ta_2 (\ta_2 \ta_9-6 \ta_4 \ta_7)) \ta_3+\ta_2^4 (19 \ta_4^4-36 \ta_2 \ta_6 \ta_4^2+4 \ta_2 (3 \ta_2 \ta_8 \\ -10 \ta_5^2) \ta_4+2 \ta_2^2 (5 \ta_6^2+8 \ta_5 \ta_7-2 \ta_2 \ta_{10})))/\ta_2^8  \end{array} $             \\
\hline
$\ta_{4.5}=$&$\begin{array}{l} 20 (14 \ta_3^9-91 \ta_2 \ta_4 \ta_3^7+56 \ta_2^2 \ta_5 \ta_3^6+\ta_2^2 (198 \ta_4^2-31 \ta_2 \ta_6) \ta_3^5+\ta_2^3 (16 \ta_2 \ta_7-205 \ta_4 \ta_5) \ta_3^4+4 \ta_2^3 (-40 \ta_4^3+23 \ta_2 \ta_6 \ta_4 \\ +\ta_2 (13 \ta_5^2-2 \ta_2 \ta_8)) \ta_3^3+\ta_2^4 (45 \ta_5 (4 \ta_4^2-\ta_2 \ta_6)+4 \ta_2 (\ta_2 \ta_9-9 \ta_4 \ta_7)) \ta_3^2+\ta_2^4 (35 \ta_4^4-51 \ta_2 \ta_6 \ta_4^2+3 \ta_2 (4 \ta_2 \ta_8 \\ -19 \ta_5^2) \ta_4+\ta_2^2 (9 \ta_6^2+16 \ta_5 \ta_7-2 \ta_2 \ta_{10})) \ta_3+\ta_2^5 (4 \ta_2 \ta_5^3+(-22 \ta_4^3+22 \ta_2 \ta_6 \ta_4-4 \ta_2^2 \ta_8) \ta_5+\ta_2 ((9 \ta_4^2-5 \ta_2 \ta_6) \ta_7 \\ +\ta_2 (\ta_2 \ta_{11}-3 \ta_4 \ta_9))))/\ta_2^9   \end{array} $             \\
\hline
$\ta_{5.5}=$&$\begin{array}{l} -5 (126 \ta_3^{10}-910 \ta_2 \ta_4 \ta_3^8+560 \ta_2^2 \ta_5 \ta_3^7+10 \ta_2^2 (231 \ta_4^2-31 \ta_2 \ta_6) \ta_3^6+20 \ta_2^3 (8 \ta_2 \ta_7-123 \ta_4 \ta_5) \ta_3^5-10 \ta_2^3 (240 \ta_4^3 \\ -115 \ta_2 \ta_6 \ta_4+\ta_2 (8 \ta_2 \ta_8-65 \ta_5^2)) \ta_3^4+40 \ta_2^4 (15 \ta_5 (5 \ta_4^2-\ta_2 \ta_6)+\ta_2 (\ta_2 \ta_9-12 \ta_4 \ta_7)) \ta_3^3+5 \ta_2^4 (175 \ta_4^4 \\ -204 \ta_2 \ta_6 \ta_4^2+12 \ta_2 (3 \ta_2 \ta_8  -19 \ta_5^2) \ta_4+\ta_2^2 (27 \ta_6^2+48 \ta_5 \ta_7-4 \ta_2 \ta_{10})) \ta_3^2+10 \ta_2^5 (12 \ta_2 \ta_5^3+(-88 \ta_4^3+66 \ta_2 \ta_6 \ta_4 \\ -8 \ta_2^2 \ta_8) \ta_5+\ta_2 ((27 \ta_4^2-10 \ta_2 \ta_6) \ta_7+\ta_2 (\ta_2 \ta_{11}-6 \ta_4 \ta_9))) \ta_3+\ta_2^5 (-51 \ta_4^5+125 \ta_2 \ta_6 \ta_4^3+15 \ta_2 (14 \ta_5^2-3 \ta_2 \ta_8) \ta_4^2 \\ +5 \ta_2^2 (-13 \ta_6^2-24 \ta_5 \ta_7+3 \ta_2 \ta_{10}) \ta_4+5 \ta_2^2 (\ta_6 (5 \ta_2 \ta_8-14 \ta_5^2)+\ta_2 (3 \ta_7^2+4 \ta_5 \ta_9-\ta_2 \ta_{12}))))/\ta_2^{10}   \end{array} $             \\
\hline
\end{tabularx}
\end{table*}


\begin{thebibliography}{99}

\bibitem{WISHART}
J. Wishart, Biometrika {\bf 20A}, 32 (1928). 
\bibitem{CHAOTIC}
Y. Fyodorov and H.J. Sommers, J. Math. Phys. {\bf 38}, 1918 (1997) and references therein. 
\bibitem{MESOSCOPIC}
C.V. J. Beenakker, Rev. Mod. Phys. {\bf 69}, 731 (1997) and references therein. 
\bibitem{RANEY}
P. J. Forrester and D.-Z. Liu, {\it Raney distributions and random matrix theory}, arXiv 1404.5759 and references therein. 
\bibitem{CHIRAL}
J.J.M. Verbaarschot, in {\it Oxford Handbook of Random  Matrix Theory}, Oxford Handbooks in Mathematics, edited by G. Akemann, J. Baik and Ph. Di Francesco (Oxford University Press, USA, 2011) and references therein. 
\bibitem{VOICULESCU}
D.V. Voiculescu, K.J. Dykema and A. Nica, {\it Free random variables}, CRM Monograph Series, Vol. 1 (American Mathematical Society, Providence, 1992).  
\bibitem{BOUCHAUD}
 L. Laloux, P. Cizeau, J.-P. Bouchaud and M. Potters, Phys. Rev. Lett. 83, 1467 (1999).
\bibitem{STANLEY}
 V. Plerou, P. Gopikrishnan, B. Rosenow, L.A.N. Amaral, and H.E. Stanley, Phys. Rev. Lett. 83, 1471 (1999).
\bibitem{LIVAN}
G. Livan and L. Rebecchi, European Phys. J. {\bf B85}, 1 (2012).
\bibitem{BOUCHAUDLACES}
J.-P. Bouchaud and M. Potters, {\it Theory of Financial Risks} (Cambridge University Press, Cambridge, 2001). 
\bibitem{OURFIN}
Z. Burda et al, Quantitative Finance {\bf 11(7)}, 1103 (2011). 
\bibitem{FOSCHINI}
G.J. Foschini, Bell Labs tech. J. {\bf 1}  41 (1996)
\bibitem{TELATAR}
E. Telatar, Eur. Trans. Telecommun. {\bf 10} 585 (1999). 
\bibitem{DEBBAHBOOK}
R. Couillet and M. Debbah, {\it Random Matrix Methods for Wireless telecommunications} (Cambridge University Press, Cambridge, 2011). 
\bibitem{TULINOBOOK}
A. Tulino and  S. Verdu, Random Matrix Theory and Wireless Communications, Foundations and
Trends in Communication and Information Theory, 1, 1-182 (2004)
\bibitem{MITPAPER}
V. Dahirel et al., Proc. Natl. Acad. Sci. USA {\bf 108}, 11530 (2011). 
\bibitem{BAISIL}
Z.D. Bai, J. Silverstein,  Ann. Prob.  {\bf 32} (2004) 533. 
\bibitem{MP}
V. A. Marcenko and L. A. Pastur, Math. USSR-Sb, 1, 457-483 (1967)
\bibitem{NUMEROUS}
Zh. Bai and J.W. Silverstein, {\it Spectral Analysis of large Dimensional Random Matrices}(Science Press, Beijing, 2011) and references therein;
N. El Karoui, Annals of Statistics {\bf 36 (6)}, 2757 (2008).  
\bibitem{RAO}
N. Raj Rao, Applied Stochastic Eigen-Analysis, Ph.D. Thesis, MIT 2007. 
\bibitem{MIT}
N. Raj Rao, J. Mingo, R. Speicher and A. Edelman, Annals of Statistics {\bf 36 (6)}, 2850 (2008). 
\bibitem{MESTRE}
X. Mestre, IEEE Transactions on Informatiuon Theory, {\bf 50(11)}, 5113 (2008). 
\bibitem{GIRKO}
V.L.Girko, {\it An Introduction to Statistical Analysis of Random Arrays} (VSP Science, 1998).
\bibitem{PHDLUKA}
G. \L{}ukaszewski, Ph.D. Thesis, Jagiellonian University, 2011 (unpublished).
\bibitem{DROGOSZ}
Z. Drogosz,  yearly scientific paper,  Jagiellonian University, 2014 (unpublished).
\bibitem{JUREKZDZICH}
Z. Burda, M.A. Nowak and J. Jurkiewicz, Acta Phys. Pol. {\bf B34}, 87 (2003). 
Z. Burda, A. G\"{o}rlich, A. Jarosz and J. Jurkiewicz, Physica A, 343, 295-310 (2004);
Z. Burda and J. Jurkiewicz, Physica A 344, 67 (2004);
Z. Burda, J. Jurkiewicz and B. Waclaw, Phys. Rev. E 71 026111 (2005).
\bibitem{AMBJURMAK}
J. Ambj\o{}rn, J. Jurkiewicz, Y. Makeenko,  Phys. Lett. {\bf  B251} (1990) 517.
\bibitem{MINGO}
B. Collins, J.A. Mingo, P. \'{S}niady, R. Speicher,  Documenta Mathematica {\bf 12} (2007)1.
\bibitem{JURLUKNOW}
J. Jurkiewicz, G. \L{}ukaszewski, M.A. Nowak, Acta Phys. Polon.  {\bf  B39} (2008) 799. 

\end{thebibliography}
\end{document}